\documentclass[aps,prd,superscriptaddress,twocolumn,floatfix,nofootinbib]{revtex4}
\usepackage{graphicx}
\usepackage{natbib}
\usepackage{amssymb}
\usepackage{amsmath}
\usepackage{dcolumn}
\usepackage{rotating}
\usepackage{color}
\usepackage{hyperref}
\usepackage{appendix}
\usepackage{array}
\usepackage[note-name=Note]{notes2bib} 
\newcommand{\nc}{\newcommand}
\newcommand{\ba}{\begin{eqnarray}}
	\newcommand{\ea}{\end{eqnarray}}
\newcommand{\be}{\begin{equation}}
	\newcommand{\ee}{\end{equation}}
\newcommand{\bi}{\begin{itemize}}
	\newcommand{\ei}{\end{itemize}}
\newcommand{\by}{\begin{array}}
	\newcommand{\ey}{\end{array}}
\newcommand{\al}{\alpha}

\newcommand{\ga}{\gamma}
\newcommand{\ta}{\theta}

\newcommand{\da}{\delta}
\newcommand{\la}{\lambda}

\newcommand{\za}{\zeta}
\newcommand{\sa}{\sigma}
\newcommand{\en}{\varepsilon}

\newcommand{\un}{\upsilon}

\newcommand{\Ga}{\Gamma}

\newcommand{\Da}{\Delta}

\newcommand{\La}{\Lambda}
\newcommand{\cA}{{\cal A}}
\newcommand{\cB}{{\cal B}}
\newcommand{\cC}{{\cal C}}
\newcommand{\cD}{{\cal D}}
\newcommand{\cE}{{\cal E}}

\newcommand{\cI}{{\cal I}}
\newcommand{\cK}{{\cal K}}

\newcommand{\cN}{{\cal N}}
\newcommand{\cP}{{\cal P}}

\newcommand{\cS}{{\cal S}}
\newcommand{\cT}{{\cal T}}
\newcommand{\cO}{{\cal O}}
\newcommand{\cU}{{\cal U}}

\newcommand{\e}{\overline}

\newcommand{\w}{\widetilde}
\newcommand{\x}{\ast}

\newcommand{\ra}{\rightarrow}
\newcommand{\Ra}{\Rightarrow}
\newcommand{\im}{\Longleftrightarrow}
\newcommand{\LF}{\left(}
\newcommand{\RF}{\right)}
\newcommand{\LT}{\left[}
\newcommand{\RT}{\right]}

\newcommand{\eh}{\widehat{e}}

\newcommand{\nh}{\widehat{n}}

\newcommand{\yh}{\widehat{y}}

\newcommand{\av}{\vec{a}}
\newcommand{\bv}{\vec{b}}
\newcommand{\cv}{\vec{c}}
\newcommand{\ev}{\vec{e}}

\newcommand{\sv}{\vec{s}}

\newcommand{\uv}{\vec{u}}
\newcommand{\vv}{\vec{v}}
\newcommand{\wv}{\vec{w}}
\newcommand{\xv}{\vec{x}}
\newcommand{\yv}{\vec{y}}
\newcommand{\zv}{\vec{z}}
\newcommand{\env}{\vec{\en}}
\nc{\sav}{\vec{\sa}}
\nc{\envx}{\env^{\;\x}}
\newcommand{\dav}{\vec{\da}}

\newcommand{\muw}{\w{\mu}}
\newcommand{\nuw}{\w{\nu}}

\newcommand{\mub}{\e{\mu}}
\newcommand{\nub}{\e{\nu}}

\nc{\cBb}{\e{\cB}}
\newcommand{\mud}{\dot{\mu}}
\newcommand{\nud}{\dot{\nu}}
\newcommand{\rhod}{\dot{\rho}}

\newcommand{\2}{\frac{1}{2}}

\newcommand{\ssn}{\subsection}
\newcommand{\sssn}{\subsubsection}
\newcommand{\mx}{\mbox}
\newcommand{\mt}{\mathtt}
\newcommand{\mand}{\mx{ and }}
\newcommand{\for}{\mx{ for }}
\newcommand{\where}{\mx{ where }}
\newcommand{\with}{\mx{ with }}

\newcommand{\appr}{\mt{appr}}
\newcommand{\tot}{\mt{tot}}
\nc{\new}{\mt{new}}
\newcommand{\ie}{{\it i.e.}}

\newcommand{\vs}{\vspace{2mm}\\}
\newcommand{\non}{\nonumber\\}

\nc{\xnew}{x^{\mt{new}}}
\nc{\etax}{\eta^{\x}}
\nc{\gx}{g^{\x}}
\newcommand{\Wc}{W_{\mt{cr}}}
\newcommand{\Wm}{W_{\mt{\min}}}
\newcommand{\Qc}{Q_{\mt{cr}}}
\newcommand{\Qm}{Q_{\mt{\min}}}
\newcommand{\wm}{\wv_{\mt{\min}}}
\newcommand{\envw}{\vec{\w{\en}}}
\begin{document}
\title{
Tensor formalism for predicting synaptic connections with ensemble modeling or optimization
}
\author{Tirthabir Biswas}
\email{tirthabir.biswas@northwestern.edu}
\affiliation{{Department of Neurobiology, Northwestern University, Evanston, IL 60208, USA.}}
\affiliation{{Janelia Research Campus, Howard Hughes Medical Institute, Ashburn, VA 20147, USA.}}
\author{Tianzhi Lambus Li}
\affiliation{{Janelia Research Campus, Howard Hughes Medical Institute, Ashburn, VA 20147, USA.}}
\affiliation{{Program in Neuroscience, Harvard Medical School, Boston, MA 02115, USA.}}
\author{Selimzhan Chalyshkan}
\affiliation{{Department of Neurobiology, Northwestern University, Evanston, IL 60208, USA.}}
\affiliation{{Northwestern University Interdepartmental Neuroscience Program, Evanston, IL 60208, USA.}}
\author{Fumi Kubo}
\affiliation{{RIKEN Center for Brain Science, Saitama, 351-0198, Japan.}}
\author{James E. Fitzgerald}
\email{james.fitzgerald@northwestern.edu}
\affiliation{{Department of Neurobiology, Northwestern University, Evanston, IL 60208, USA.}}
\affiliation{{Janelia Research Campus, Howard Hughes Medical Institute, Ashburn, VA 20147, USA.}}
\affiliation{{Department of Physics and Astronomy, Northwestern University, Evanston, IL 60208, USA.}}
\affiliation{{Department of Engineering Sciences and Applied Mathematics, Northwestern University, Evanston, IL 60208, USA.}}
\affiliation{{NSF-Simons National Institute for Theory and Mathematics in Biology, Chicago, IL 60611, USA.}}
\date{\today}
\begin{abstract}

Theoretical neuroscientists often try to understand how the structure of a neural network relates to its function by focusing on structural features that would either follow from optimization or occur consistently across possible implementations. 
Both optimization theories and ensemble modeling approaches have repeatedly proven their worth, and it would simplify theory building considerably if predictions from both theory types could be derived and tested simultaneously. Here we show how tensor formalism from theoretical physics can be used to unify and solve many optimization and ensemble modeling approaches to predicting synaptic connectivity from neuronal responses. 
We specifically focus on analyzing the solution space of synaptic weights that allow a threshold-linear neural network to respond in a prescribed way to a limited number of input conditions. 
For optimization purposes, we compute the synaptic weight vector that minimizes an arbitrary quadratic loss function.
For ensemble modeling, we identify synaptic weight features that occur consistently across all solutions bounded by an arbitrary ellipsoid. 
We derive a common solution to this suite of nonlinear problems by showing how each of them reduces to an equivalent linear problem that can be solved analytically. 
Although identifying the equivalent linear problem is nontrivial, our tensor formalism provides an elegant geometrical perspective that allows us to solve the problem approximately in an analytical way or exactly using numeric methods. 
The final algorithm is applicable to a wide range of interesting neuroscience problems, and the associated geometric insights may carry over to other scientific problems that require constrained optimization. We conclude by applying and testing our ensemble modeling framework to whole-brain recordings of larval zebrafish performing optomotor and optokinetic responses.
\end{abstract}
\maketitle
\section{INTRODUCTION}

Quantitatively linking the structure of a biological neural network to its function is a central aim of neuroscience. 
The network of synaptic connections can be measured directly in small circuits and small brains~\cite{bargmann2013connectome,galili2022connectomics,friedrich2021dense,abbott2020mind}, and one approach is to ask how patterns of synaptic connectivity lead to dynamical patterns of neural activity and behavior~\cite{parmelee2022sequence,curto2019relating, litwin2019constraining}.
However, it remains infeasible to densely reconstruct synaptic connectivity in larger brains, and network reconstruction alone is insufficient to predict the computational and dynamical properties of the biological system~\cite{bargmann2013connectome,lappalainen2023connectome}.
This paper furthers our efforts \cite{Biswas22} on the inverse problem of predicting synaptic connectivity from the network's functional responses ~\cite{friston2003dynamic,schneidman2006weak,pillow2008spatio}. A significant appeal of the inverse problem is that various techniques make large-scale recordings possible in many brains~\cite{abbott2020brain}. A major challenge of this approach is that many different synaptic connections can often underlie the same functional response properties~\cite{huang2016linking,Biswas20, Biswas22}. Thus a  method that can correctly predict synaptic connectivity, or some of their key features, would have great theoretical and practical value.

Two basic ideas have allowed neuroscientists to start to identify structural principles that organize biological neural networks. 
The first idea is that biology is optimized through natural selection and learning to find solutions that are exquisitely accurate and efficient~\cite{barlow1961possible, sterling2015principles, richards2019deep, clark2024optimization}.
This has motivated theorists to derive optimal solutions for various objective functions that could matter biologically, with notable successes including postdictions for the adaptation of functional connection strengths in retina~\cite{atick1990towards}, the anatomical layout of neurons in the worm brain~\cite{chen2006wiring}, and the canonical neural network architecture of olfaction~\cite{wang2021evolving}. 
The second idea is that natural selection and learning don't require biological solutions to be optimal, just good enough, leaving many plausible solutions~\cite{Prinz, gutenkunst2007universally}. 
This has motivated theorists to characterize and study ensembles of solutions that could work, with notable successes including explanations for individual variability in pyloric circuits~\cite{Marder}, the identification of circuit architectures for persistent oculomotor activity~\cite{Fisher}, and a candidate formalization of biological emergence~\cite{transtrum2015perspective}. Which biological neural networks are near optimal, variable, or haphazard is largely unknown.

Breakthrough technologies increasingly provide large-scale datasets that can sharpen our theoretical understanding of neural network structure and function~\cite{Biswas20}.
Whole-brain functional imaging in small organisms like \emph{C. elegans}~\cite{Schrodel}, larval zebrafish~\cite{Ahrens}, and larval \emph{Drosophila}~\cite{Lemon} grants neuroscientists access to the activity of all the neurons that may be relevant for accomplishing a given set of tasks. There are no unobserved neurons, which makes structurally insightful neural network modeling more tractable~\cite{Fisher, Naumann, Biswas22}. Such models could be tested with closed-loop optogenetic stimulation experiments~\cite{grosenick2015closed} or electron-microscopy-based circuit reconstruction~\cite{seung2009reading}. Indeed, the field of connectomics is providing remarkable opportunities for theory testing by producing dense electron microscopy measurements of neural network structure in \emph{C. elegans}~\cite{Varshney}, larval zebrafish~\cite{Hildebrand, svara2022automated}, larval \emph{Drosophila}~\cite{eichler2017complete}, and adult \emph{Drosophila}~\cite{Bock, Scheffer20, dorkenwald2023neuronal}. This connectomic knowledge also provides valuable information on how to constrain neural network models. For instance, one could make robust predictions by combining functional and structural information to find neural networks that function appropriately and lie near the observed connectome~\cite{lappalainen2023connectome,Wanner,mi2021connectome}. 

A unified theoretical framework for neural network optimization and ensemble modeling would allow theorists to best capitalize on this experimental progress. Rather than having to toil through disjoint concepts and separate laborious calculations, a unified framework would allow theorists to rapidly make predictions from many different theories. Testing all of these predictions against experimental data would allow the field to rapidly map the domain of applicability of various optimization theories and ensemble modeling techniques~\cite{Biswas20}. For instance, linear neural network models are often simple enough to calculate entire solution spaces~\cite{atick1990towards,Baldi,hastie2009elements}, and it is straightforward to use this solution space to find common structure across solutions~\cite{Biswas22}, to optimize various quadratic loss functions~\cite{Fisher, Naumann,hastie2009elements}, or to find solutions near empirical constraints~\cite{doi2012efficient}. Recent theoretical progress suggests that it may be possible to generalize some of these results to threshold-linear neural network models~\cite{Biswas22, CurtoP, Morrison}, which improve on the linear model's biological relevance and computational power while retaining significant analytical tractability. This model class neglects many biological details of real neural networks, but some of these details  may not be needed to make accurate predictions~\cite{Billeh,Biswas20,transtrum2014model,lappalainen2023connectome,Wanner}

Here we will present a unified theoretical formalism for calculating many structural predictions with threshold-linear neural networks. As in our earlier work~\cite{Biswas22}, we will focus on the inverse problem of predicting synaptic connectivity from neuronal activity, and throughout we will assume knowledge of steady-state activity patterns for all neurons under a number of conditions that does not exceed the number of neurons. These assumptions are tailored to whole-brain imaging data and allow us to utilize our previously derived solution space. However, the formalism introduced here will allow us to relax downstream assumptions that we previously needed to make structural predictions with ensemble modeling. It will also allow us to address a much broader class of problems. For example, our formalism will allow us to identify the element of the solution space that minimizes an arbitrary quadratic function of the weights. Moreover, our formalism can pinpoint structural features that occur across all solutions that satisfy an arbitrary quadratic bound. Our ability to treat arbitrary quadratic functions opens myriad theoretical possibilities. For instance, the quadratic functions being optimized and/or bounding the solutions can represent the weight norm, the distance to connectomic data, the generalization error, and so forth. 

Our mathematical formalism relies on three key insights. 
First, discerning whether a synaptic weight is non-zero across an elliptically bounded set of weight vectors amounts to finding the smallest ellipse (quadratic cost) that permits the weight to be zero. This maps each ensemble modeling problem onto a related optimization problem. This realization also allows us to make new types of predictions that were not discussed in our previous work~\cite{Biswas22}, such as making predictions about a pattern of synapses (not just a single synaptic connection), or about how a neuron will respond to a new pattern  (active or inactive). Such activity predictions allow us to test our ensemble modeling framework without synapse-resolution anatomical data, as illustrated through an application to the neuronal responses underlying the optokinetic~\cite{kubo2014functional,svara2022automated} and optomotor responses~\cite{Naumann} of larval zebrafish.
Second, each threshold-linear optimization problem reduces to a linear one once we know which inactive neurons are at threshold in the optimal solution. These neurons impose the same equality constraints that they would in a linear neural network; other inactive neurons are automatically sub-threshold.  
Third, the tensor formalism introduced by Einstein for relativistic calculations in theoretical physics~\cite{misner1973gravitation,synge1978tensor,wald2010general} facilitates the analysis of threshold-linear optimization problems. In particular, the metric tensor makes it easy to minimize arbitrary quadratic loss functions in non-orthogonal coordinate systems, and writing synaptic weight vectors in a coordinate system that mixes contravariant and covariant basis vectors allows us to assess whether we've correctly mapped the threshold-linear problem onto its linear counterpart. The tensor formalism also allows us to approximate the exact solution with an interpretable analytic formula that depends on how neural activity correlates within and across patterns.


We begin in Section \ref{sec:toy} by illustrating different predictions from ensemble modeling and constrained optimization in a simple toy problem. In Section \ref{sec:problem}, we then state the problem we want to solve in its most general form. We re-derive the solution space of ~\cite{Biswas22} in the tensor formalism in Section \ref{sec:solution-space}, and Section \ref{sec:linear} uses this formalism to derive results for linear neural networks. We expect tensor analysis to be new to some readers, so Appendix~\ref{sec:Table} compactly summarizes the most important relations and results with easy-to-reference tables. Furthermore, Appendix~\ref{sec:tensors} briefly reviews the tensor formalism and metric spaces, including coordinate and basis transformations, the metric tensor, and induced geometries on subspaces. Section \ref{sec:nonlinear} generalizes the results derived in Section \ref{sec:linear} to the nonlinear case. In Sections~\ref{sec:linear} and \ref{sec:nonlinear}, we progressively solve more complex problems by reducing them to simpler problems. Relatedly, Appendix~\ref{sec:spheretoellipse} shows  how our results and formalism  generalize as we move from optimizing spherical $L^2$ norm to optimizing arbitrary elliptic quadratic functions. Section~\ref{sec:activity} further generalizes our formalism to include additional structural constraints, more complex structural predictions, and  predictions about neuronal activity. Section~\ref{sec:zebrafish} illustrates an application of our formalism to neuronal activity data from larval zebrafish performing optokinetic and optomotor behaviors. Finally, Sections~\ref{sec:discussions} and \ref{Methods} are the Discussion and Methods. Appendices~\ref{sec:average} and~\ref{sec:Qcr} facilitate conceptual interpretations of several key results. Appendix~\ref{app:runtimes} compares the numerical run times of several algorithms discussed in the paper. 

\section{A simple illustrative problem}
\label{sec:toy}
\begin{figure}[!thbp]
	\centering
\includegraphics[width=0.46\textwidth,angle=0]{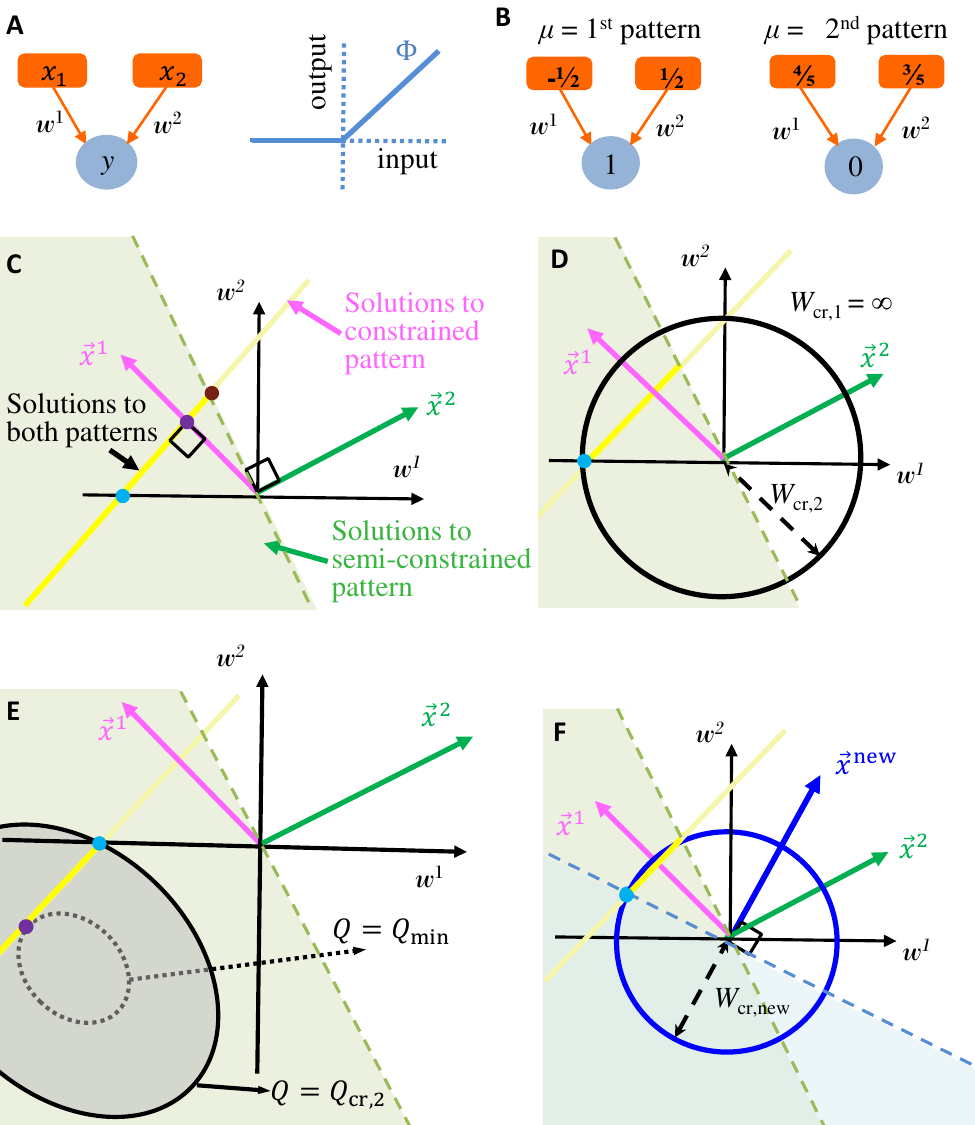}
\caption{{\bf Toy Examples.} {\small ({\bf A}) We consider a simple feedforward network where two input neurons, $x_1$ and $x_2$, drive a target neuron, $y$, via synaptic weights, $w^1$ and $w^2$. We have depicted the threshold-linear transfer function that converts the input drive to the output activity of the target neuron.  ({\bf B}) Here we specify two input-output response patterns in the feedforward network. ({\bf C}) We show the set of weights, deep yellow ray, within the 2D weight space that can  correctly reproduce the prescribed input-output patterns. The pink and green arrows depict the input vectors representing the constrained (first) and semi-constrained (second) patterns, respectively. The brown dot represents the solution one would obtain in a linear theory, the blue dot represents the solution with maximal sparsity, and the purple dot represents the minimum $L^2$ norm solution, which is closest to the origin. ({\bf D}) We illustrate that $w^2$ is consistently positive when the allowed weight vectors are confined within the black circle with radius $W_{\mt{cr},2}$ that passes through the blue dot where $w^2=0$. We would therefore make the experimental prediction that this synapse must exist, and it is excitatory, if the norm of biological weights is bounded above by $W_{\mt{cr},2}$. Irrespective of the weight-norm, the $w^1$ synapse is present and inhibitory in all solutions, $W_{\mt{cr},1}=\infty$.  ({\bf E}) Here we show how we can make similar predictions for not just spherically bounded regions but also when the weight space is bounded by arbitrary ellipses/quadratic functions ($Q(\{w^m\}$). In the depicted figure, as the bounding ellipse grows while preserving its shape and orientation, at some point (purple dot) a solution becomes available. Thereafter, while more solutions become  possible, all solutions have negative $w^2$ as long as the weights are confined within the bold ellipse that passes through the blue dot where $w^2=0$. 
({\bf F}) The light blue region depicts weights that would not lead the postsynaptic neuron to respond to the new input (blue vector). If the weight vectors are confined within the deep blue circle,
none of the deep yellow line segment would lie within this light blue region. 
We could therefore predict that the target neuron will  respond to this new input. 
\label{fig:2dtoy}}}
\end{figure}

To illustrate how neuronal response patterns can be used to obtain  structural predictions, let us look at a simple  toy problem
where a target neuron ($y$) receiving inputs from two neurons ($x_1$ and $x_2$) responds according to
\be
y=\Phi(w^1x_1+w^2x_2)
\ee
(Fig.~\ref{fig:2dtoy}A), where $w^1$ and $w^2$ are the synaptic weights from the input neurons, $x_1$ and $x_2$, onto the target neuron~\footnote{As we progress, it will become clear that quantities associated with upper or lower indices will transform differently under coordinate transformations. This physics-style notation helps both in computing and interpreting our results.}, and $\Phi(s)=\max(s,0)$ is a threshold-linear transfer function that relates the postsynaptic neuron's firing rate to the input drive it receives from the presynaptic neurons. 
Let us consider two input-output patterns, $\{x^\mu{}_m, y^\mu\}$, where $\mu=1,2$ labels the two patterns, $m=1,2$ labels the two input neurons, and
\be
\LF
\begin{array}{cc}
	x^1{}_1 & x^1{}_2\\
	& \\
	x^2{}_1& x^2{}_2
\end{array}\RF\
=\LF
\begin{array}{cc}
	-\frac{1}{2} & \2\\
	& \\
	{4\over 5}& {3\over 5}
\end{array}\RF\ ,\ 
\LF
\begin{array}{c}
y^1\\
\\
y^2
\end{array}
\RF
=\LF
\begin{array}{c}
1\\
\\
0
\end{array}
\RF
\ee
(Fig.~\ref{fig:2dtoy}B). 

We will first review how we can find the solution space, the set of all weight vectors $\wv=(w^1,w^2)$ that generate the aforementioned target responses~\cite{Biswas22}, by illustrating it explicitly for the above input-output transformation (Fig.~\ref{fig:2dtoy}C). We see that since $y$ responds to the first pattern, represented by the input activity pattern vector in pink, the weights must be {\it constrained} to lie on the yellow line, satisfying the equation
\be
x^1{}_1w^1 + x^1{}_2w^2=y^1\Ra -{w^1\over 2}+{w^2\over 2}=1\ .
\ee
On the other hand, the fact that the target neuron doesn't respond to the second pattern, represented by the input activity pattern vector in green, only implies that  the weights have to be {\it semi-constrained}, \ie, have to reside in the light green region  in Fig.~\ref{fig:2dtoy}C, because the thresholding function $\Phi$ sets anything negative to zero:
\be
x^2{}_1w^1 + x^2{}_2w^2\leq 0\Ra {4w^1\over 5}+{3w^2\over 5}\leq 0\ .
\ee
Accordingly, the solution space is given by the deep yellow ray representing the intersection of the yellow line and light green half plane. In contrast, if we had a linear transfer function, $\tilde\Phi(x) = x$, then the weights would be constrained to lie on the dashed green line by the green pattern, and we would have had a unique solution where the yellow and dashed green lines intersect
(brown dot). 

The threshold nonlinearity adds degeneracies. In fact, if we start from the solution space of a linear theory, every time we have a null target response, a semi-constrained dimension will appear in the solution space because of the threshold nonlinearity~\cite{Biswas22}. Despite these degeneracies, certain model features can be consistent across the ensemble. In our toy model,  as can be readily seen from  Fig.~\ref{fig:2dtoy}C, the $w^1$ synapse must be present and negative (\ie, be inhibitory). Moreover,  such degeneracies can be broken if we assume additional biological principles or information. For instance, one may postulate that the $L^2$ weight-norm is minimized, leading to a unique solution marked by the purple dot, $\wv_{L^2}=(-1,1)$. Another popular optimization objective is to minimize the number of synapses, or the $L^0$ norm. The blue dot representing $\wv_{L^0}=(-2,0)$, with only an inhibitory $w^1$ synapse, is the unique solution that minimizes this norm. To summarize, different candidate principles and transfer functions lead to different structural predictions.

Interestingly, even conservative biological constraints, such as the $L^2$ weight-norm being bounded by some  value, $ ||\wv||\leq W$, can lead to additional structural predictions at the ensemble level. In the toy model, as long $W<W_{\mt{cr},2}=2$, we can conclude that the $w^2$ synapse must be present and positive (\ie, be excitatory) (Fig.~\ref{fig:2dtoy}D). For every synapse one can find a $W$-critical value of the weight norm, $\Wc$, below which the synapse is {\it certain} to exist. One can then  also predict its sign. Thus, even in the absence of a known norm bound, $W$, one can compute a sequence of synapses ordered according to their $W$-critical values, thereby ordering them by when they become indispensable for the observed network function with decreasing $W$. In certain special situations,  such as for $w^1$ in this example, one can make very robust predictions that do not depend on $W$. We will indicate that by $W_{\mt{cr},1}=\infty$. 

In Fig.~\ref{fig:2dtoy}E we show how a certain synapse can emerge even if the weight vectors are bounded by an arbitrary ellipse,
\be
Q(w^1,w^2) \equiv {(w^1-f^1)^2\over (h^1)^2}+{(w^1-f^2)^2\over (h^2)^2} = 1\ ,
\ee
where $f^1, f^2, h^1$, and $h^2$  are fixed parameters, or equivalently if the weights are bound by the maximal allowed value of the quadratic function $Q$.
As one can see from the figure, as long as, all the weights lie inside the bold ellipse, $w^2$ must exist and be negative.

Finally, one can also use ensemble modeling to make predictions about neural activity measurements. For instance, one can find a different set of $W$-critical values that allow one to predict whether a given neuron will respond to some arbitrary new pattern or not. In our example, for the new pattern in Fig.~\ref{fig:2dtoy}F (blue arrow), as long as the weights reside within the blue bounding circle, we can predict that $y$ will respond. This is because the circle passes through the intersection of the boundary of the light-blue region and the deep yellow ray of solutions, ensuring that no allowed solutions are in the light-blue region. The projection of the solutions along the new blue pattern vector is therefore always positive. 

In previous work, we developed a geometric formalism to make synapse-level predictions when the input patterns are orthogonal and uncorrelated~\cite{Biswas22}. In this paper we will generalize the geometric formalism to be able to make predictions for realistic patterns, which are typically non-orthogonal and correlated, for both structural connectivity as well as new activity patterns. We will also be able to include a more general class of optimization and ensemble modeling problems.
\section{General Problem}
\label{sec:problem}
Our goal in this paper will be to generalize the identification of structure-function links that we described in the simple 2D toy problem to arbitrary sized networks, neural patterns, and more general optimization and ensemble modeling problems. In previous work~\cite{Biswas22}, we've shown that once one knows how to make such structure-function links in a feedforward network, it also becomes possible to do so in recurrent networks, at least, in the absence of noise. This continues to hold true even as we generalize our framework.
For technical simplicity  we will therefore focus on threshold-linear feedforward networks. 

\begin{figure}[!thbp]
\centering
\includegraphics[width=0.46\textwidth,angle=0]{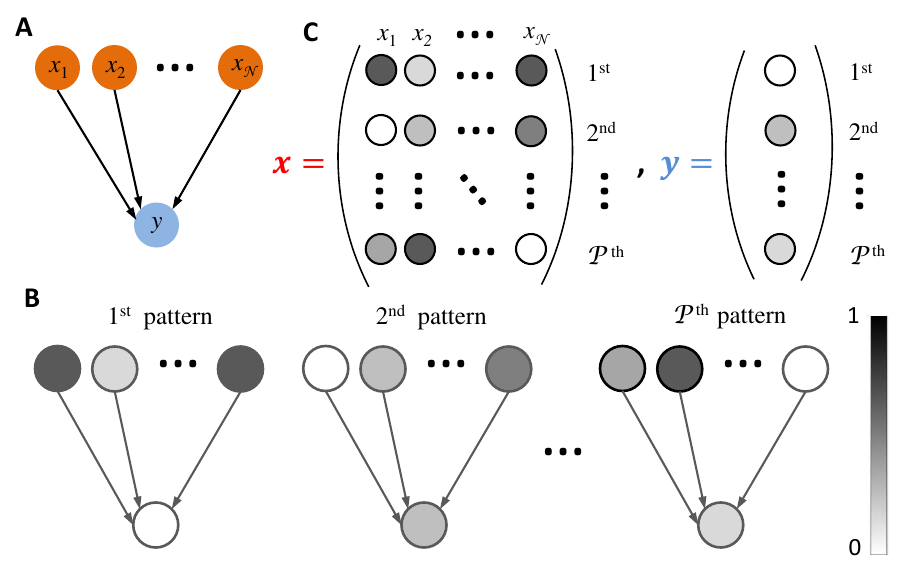}
\caption{{\bf General Problem.} {\small ({\bf A}) We show a  cartoon of a blue target neuron, $y$, being driven by a population of input neurons in red. ({\bf B}) The figure depicts specific input patterns (activities in grey scale) evoking different response in the $y$. ({\bf C}) We specify input-output response patterns that the above network elicits in matrix form. 
\label{fig:formulation}}}
\end{figure}

Let us consider a feedforward threshold-linear network (Fig.~\ref{fig:formulation}A) where the target neuron $y$ responds to its inputs according to 
\begin{align} 
y = \Phi\left( \sum_{m=1}^{\cN} x_{m} w^{m}\right)\ . \label{FF} 
\end{align}
The driven neuron integrates its presynaptic signals weighted by the synaptic weights, $w^m$, and we assume  $m=1,\dots,\cN$. We suppose that the network functionally maps $\cP$ input patterns, \( x^{\mu}{}_m\), to driven signals, \(y^{\mu} \ge 0\), where \(\mu = 1,\cdots,\cP\) labels the patterns (Fig.~\ref{fig:formulation}B). We assume throughout that \(\cP\le\cN\), as this simplifies the problem and the number of known response patterns is typically small. The network weights must therefore satisfy
\begin{align} 
y^{\mu} = \Phi\left( \sum_{m=1}^{\cN}x^{\mu}{}_{ m} w^{m}\RF \label{SteadyStateEqn} \ .
\end{align}
Fig.~\ref{fig:formulation}C depicts the matrix representation of the input and output responses.
Our first goal will be to find all weight vectors that satisfy the above equations. 

Next we will consider a quadratic optimization function,
\be
Q^2(\vec w)=\sum_{m,n=1}^{\cN}q_{mn}w^mw^n+\sum_{m=1}^{\cN}b_mw^m\ ,
\label{quadratic}
\ee
and find the weight vector, $\wv_Q$, in the solution space that minimizes it (purple dot in Fig.~\ref{fig:2dtoy}E). Setting $q_{mn}=\da_{mn}$ and $b_m=0$, one recovers the $L^2$ norm. However, there may be situations where $y$ is receiving inputs from different types of neuronal populations so that it may become more sensible to assign the cost of   different types of synaptic strengths differently leading to an elliptic optimization function. Some problems demand that the weight vector be near some preferred location other than the origin (\emph{e.g.}, a set of measured weights or the weights that minimizes generalization error), or preferentially close in certain directions than others. A general quadratic function such as Eq.~(\ref{quadratic}) can incorporate these features.
Fig.~\ref{fig:2dtoy}E depicts the minimization of such a general quadratic function whose constant surfaces represent off-centered ellipses in the two dimensional  weight-space (ellipsoids in a higher dimensional settting). The picture shows the smallest ellipse that contains a solution, the purple dot.

Finally, for each synapse $p$ we will compute a $Q$-critical, $Q_{\mt{cr},p}$, such that as long as $Q$ is bounded by this value the given synapse is always present in the ensemble of weight vectors that satisfies Eq.~(\ref{SteadyStateEqn}). We will identify the sign of the synapse as well. For the special case, when the quadratic function is just the $L^2$ norm, we will refer to $Q$-critical as $W_{\mt{cr},p}$. In the example of  Fig.~\ref{fig:2dtoy}E, as long as the weight vectors have $Q<Q_{\mt{cr},2}$, \ie, they are within the grey region, we know $w^2$ must be present and be inhibitory. We will also be able to generalize the notion of $Q$-critical by calculating a $Q_{cr,\nh}$, for every new input vector direction, $\nh$, such that if $Q<Q_{cr,\nh}$, we will be able to predict whether the target neuron definitively responds (or definitively doesn't) to this new input pattern. Note that mathematically,  $Q_{\mt{cr},p}=Q_{cr,\eh_p}$, where $\eh_p$ denotes the unit vector along the $p^{th}$ synapse direction in weight space.
\section{Solution Space} 
\label{sec:solution-space}	
\begin{figure*}[!htbp]
	\centering
	\includegraphics[width=0.7\textwidth,angle=0]{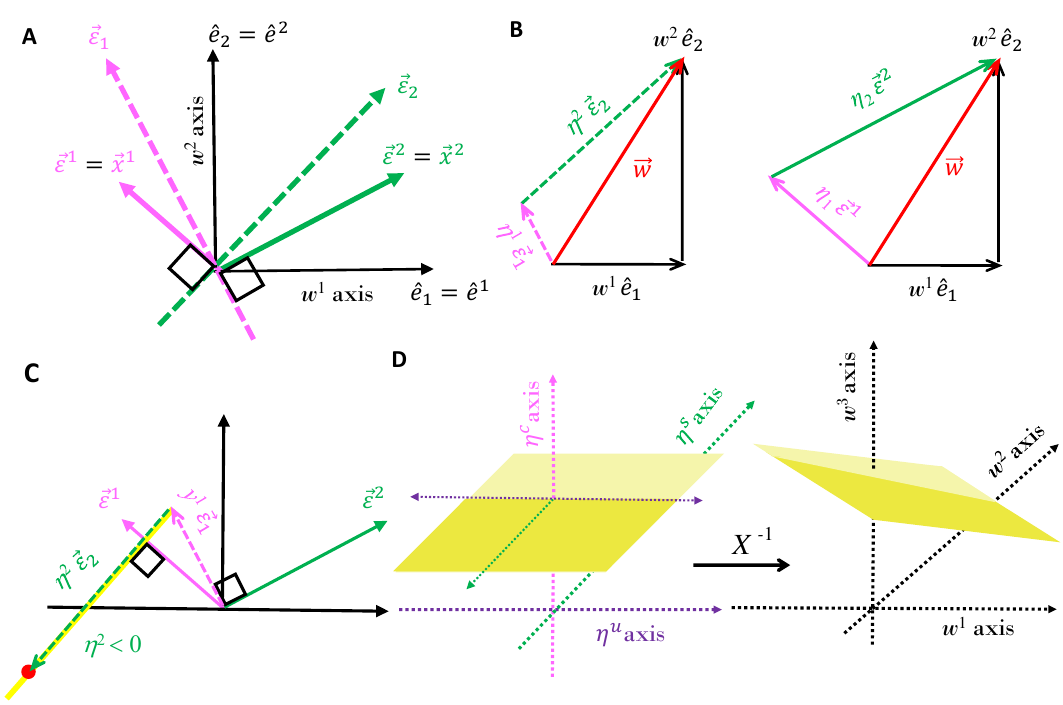}
	\caption{{\bf Coordinate  and basis transformations in weight space to illustrate the solution space from multiple prespectives.} {\small ({\bf A})  We illustrate the orthogonality relationship between the natural (dashed) and complementary (bold) basis vectors within the context of the toy problem discussed in Fig.~\ref{fig:2dtoy}A-D. ({\bf B}) We illustrate how any weight vector in general can be expressed in terms of the synaptic, natural or complementary coordinate system. ({\bf C}) We show several geometric aspects of the solution space (deep yellow line). Firstly, it is perpendicular to the constrained complementary basis vector, $\env^{\,1}$. Secondly, it can be expressed as a linear combination of the natural basis vectors where components along constrained directions are fixed to target neuron's response values, while the components along semi-constrained directions are non-positive.   ({\bf D}) On the left we depict the solution space in terms of natural $\eta$-coordinates. The case depicted here contains one constrained, one semi-constrained, and  one unconstrained dimension. The constrained pattern restricts the solutions to the yellow hyperplane, while the semi-constrained pattern further restricts them to the deep yellow solution space containing all the points that have $\eta^s\leq 0$. On the right we depict how the patch of solution space on the left  transforms into synaptic coordinates.
			\label{fig:solutions}}}
\end{figure*}
Our goal in this section is to find the ensemble of all weight vectors that satisfy Eq.~(\ref{SteadyStateEqn}). We will start by introducing a set of weight combinations, $\eta$'s, that provides a simple and intuitive way to characterize the solution space. In all our discussions the input neuronal response matrix, $x^{\mu}{}_m$, will be assumed to be fixed. This enables us to define $\cP$ number of {\it natural variables} referred by $\eta$, each of which controls the target response to a single stimulus condition:
\be
y^{\mu}=\Phi(\eta^{\mu})\ ,\where\ \eta^{\mu}\equiv \sum_{m = 1}^{\cN}x^{\mu}{}_m w^m\ .
\label{omega-defn}
\ee
Further, we can extend this set of natural variables to a linearly independent set of $\cN$ such variables, so that $w\rightarrow \eta$ now represents a coordinate transformation. To see how this can be done, we will henceforth make the simplifying assumption that the $\cP\times\cN$ matrix has the maximal rank, $\cP$, although we anticipate that much of our framework, results, and insights will apply more generally. If $x$ has maximal rank, then one can define $\cP$ linearly independent vectors,
\be
\env^{\,\mu}\equiv \sum_{m = 1}^{\cN} x^{\mu}{}_{m}\ev^{\,m} = \vec x^{\,\mu}\ ,
\label{env-up}
\ee
(e.g., Fig.~\ref{fig:2dtoy}C-F), where $\{\ev^{\,m}\}$ defines the different synapse directions and forms an orthonormal~\footnote{Later, while solving the most general optimization  problem it will be convenient to introduce a more nontrivial inner product structure where $\{\ev^{\,m}\}$  will no longer be an orthonormal set, but will still remain a basis. The solution space of course does not change if we change what is being optimized.}  set of basis vectors in the weight space. We want to alert the readers that $\{\env^{\,\mu}\}$ is not the basis for the coordinates, $\{\eta^{\mu}\}$, and we will refer to them as the {\it complementary} basis. Their relation will be explained shortly. On the other hand, we will soon see how the response of the target neuron to the various input patterns will constrain the projections of the weight vector along the $\env^{\,\mu}$ directions, thereby providing an intuitive geometric characterization of the solution space.  Also, since $x$ has maximal rank, its kernel will be an \((\cN-\cP)\)-dimensional linear subspace that can be spanned by \((\cN-\cP)\) orthonormal basis vectors, which will be denoted by $\env^{\,\mu}$ with  $\mu=\cP+1,\dots, \cN$. We can now extend $x$  to an $\cN\times\cN$ matrix, $X$, as follows:
\ba
X^{\mu}{}_m&=&x^{\mu}{}_{m}\ \for\ \mu=1,\dots, \cP,\non
X^{\mu}{}_{m}&=&\en^{\mu}{}_{m}\ \ \ \for\ \mu=\cP+1,\dots, \cN,
\ea
where $\en^{\mu}{}_{m}$ is the $m^{th}$ component of the null vector $\env^{\,\mu}$. By construction, $\{\env^{\,\mu},\ \mu=1,\dots,\cN\}$ forms a complete basis for the weight space. The extended response matrix $X$ defines the basis (coordinate) transformation connecting synaptic directions, $\{\ev^{\,m}\}$, and  $\{\env^{\,\mu}\}$ directions,  
\be
\env^{\,\mu}\equiv \sum_{m = 1}^{\cN} X^{\mu}{}_{m}\ev^{\,m} \ \Leftrightarrow\ev^{\,m}=\sum_{\mu = 1}^{\cN}\left(X^{-1}\right)^{m}{}_{\mu}\env^{\,\mu}\ .
\label{complementary-transform}
\ee
Similarly, the natural coordinates transform as
\be
\eta^{\mu}= \sum_{m = 1}^{\cN}X^{\mu}{}_{m}w^m \Leftrightarrow w^m=\sum_{\mu = 1}^{\cN} \left(X^{-1}\right)^{m}{}_{\mu}\eta^{\mu}\ .
\label{coordinatetransformation}
\ee

At this point one can enumerate the entire ensemble of weights that satisfies the input-output transformation,  Eq.~(\ref{omega-defn}), as was already discussed in~\cite{Biswas22}. Every time we have a positive response, the corresponding coordinate must be fixed, but when the target neuron doesn't respond to an input pattern, the corresponding coordinate can take any non-positive value.  Accordingly, we refer to these patterns as {\it constrained} and {\it semi-constrained}, respectively ~\cite{Biswas22}. The remaining natural coordinates from $\mu=\cP+1,\dots,\cN$ can take any value, and we therefore refer to them as {\it unconstrained} coordinates. Thus, if  $\cC, \cS$ and $\cU$ are the set of constrained, semi-constrained, and unconstrained coordinates, respectively, then the solution space is given by
\ba
\eta^{\mu}= y^{\mu} &\for&\ \mu\in \cC,\non
-\infty<\eta^{\mu}\leq 0 &\for&\ \mu\in \cS,\non
-\infty<\eta^{\mu}<\infty & \for&\ \mu\in \cU \ .
\label{solnspace} 
\ea
The corresponding synaptic coordinates can straightforwardly be obtained from Eq.(\ref{solnspace}) via the inverse coordinate transformation prescribed in Eq.~(\ref{coordinatetransformation}). For later convenience we will slightly abuse our notations and use the same letters, $\cC, \cS$ and $\cU$, to also mean the cardinality of each of these sets, so $\cS=\cP-\cC$, and $\cU=\cN-\cP$. For various geometric proofs that follow, it will be essential to be able to represent solutions as vectors in the weight space and to take projections along various directions. For this purpose we need to introduce a new set of basis vectors, $\{\env_{\nu},\ \nu=1,\dots,\cN\}$, that obeys the following orthogonality relations:
\be
\env^{\,\mu}\cdot\env_{\nu}=\da^{\mu}{}_{\nu}\ .
\label{orthogonality}
\ee
Here $\da$ is the Kronecker delta function, which is one when the indices are the same and zero otherwise. We will refer the basis $\{\env_{\nu}\}$ as natural, as they will soon be associated with the natural coordinates. We note that in physics any two set of basis vectors obeying orthogonality relations such as Eq.~(\ref{orthogonality}) are referred to as dual to each other. Thus, the natural and complementary bases are dual. One immediately recognizes that the dual of the synaptic basis vectors are identical to themselves:
\be
\ev_m=\ev^{\,m}\ .
\ee
Any arbitrary synapse vector can now be represented in terms of the synaptic basis, $\{\ev^{\,m}\}$, natural basis,  $\{\env_{\mu}\}$, or complementary basis, $\{\env^{\,\mu}\}$:
\be
\vec{w}=\sum_{m = 1}^{\cN}w^m \ev_{m}=\sum_{\mu = 1}^{\cN}\eta^{\mu} \env_\mu=\sum_{\mu = 1}^{\cN}\eta_{\mu} \env^{\,\mu}\ ,
\label{weight-vector}
\ee
where the components in one basis can be obtained by taking projections along its dual counterparts by virtue of the orthogonality relations Eq.~(\ref{orthogonality}), a major reason why complementary  basis vectors are so useful:
\be
w^m=\ev^{\,m}\cdot\wv\ ;\ \eta^{\mu}=\env^{\,\mu}\cdot\wv\ ;\ \eta_{\mu}=\env_{\mu}\cdot\wv\ .
\ee
Specifically, the ensemble of weight-vectors that solves Eq.~(\ref{SteadyStateEqn}) can now be written straightforwardly in the natural basis,
\be
\vec{w}=\sum_{\mu \in \cC}y^{\mu} \env_\mu+\sum_{\mu \in \cS\cup\cU}\eta^{\mu} \env_\mu,
\ee
where the natural coordinates, $\{\eta^{\mu}\}$, are given by Eq.~(\ref{solnspace}).  

This provides two alternate ways to characterize the solution space. Every time we have a constrained pattern, the solution space gets restricted to a hyperplane satisfying $\env^{\,\mu}\cdot \wv=y^{\mu}$, which is perpendicular to the corresponding basis vector, $\env^{\,\mu}$. Thus, the constrained patterns together restrict the solution space to an $(\cN-\cC)$-dimensional space of the intersection of all the constraining hyperplanes. The null responses, on the other hand, only restrict the weight vectors to reside on the side of the hyperplane $\env^{\,\mu}\cdot \wv=0$ that satisfies the inequality in Eq.~(\ref{solnspace}). While the geometry of the solution space is illuminated by considering projections along the complementary basis vectors $\{\env^{\,\mu}\}$, the  weight vectors belonging to the solution space are most readily parametrized in terms of the natural basis $\{\env_{\mu}\}$ \footnote{The covariant basis vectors also carry biological information. For $\mu=1,\dots,\cP$, the $\mu^{th}$ covariant basis vector is the direction in weight subspace spanned by the patterns that is orthogonal to all the patterns except the  $\mu^{th}$  pattern. Therefore, it is the direction for which all other patterns provide no constraints.
}. 

In Fig.~\ref{fig:solutions} we illustrate some of these geometric relationships. Fig.~\ref{fig:solutions}A shows the three sets of vectors associated with the synapse basis, natural basis and complementary basis. Fig.~\ref{fig:solutions}B illustrates how any given weight vector can be expressed in any of this basis with the help of appropriate coordinates. Specifically, in Fig.~\ref{fig:solutions}C we depict how weights belonging to the solution space can most naturally be expressed in the natural basis. Nevertheless, the complementary basis provides intuition behind the solution space geometry. Fig.~\ref{fig:solutions}D depicts the solution space in terms of natural coordinate system, while Fig.~\ref{fig:solutions}E depicts it in the physical synapse basis.

Finally, one can see that the natural basis vectors transform inversely, as compared to the complementary  basis vectors given by Eq.~(\ref{complementary-transform}),
\be
\env_\mu\equiv \sum_{m = 1}^{\cN}\ev_{m} \left(X^{-1}\right)^{m}{}_{\mu}\Leftrightarrow\ev_{m}=\sum_{\mu = 1}^{\cN}\env_\mu X^{\mu}{}_{m}\ ,
\label{transformation-basis}
\ee
The coordinate transformations are similarly inverted:
\be
\eta_{\mu}= \sum_{m = 1}^{\cN}w_m \left(X^{-1}\right)^{m}{}_{\mu} \Leftrightarrow w_m=\sum_{m = 1}^{\cN} \eta_{\mu}X^{\mu}{}_{m}\ ,
\ee
where $w^m=w_m$. In the language of tensor calculus, $\eta^\mu$ and $\eta_\mu$ are respectively called contravariant and covariant coordinates, see Appendix~\ref{sec:tensors} for details. This inverse relationship between transformation property of the two different basis vectors and their coordinates will prove extremely useful and insightful in the calculations that follow. For example, this inverse relationship means that the vector itself doesn't change depending on what coordinate system one uses (the transformation of the coordinates and the basis vectors cancel each other), and one can choose the coordinate system that is most convenient for computing a given vector. Similarly, one can compute scalar {\it invariants} that can provide coordinate independent intuition behind essential geometric or algebraic quantities that help us link structural features with activity patterns. 
\section{Structural predictions in a linear theory}
\label{sec:linear}
\begin{figure}[!thbp]
	\centering
	\includegraphics[width=0.46\textwidth,angle=0]{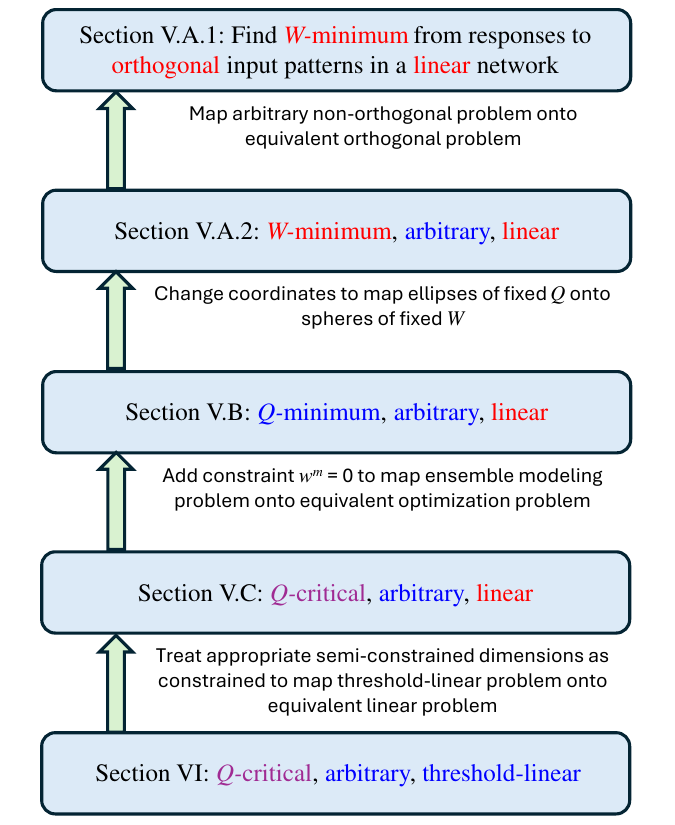}
	\caption{{\bf Conceptual flow of progressively complex computations.} {\small We schematically show how the complex problem of finding $\Qc$ for synapses in threshold-linear networks from arbitrary neuronal response patterns can ultimately be mapped to solving appropriate linear problems with orthogonal input patterns. We start from this  simple problem in section~\ref{sec:minL2} and end with the most complex problem of finding $\Qc$ in section~\ref{sec:nonlinear}.
			\label{fig:table}}}
\end{figure}

In this section our goal will be to slowly build towards a general framework that can make structural predictions based on different principle driven optimization and ensemble modeling approaches in a linear theory. In the next section we will generalize our results to threshold-linear transfer functions. Fig.~\ref{fig:table} shows the conceptual flow of our derivations.
\ssn{Minimum $L^2$ norm}
\label{sec:minL2}
Let us start by considering the very simple problem of finding the minimum $L^2$ weight norm 
\be
 ||\wv||^2\equiv\sum_{m}(w^m)^2\ ,
\ee
required to produce a specified input-output transformation in a linear network,
\be
y^{\mud}=x^{\mud}{}_mw^m=\env^{\,\mud}\cdot\wv=\eta^{\mud}\ ,
\label{linear-eqns}
\ee
where we have now introduced the Einstein summation convention, according to which repeated indices, with one superscript and one subscript, will be assumed to be summed over their entire available range. Here, this convention implies that $m$ is summed over $1,\dots,\cN$. Also, for convenience we have introduced indices $\mud,\nud$ to include all the constrained dimensions. There are no semi-constrained dimensions in a linear theory, and we will label the  unconstrained dimensions by $\mub, \nub$.

This problem can be solved very conveniently in the natural coordinate system. To recast the problem in terms of the natural coordinates, we define a metric tensor $g$ via its components according to  
\ba
g_{mn}&\equiv&\ev_m\cdot\ev_n=\da_{mn}\nonumber\\
\Ra \ga_{\mu\nu}&\equiv&\env_{\mu}\cdot\env_{\nu}= (X^{-1})^m{}_{\mu}(X^{-1})^n{}_{\nu}\da_{mn}.
\ea
Please see Appendix~\ref{sec:tensors} for more details on how the metric components transform under change of coordinates. The metric provides a way to compute distances between any two points in any coordinate system, and in particular the $L^2$ weight norm is just  the Euclidean distance of a point in the weight space from the origin:
\be
 ||\wv||^2=g_{mn}w^mw^n=\da_{mn}w^mw^n=\ga_{\mu\nu}\eta^{\mu}\eta^{\nu}\ .
\ee
The metric also provides us with coordinate invariant inner products between any two vectors: 
\be
\uv\cdot\wv=u^mg_{mn}w^n=u^m \delta_{mn}w^n=\un^{\mu}\ga_{\mu\nu}\eta^{\nu}\ ,
\label{inner-products}
\ee
where  $\wv$  is given by Eq.~(\ref{weight-vector}), and we have defined,
\be
\vec{u}=u^m \ev_m=\un^{\mu} \env_\mu\ .
\ee
Such inner products will let us compute projections and angles between vectors that lend geometric interpretability to the results. 
\sssn{Orthonormal patterns}
If the patterns are orthonormal, then the basis transformation between the synaptic and natural coordinates mediated by $X$ is just a rotation that preserves the {\it form} of the inner product, $\ga_{\mu\nu}=\da_{\mu\nu}$. Accordingly, for any weight-vector in the solution space we have
\be
 ||\wv||^2=\da_{\mu\nu}\eta^{\mu}\eta^{\nu}=\sum_{\mud=1}^{\cP}(y^{\mud})^2+\sum_{\mub=\cP+1}^{\cN}(\eta^{\mub})^2\ .
\ee
The above expression is minimized when we set $\eta^{\mub}=0$ for all the unconstrained coordinates. Thus, the synapse vector with minimum norm that solves Eq.~(\ref{linear-eqns}) is just given by~\footnote{Both $\wv_{\min}$ and $\yv$ should be thought of as $\cN$-dimensional vectors whose components along $\env_{\mub}$ are set to zero.}
\be
\wv_{\min}=w^m\ev_m=\yv\equiv y^{\mud}\env_{\mud}\ ,
\label{wmin}
\ee
and the minimum norm is
\be
\Wm^2=||\yv||^2=\sum_{\mud=1}^{\cP}(y^{\mud})^2\ ,
\label{ortho-linear}
\ee
because the $\env_{\mud}$'s are orthonormal to each other. 
\sssn{Nonorthogonal patterns}
Let us next consider the case where the patterns are not necessarily orthogonal. Let us call the subspace spanned by the complementary basis vectors, $\env^{\,\mud}$, as the constrained subspace. By definition the null basis vectors, $\env^{\,\mub}$, are all orthogonal to the constrained subspace and forms the basis for the orthogonal complement of the constrained subspace. A convenient consequence of this is that the metric is block diagonal, \ie, $\ga_{\mud\nub}=0$. 
Also, the natural basis vectors, $\env_{\mud}$'s, which by definition are orthogonal to all  $\env^{\,\mub}$'s, must therefore reside in the constrained subspace and form another  basis for it. 

Now, we can always find an orthonormal set of basis vectors spanning the constrained subspace\footnote{Such an orthonormal set is of course not unique, but given any orthonormal basis, one can find the specific transformation. In any case, the final results will not depend on the choice of this orthonormal set, but it is introduced for the purpose of the mathematical derivation.}, let's call them $\sav^{\,\mud}$. Suppose the two different sets of basis vectors are related via
\be
\sav^{\,\mud}=\La^{\mud}{}_{\nud}\env^{\;\nud}\Leftrightarrow \sav_{\mud}\equiv(\La^{-1})_{\mud}{}^{\nud}\env_{\nud}\ ,
\ee
effectively defining a basis transformation. Due to orthonormality of $\sav_{\mud}$, its dual, $\sav^{\,\mud}$,  must be the same,  $\sav^{\,\mud}= \sav_{\mud}$, and it satisfies,  
\be
\sav^{\,\mud}\cdot\sav_{\nud}=\da^{\mud}{}_{\nud}\ .
\ee
In a linear network we can find the way the target neuron would have responded to these orthonormal patterns from the responses we have already observed:
\be
\w{y}^{\mud}\equiv \sav^{\,\mud}\cdot\wv=\La^{\mud}{}_{\nud}y^{\nud}=\La^{\mud}{}_{\nud}x^{\nud}{}_{m}w^m\equiv\w{x}^{\mud}{}_{m}w^m\ .
\label{new-equations}
\ee
In other words, the constraint equations in Eq.~(\ref{linear-eqns}) are equivalent to a new set of equations, Eq.~(\ref{new-equations}), representing ``fictitious'' responses, $\w{y}^{\mud}$'s, to a different set of orthonormal patterns, $\w{x}^{\mud}{}_{m}$'s. We can thus use the previously derived results, Eqs.~(\ref{wmin}, \ref{ortho-linear}), to obtain the weight vector where the norm is minimized:
\be
\wv_{\min}=w^m\ev_m=\w{y}^{\mud}\sav_{\mud}\Ra W_{\min}^2=\left|\left|\w{y}^{\mud}\sav_{\mud}\right|\right|^2=\sum_{\mud}(\w{y}^{\mud})^2\ .
\ee
What  this means is that what  determines $\Wm$ is the average squared response the target neuron makes if it receives input patterns over all directions in the constrained subspace\footnote{To compute the average one has to evaluate $\int d^{\cP-1}x\  (\wv\cdot\ \vec{x})^2$, where the integral is over all possible unit directions, $\vec{x}$.  Spherical symmetry then dictates that such an integral is proportional to  $||\wv||^2$, which in turn can be written as the sum of squares of its components in any orthonormal basis. For more details, please see Appendix~\ref{sec:average}.}. In particular, this means that if the target neuron responds very differently to closely correlated patterns, one expects large responses to some pattern directions that are approximately orthogonal to the correlated patterns, and this effect has to be factored in while calculating $\Wm$. 

The above results are in terms of fictitious patterns. However, note that both $\wv_{\min}$ and $\Wm$ are expressed in a coordinate invariant form\footnote{Here by coordinate invariance we mean only with respect to transformations in the constrained subspace. One can also convince oneself that the statement, $\wv_{\min}=\yv$, is independent of any coordinate system. Thus its norm can be calculated in any coordinate system to give us $\Wm$.} because the transformations of upper and lower indexed quantities precisely cancel each other. In other words,
\ba
\wv_{\min}&=&\w{y}^{\mud}\sav_{\mud}=y^{\mud}\env_{\mud}=\yv\nonumber\\
\Ra W_{\min}^2&=&\left|\left|y^{\mud}\env_{\mud}\right|\right|^2=y^{\mud}y^{\nud}\ga_{\mud\nud}\ .
\label{wnorm-general}
\ea
To obtain $\Wm$ in terms of the observed data, we need to use another property of the metric. In keeping with our convention of upper and lower indices based on their transformation properties, we will denote the metric components associated with the complementary  basis by $\ga^{\mu\nu}\equiv \env^{\;\mu}\cdot\env^{\;\nu}$. One can then show that the matrix defined by the natural metric components is just the inverse of the matrix defined by the metric components in the complementary basis:
\be
\ga_{\mu\nu}\ga^{\nu\rho}=\da_{\mu}{}^{\rho}\ ,\where \ga^{\mu\nu} = X^{\mu}{}_mX^{\nu}{}_n\da^{mn}\ .
\ee
Now a further simplification occurs because $ \env^{\;\mud}$'s are orthogonal to all the $ \env^{\;\mub}$'s. So, just like $\ga_{\mu\nu}$, the metric components $\ga^{\mu\nu}$ obey a block diagonal structure:
\be
\ga^{\mub\nud}=0\ ,\  \ga_{\mud\nud}\ga^{\nud\rhod}=\da_{\mud}{}^{\rhod}\ ,\mand \ga_{\mub\nub}\ga^{\nub\bar{\rho}}=\da_{\mub}{}^{\bar{\rho}}\ .
\ee
Moreover, one can rewrite $\ga^{\mud\nud}$  solely in terms of neuronal responses:
\be
\ga^{\mud\nud}= \env^{\;\mud}\cdot\env^{\;\nud} = x^{\mud}{}_mx^{\nud}{}_n(\ev^{\,m}\cdot\ev^{\,n})=x^{\mud}{}_mx^{\nud}{}_n\da^{mn}\ .
\ee
Thus the metric components in the  complementary basis are simply the correlations between different input patterns, and to compute the minimum norm via Eq.~(\ref{wnorm-general}), one needs to weigh the target responses by elements of the inverse correlation matrix\footnote{We remind the readers that because the metric is block diagonal with respect to the constrained and unconstrained coordinates, the constrained submatrix of the inverse metric is the same as the inverse of the constrained sub-metric.}, $\ga_{\mud\nud}$.

Finally, to compute the synapse values that attains the minimum norm one needs to utilize one last property of the metric: the metric components can be used to go from the natural coordinates to complementary  coordinates,
\be
\eta_{\mu}=\ga_{\mu\nu}\eta^{\nu}\ ,\mand \eta^{\mu}=\ga^{\mu\nu}\eta_{\nu}\ .
\ee
Simple tensorial algebra then yields,
\ba
w^m&=&\yv\cdot\ \ev^{\;m}=y^{\mud}(\env_{\mud}\cdot\ \ev^{\;m})=y^{\mud}\ga_{\mud\nud}(\env^{\,\nud}\cdot\ \ev_{n})\da^{mn} \nonumber \\
&=&y^{\mud}x^{\nud}{}_n\ga_{\mud\nud}\da^{mn}
\ea
This is the weighted correlation between the activities of the pre- and postsynaptic neurons, $x$'s and $y$. 

To summarize, we have used tensor analysis to derive a simple analytical formula for the minimum $L^2$ norm needed to implement a given linear input-output transformation and the synapse vector that achieves this. We note how the metric-based geometric formalism lets us generalize the results from orthogonal patterns to non-orthogonal patterns in a very interpretable manner in terms of  various correlations involving input and output neuronal responses~\footnote{Note that $\ga^{\mud\nud}=x^{\mud}{}_m \da^{mn}x^{\nud}{}_n=(xx^T)_{\mud\nud}$.}:
\ba
W_{\min}^2=y^{\mud}y^{\nud}\da_{\mud\nud}&\rightarrow& y^{\mud}y^{\nud}\ga_{\mud\nud}\non
w^m=y^{\mud}x^{\nud}{}_n\da_{\mud\nud}\da^{mn}&\rightarrow& y^{\mud}x^{\nud}{}_n\ga_{\mud\nud}\da^{mn}
\label{Ww-quad}
\ea
It is instructive to rewrite the above formulae in terms of the $\cP\times \cN$ input matrix, $\cP\times 1$ output response, and $\cN\times 1$ weight vector:
\ba
W_{\min}^2=y^T(x x^T)^{-1}y\ ,\mand w_{\min}=x^T(x x^T)^{-1}y\ .
\ea
We thereby recover the standard formulas usually derived with linear algebra in statistical learning theory~\cite{hastie2009elements}. 

\ssn{General quadratic cost function}
\sssn{From sphere to ellipse}
We have seen how we can find the unique weight vector that minimizes the norm,
\be
\left|\left|\wv\right|\right|^2=\da_{mn}w^mw^n\ ,
\ee
while satisfying the constraints given by Eq.~(\ref{linear-eqns}). 
We have also seen that if we define an inner product structure using the metric, $g_{mn}=\da_{mn}$, 
\be
\uv\cdot\wv\equiv \da_{mn}u^mw^n\ ,
\non
\ee
then the above problem can be completely recast in a coordinate invariant language, that of finding the synapse vector with minimum norm, $ ||\wv||$, as defined by the above inner product,
\be
||\wv||^2=g_{mn}w^{m}w^{n}\ ,
\ee
 that satisfies
\ba
\env^{\,\mu}\cdot\wv&=&y^{\mu}\ .
\ea

The solution to the problem can also be written in a tensorial  language, \ie, in terms of vectors and the metric tensor that obey specific coordinate transformation properties (Appendix~\ref{sec:tensors}), as:
\ba
w^m&=&\yv\ \cdot\ \ev^{\,m}=y^{\mud}x^{\nud}{}_n\ga_{\mud\nud}\da^{mn}\ ,\mand \nonumber\\
\Wm^2&=&||\yv||^2=y^{\mud}y^{\nud}\ga_{\mud\nud}\ .
\label{L2summary}
\ea

One advantage of recasting the problem in this way is that the results straightforwardly generalize  to a  quadratic cost function, 
\be
Q^2\equiv q_{mn}w^mw^n\ ,
\label{Qfunction}
\ee
by simply defining a new metric 
\be
g_{mn}=\ev_{m}\cdot \ev_{n}=q_{mn},
\ee
where without loss of generality we have assumed $q_{mn}$ to be symmetric, and to ensure $Q$ always stays positive we have also assumed $q_{mn}$ to be elements of a positive semi-definite matrix. These assumptions make it possible to identify $q_{mn}$ as components of a metric in the weight space. We note that since the metric is no longer orthonormal in the synapse basis, $\ev^{\,m}\neq \ev_m$. Instead, they are duals to each other and can be obtained using the metric components:
\be
\ev_m=g_{mn}\ev^{\,n}\ ,\mand \ev^{\;m}=g^{mn}\ev_n\ ,
\ee 
where, as usual, the upper and lower metric components obey the inverse relationship,
\be
g_{mn}g^{np}=\da_{m}{}^{p}\ .
\ee
The inner product between two vectors in the synapse space is now given by,
\be
\uv\cdot\wv\equiv g_{mn}u^mw^n\ ,
\ee
which effectively replaces $\da_{mn}\ra g_{mn}$ in Eq.~(\ref{L2summary}). Similarly, the metric components in the complementary basis now read
\be
\ga^{\mu\nu}=\env^{\,\mu}\cdot\env^{\,\nu}=X^{\mu}{}_mX^{\nu}{}_ng^{mn}\ ,\mand \ga_{\mu\nu}\ga^{\nu\rho}=\da_{\mu}{}^{\rho}\ .
\label{gamma-components}
\ee

One can then calculate the synapse values that minimizes $Q^2$ as
\be
w^m=\yv\cdot\ev^{\,m}=y^{\mud}x^{\nud}{}_n\ga_{\mud\nud} g^{mn}\ ,
\label{w-linear}
\ee
while the minimum $Q^2$ is given by
\be
Q_{\min}^2=||\yv||_q=y^{\mud}y^{\nud}\ga_{\mud\nud}\ ,
\label{Qlinear}
\ee
where the subscript $q$ in $||\yv||_q^2$ signals that this is not the Euclidean norm but rather given by the metric associated with the quadratic function $Q^2$, $g_{mn}=q_{mn}$.

A more detailed derivation of Eqs.~(\ref{w-linear}), and (\ref{Qlinear}) is presented in Appendix~\ref{sec:spheretoellipse}. We note that while this expression is identical to Eq.~(\ref{Ww-quad}), the metric components, $\ga_{\mud\nud}$, are computed differently. In particular, the induced metric components in the constrained subspace that are needed to obtain $\wv_{\min}$ and $\Qm$ in terms of the neuronal activities can now be computed from the knowledge of the input responses and the cost function according to,
\be\ga^{\mud\nud}=\env^{\,\mud}\cdot\env^{\,\nud}=x^{\mud}{}_mx^{\nud}{}_ng^{mn}\ ,\mand\ \ga_{\mud\nud}\ga^{\nud\rhod}=\da_{\mud}{}^{\rhod}.
\label{induced-gen}
\ee

As in the case of optimizing the $L^2$ norm, one can rewrite the results in terms of matrices and column vectors in this general case as well:
\ba
Q_{\min}^2&=&y^T(x\ Q^{-1}x^T)^{-1}y\ ,\mand \non
w_{\min}&=&Q^{-1}x^T(x\ Q^{-1}x^T)^{-1}y\ .
\label{wmin-gen}
\ea
Here $Q$ denotes the matrix with matrix elements given by the coefficients in the cost functions, $q_{mn}$.
\sssn{Off-centered ellipse}
Let us look at one final generalization relevant for biology. Consider a more general cost function that includes linear terms along with quadratic terms:
\be
Q^2=q_{mn}w^mw^n-2w^mb_m+Q_0^2\ .
\ee 
We can rewrite this as
\be
Q^2=g_{mn}(w^m-b^m)(w^n-b^n)+(Q_0^2-g_{mn}b^mb^n)\ ,
\ee
where we have again identified $g_{mn}=q_{mn}$ and defined
\be
b^m\equiv g^{mn}b_n\ .
\ee
Introducing shifted coordinates
\be
w^{'m}\equiv w^m-b^m\ ,
\ee
the problem can essentially be recast into the minimization problem we just solved, \ie,  minimize
\be
Q^2=g_{mn}w^{'m}w^{'n}=||\wv'||_Q^2\ ,
\ee
subject to the constraints,
\be
x^{\mud}{}_mw^{'m}=y^{'\mud}\equiv y^{\mud}-x^{\mud}{}_mb^m\ ,\mx{or } \env^{\,\mud}\cdot \wv'=y^{'\mud}\ ,
\ee
Following Eq.~(\ref{w-linear}) the solution is given by 
\ba
w^{'m}&=&\yv^{\,'}\cdot\ev^{\,m}=(y^{\mud}-x^{\mud}{}_pb^p)x^{\nud}{}_n\ga_{\mud\nud} g^{mn}\non
\Ra w^m&=&(y^{\mud}-x^{\mud}{}_pb^p)x^{\nud}{}_n\ga_{\mud\nud} g^{mn}+b^m\ .
\label{wmin-general}
\ea
Similarly, the minimum $Q^2$ value is given by
\be
Q_{\min}^2=Q_0^2-g_{mn}b^mb^n+(y^{\mud}-x^{\mud}{}_mb^{m})(y^{\nud}-x^{\nud}{}_nb^{n})\ga_{\mud\nud}\ ,
\label{Qmin-general}
\ee
where we have just added the constant piece $Q_0^2-||\bv||_Q^2$ to the $Q^2_{\min}$ thus obtained.
Rewriting this in matrix notation we have,
\ba
Q_{\min}^2&=&Q_0^2-(Q^{-1}b)^Tb\non
&+&(y-x\ Q^{-1}b)^T(x\ Q^{-1}x^T)^{-1}(y-x\ Q^{-1}b)\ ,\mand \non
w_{\min}&=&Q^{-1}x^T(x\ Q^{-1}x^T)^{-1}(y-x\ Q^{-1}b)+Q^{-1}b\ ,
\ea
where $b$ is the column vector with elements, $b_m$, that appear in the optimization function.
\ssn{Computing $Q$-critical values from $\Qm$}\label{sec:ensemble-modeling}
We have seen how our characterization of the solution space can enable us to find the unique weight vector that minimizes $Q^2$ while reproducing the specified input-output transformation. However, there could be biological reasons that warrants taking a more conservative approach where we only assume that the $Q^2$ is bounded. For instance, we know that strengthening synapse comes with a biological cost, and it is reasonable to assume that the strengths of all the incoming synapses onto a given neuron is bounded in some fashion. Bounding the $L^2$ norm of the weight vector may be a natural way to incorporate such a biological constraint. However, we also know that different neuronal populations often can synapse onto a given neuron. In this case it may be more appropriate to weigh the cost of different incoming synapses differently and consider a more general quadratic cost function that is bounded, 
\be
Q^2= \sum_{\al}\LT\sum_{m\in \cA_{\al}} \la_{\al}^2 (w^m)^2\RT\ ,
\ee
where $\al$ labels the different neuronal populations and $\cA_{\al}$ the set of neurons in $\al$ population.
Constraining an even more general quadratic function may become relevant for certain biological situations. For example, brains need to both memorize the idiosyncrasies of specific examples and generalize well to unseen examples that have the same statistics~\cite{sun2023organizing}. In this case, it may be more pertinent to characterize all weight vectors that precisely reproduce the memorized examples and bound the generalization error, $G$,  which can be computed straightforwardly~\cite{hastie2009elements} from the average correlation function among the neurons as
\be
G=\2(G_{mn}w^mw^n-2G_mw^m+G_0^2)\ ,
\ee
where 
\be
G_{mn} = \left<x_mx_n\right>\ , G_m=\left<x_m y\right>\ ,\mand G_0^2=\left<y^2\right> \ .
\ee 

Let us therefore consider an ensemble of solutions satisfying,
\be
Q^2<Q^2_{\max}\ .
\label{Qbound}
\ee
For the special case when $Q^2= L^2$ and the input patterns were orthonormal, in~\cite{Biswas22} it was shown how the characterization of the solution space allowed one to find structural features that were consistent across the entire ensemble of solutions and therefore could lead to robust predictions. Specifically, for every synapse, ~\cite{Biswas22} derived a critical value, $\Wc$, such that if one knew that $ ||\wv||<\Wc$, then one could be certain that the synapse exists in the network. Moreover, one could determine whether the synapse is excitatory or inhibitory, because its sign becomes consistent across the entire ensemble. These were termed certain synapses and an analytic condition was derived. Here we will generalize the framework by specifying a way to calculate $\Qc$ for arbitrary patterns, such that if $Q^2<\Qc^2$, the synapse become certain. This will be achieved by providing a link between the optimization problem and the ensemble modeling approach. 

Suppose we are interested to find the $Q_{\mt{cr},p}$ corresponding to the $p^{th}$ synapse. 
The $w^p=0$ hyperplane divides this solution space into positive and negative $w^p$ values. We also know that  Eq.~(\ref{Qbound}) describes higher dimensional elliptic regions. As long as $Q<\Qm$ there exists no solutions at all, and this ellipse doesn't intersect with the solution space. At  $Q=\Qm$, the ellipse touches the solution space at a unique point, $\wv_{\min}$. Suppose this weight vector has $w_{\min}^p>0$. Then, until we increase $Q$ such that the projection of the elliptic region on the solution space touches the  $w^p=0$ hyperplane, all the solutions within the ellipse will have a positive sign. In other words, this synapse must exist and be excitatory until $Q=Q_{\mt{cr},p}$, the value when the expanding ellipse touches the  $w^p=0$ hyperplane within the solution space. This point represents the solution with minimum $Q$ in the absence of the $p^{th}$ synapse. One can indeed repeat the arguments if $w_{\min}^p<0$ leading to a consistent negative sign as long as $Q<Q_{\mt{cr},p}$. 

Thus as long as $w_{\min}^p\neq 0$, determining $Q_{\mt{cr},p}$ for the $p^{th}$ synapse in the ensemble modeling approach can be converted to an optimization problem of finding $\Qm$ for a reduced network where the $p^{th}$ input neuron is absent
Or equivalently, one could impose an additional constraining pattern, $x^{\mu=\cP+1}_{m}=\da^p{}_m$ to which  the target response is zero, $y^{\cP+1}=0$:
\be
x^{\mu=\cP+1}_{m}w^m=\da^p{}_mw^m=w^p=0\ .
\ee
In Appendix~\ref{sec:Qcr} we have derived an expression of $Q_{\mt{cr},p}$ in terms of various correlations involving the pre- and post-synaptic neuronal activities. Finally, we note that if $w_{\min}^p=0$, then $\Qm=Q_{\mt{cr},p}$, and the synapse can always be eliminated.
\section{Threshold-linear networks}
\label{sec:nonlinear}
In the previous section, we have seen how one can make various structural predictions using either  optimization principles or an ensemble modeling approach from steady state activity data. To obtain these predictions we assumed a linear transfer function. We are now going to show how the problem with a threshold-linear transfer function can effectively be reduced to solving a linear problem. 
\begin{figure}[!thbp]
	\centering
\includegraphics[width=0.46\textwidth,angle=0]{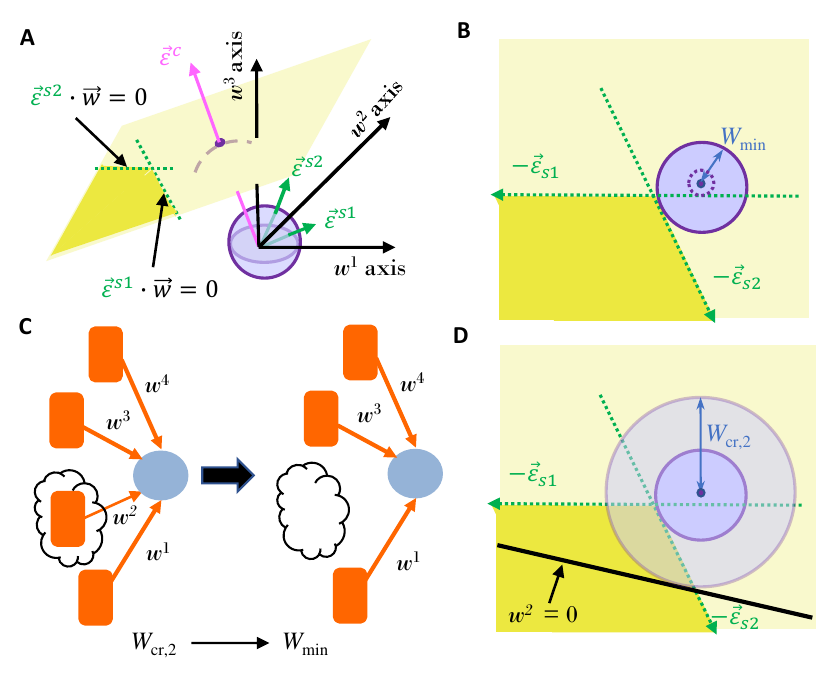}
	\caption{{\bf Optimization and ensemble modeling in threshold-linear theory.} {\small ({\bf A}) We depict a three dimensional weight space. The case depicted here contains one constrained and two semi-constrained dimensions. The constrained pattern restricts one to the light yellow hyperplane, while the semi-constrained patterns further restricts one to the deep yellow solution space. The dotted green lines represent intersection of planes passing through origin and perpendicular to the basis vectors $\env^{\,s1}$ and $\env^{\,s2}$, with the light yellow plane. The violet sphere represents all weights below a given weight norm. As the norm bound grows, the expanding sphere touches the light yellow plane at the purple dot which, in this case, lies outside the solution space.  ({\bf B}) We depict the projection of the weight space onto the two dimensional semi-constrained hyperplane. As the  weight bound grows, the bounding dashed purple circle will first touch the $\eta^{s2}=0$ line but it is outside the solution space as $\eta^{1s}>0$. Accordingly, this point has to be excluded while computing the $\Wm$ using the combinatorial method. The expanding circle, depicted in bold purple, eventually touches the solution space at $\eta^{1s}=0$ line. ({\bf C}) We illustrate how one can obtain $Q_{\mt{cr},2}$ for a given synapse ($w^2$ in this case) relevant for an ensemble modeling approach from $\Qm$ for a reduced network where the given synapse ($w^2$ here) is absent. ({\bf D}) We show the same semi-constrained plane as in panel ({\bf B}) with the addition of the black line that marks where the $w^2=0$ plane intersects the solution space. As the weight bound continues to enlarge, it (dashed purple circle) eventually touches this black line at the point whose norm is now simultaneously $\Wm$ for the reduced network without $w^2$, and $W_{\mt{cr},2}$. 
			\label{fig:nonlinear}}}
\end{figure}
\ssn{Combinatorial algorithm}
We start by reminding that even for a threshold-linear network,  the solution space is a connected region in  weight-space~\cite{Biswas22}. In terms of the $\eta$-coordinates, the solution space is constrained to lie in the {\it flexible subspace} defined by $\{\eta^{\mu}=y^{\mu}|\ \mu\in \cC\}$. It is further bounded by hyperplanes  $\{\eta^{\mu}=0|\ \mu\in \cS\}$. Now consider the surface characterized by a constant value of $Q$, we will refer to them as $Q$-surface.  As the value of $Q$ increases and the surface grows (Fig.~\ref{fig:nonlinear}A), first there comes a moment when it just touches the flexible subspace. If all the semi-constrained coordinates at the point of contact are non-positive, $\eta^{\mu}\leq 0$, then this point represents the minimum norm solution. If not, then one has to consider larger and larger $Q$-surfaces until it just touches the solution space at a point, which will be on one of the boundaries of the solution space, where each boundary is specified by a subset $\cB\subset \cS$ such that $\eta^{\mu}=0$ for all $\mu\in \cB$ and $\eta^{\mu}\ne 0$ for all $\mu\in \overline{\cB}\equiv\cS-\cB$, see Fig.~\ref{fig:nonlinear}B. 

This conceptual picture provides a way to determine $\Qm$ in two stages. For every boundary we can find $Q_{\cB}$, the value of $Q$ at which the  $Q$-surface touches a given  boundary $\cB$. This is nothing but solving the linear problem where we treat the semi-constrained dimensions that are set to zero as constrained, and the rest of the semi-constrained dimensions, as unconstrained. Next, we can find the minimum among all these  $Q_{\cB}$'s yielding the true minimum, $\Qm$, as well as the $\wv$ where the minimum occurs. The same algorithm can be used to compute $Q$-critical, since it can be recast as a $\Qm$ problem (Sec.~\ref{sec:ensemble-modeling}), where we either delete one input neuron from the network (Fig.~\ref{fig:nonlinear}C) or equivalently impose one additional equality constraint (Fig.~\ref{fig:nonlinear}D)

However, there are  two issues with this algorithm that one has to address. First, the number of boundaries grows exponentially as $2^{\cS}$. Equivalently, there are $2^{\cS}$ ways to assign subsets of semi-constrained dimensions as constrained when converting the threshold-linear optimization problem to a linear optimization problem. This makes the combinatorial algorithm untenable as the network size grows. Secondly, we note that for a given boundary choice, it may turn out that the point where the growing $Q$-surface touches the hyperplane specified by $\eta^{\mu}=0\ \forall\ \mu\in \cB$, is not actually in the solution space, \ie, some of the semi-constrained coordinates that do not belong to $\cB$ may have positive values. We will have to discard these $Q_{\cB}$'s in our combinatorial search algorithm. For e.g. in Fig.~\ref{fig:nonlinear}B (top), as the circular bound grows, it first touches the $\eta^{1s}=0$ line, but the contact point lies outside the solution space as $\eta^{2s}>0$. Thus this contact point has to be ignored.

This problem setup makes it clear that computing $\Qm$ for a threshold-linear network is equivalent to minimizing $Q$ subject to the equality contraints that $\eta^{\mu}=y^{\mu}$ for all constrained dimensions and the inequality constraints that $\eta^{\mu}\leq 0$ for all semi-constrained dimensions. Since $Q$ is a quadratic function and all of the constraint equations are linear, this is a quadratic programming problem that can be solved numerically with methods from convex optimization~\cite{boyd2004convex, wright1999continuous}. The combinatorial algorithm explained here is thus wildly inefficient, and it's akin to a brute force scan through the possible sets of active inequality constraints in the Karush-Kuhn-Tucker conditions~\cite{karush1939minima,kuhn1951proceedings}. However, the geometric insights that underlie the combinatorial algorithm can be pushed further, and we have discovered a geometric algorithm that can identify the correct boundary via a polynomial search, as well as interpretable analytical formulas that often approximates the solution well. 
We want to emphasize that our goal is not to try to find a numerical algorithm that is more efficient than quadratic programming but rather obtain semi-analytic results that lends biological interpretability in conjunction with a tractable numerical prescription. 

\ssn{$\Qm$ at a given boundary}
In this subsection we will solve the first step of finding where and when the Q-surface touches a given boundary defined by
\be
\eta^{\mu}=0|\ \mu\in \cB\subseteq\cS\ ,\mand \eta^{\mu}=y^{\mu}|\ \mu\in \cC\ .
\ee
This step will be needed both for the combinatorial and the geometric algorithms.
We start by noting that the problem can be formulated as finding the minimum value of 
\be
Q^2=||\wv-\bv||_Q^2
\ee
subject to the constraints
\be
\wv\cdot\env^{\,\mud}=y^{\mud}\ ,
\ee
where we have defined $y^{\mud}=0$ for $\mud\in\cB$. For simplicity, in this section we will consider the special case when $\bv=0$. 
For convenience, let us now refer to all the constrained dimensions and the semi-constrained dimensions that are set to zero at the specified boundary by the dotted indices, \ie, $\mud,\nud\in \cC'\equiv \cB\cup\cC$. These dimensions all behave as effectively constrained. All the other indices, \ie, the unconstrained dimensions and the semi-constrained dimensions whose coordinates are non-zero at the boundary, will be labeled by $\mub, \nub$, as these are effectively unconstrained. This is thus equivalent to solving a linear problem, whose results were given by Eqs.(\ref{wmin-general}),  (\ref{Qmin-general}), with correctly identified indices. 

We will now elaborate on one critical subtlety that one has to be mindful of while computing $\Qm$ and $\wm$. In the linear theory $\wm$ was obtained in terms of natural basis vectors, $\env_{\mud}$, that were dual to complementary basis vectors, $\env^{\,\mud}$. By construction, the natural basis vectors, $\env_{\mud}$, 
resided in the space spanned by $\{\env^{\,\mud}\}$, because $\env^{\,\mub}$ spans the kernel of $x$ and are orthogonal to the $\env^{\,\mud}$'s. In the threshold-linear case, the  $\env^{\,\mub}$ that correspond to semi-constrained dimensions, $\mub\in\cBb\equiv \cS-\cB$, no longer necessarily reside in the orthogonal complement of the  $\env^{\,\mud}$ vectors. Therefore, $\env_{\mud}$ doesn't necessarily belong to the linear span of $\{\env^{\,\mud}\}$~\footnote{Another way to see the problem is to realize that $\ga^{\mu\nu}$ is not block diagonal. And therefore  the induced metric,  $\ga_{\cB,\mud\nud}$,  defined to be the inverse of submatrix $\ga^{\mud\nud}$, is not a submatrix of $\ga_{\mu\nu}$.}. 

For the linear derivation to carry over to the threshold-linear theory, one thus has to introduce a new set of basis vectors, $\env_{\mud,\cB}$, that reside in the space spanned by $\{\env^{\,\mud}\}$ but still obey the orthogonality relation,
\be
\env^{\,\mud}\cdot \env_{\nud,\cB}=\da^{\,\mud}{}_{\nud}\ .
\ee
This induced basis in the effectively constrained subspace, $\env_{\mud,\cB}$, and the associated induced metric, $\ga_{\cB,\mud\nud}$, can be computed via
\ba
\ga_{\cB}^{\mud\nud}&\equiv&\ga^{\mud\nud}=\env^{\,\mud}\cdot\env^{\,\nud}= g^{mn}x^{\mud}{}_mx^{\nud}{}_n\ ,\non
\ga_{\cB,\mud\nud}\ga_{\cB}^{\nud\rhod}&=&\da_{\mud}{}^{\rhod}=\ga_{\cB,\mud\nud}\ga^{\nud\rhod}\ ,\non
\env_{\mud,\cB}&=&\ga_{\cB,\mud\nud}\env^{\,\nud}\neq \env_{\mud}\ , \non
\env^{\,\mud}&=&\ga_{\cB}^{\mud\nud}\env_{\nud,\cB}\ .
\label{induced-basis}
\ea
The induced basis vectors let us compute components of vectors in the constrained subspace along the complementary basis vectors easily. They also provide the relevant basis for the natural coordinates. In particular, as in the previous section, one can now determine the minimizing weight vector as, 
\be
\wm=\yv\equiv y^{\mud}\env_{\mud,\cB}\ ,
\label{Awmin}
\ee
and the minimum cost as~\footnote{We point out that naively using the contravariant sub-metric components while computing $\wm$ and $\Qm$ would lead to incorrect results, $\wm\neq y^{\mud}\env_{\mud}$, and $Q_{\min}^2\neq y^{\mud}y^{\nud}\ga_{\mud\nud}$.}, 
\be
Q_{\min}^2=y^{\mud}y^{\nud}\ga_{\cB,\mud\nud}\ .
\label{Qmin-boundary}
\ee
The synapse values can be computed as,
\be
w^m=y^{\mud}x^{\nud}{}_n\ga_{\cB,\mud\nud}g^{mn}\ .
\label{wmin-boundary}
\ee

We must check that the semi-constrained coordinates that are effectively unconstrained have negative values to determine if $\wm$ belongs to the solution space. In other words, $\forall\ \mub\in \cS$, we check if
\be
\env^{\;\mub}\cdot\yv=y^{\nud}\env^{\;\mub}\cdot\env_{\nud,\cB}=y^{\nud}\ga_{\cB,\nud\rhod}\env^{\;\mub}\cdot\env^{\;\rhod}=\ga^{\mub\rhod}\ga_{\cB,\rhod\nud}y^{\nud}<0\ ,
\label{L2inequality}
\ee
where we have used Eqs.~(\ref{Awmin}), (\ref{induced-basis}), and (\ref{gamma-components}).
\ssn{Identifying the boundary}
In this subsection we will provide a geometric algorithm to determine the boundary where the bounding $Q$-surface first touches the solution space, or in other words, find the set of semi-constrained dimensions that should be effectively constrained while obtaining $\Qm$. We will also use the geometric insights we derive to propose an analytical approximation that usually works well in practice. 
\sssn{Projection along the semi-constrained space}
We start by reminding that any weight vector in the solution space can be decomposed as 
\be
\wv=y^{\mud}\env_{\mud}+\eta^{\muw}\env_{\muw}+\eta^{\mub}\env_{\mub}\equiv\wv_{d}+\wv_u\ ,
\label{w-decomposition}
\ee
where we are now using the convention that $\mud\in \cC\ ,\muw\in\cS$, and $\mub\in \cU$. Moreover, we have  defined $\wv_d$ and $\wv_u$ to lie in the data-constrained subspace (spanned by the constrained and semi-constrained natural basis vectors) and unconstrained subspace, respectively. We note that the unconstrained subspace is orthogonal to the data-constrained subspace. Therefore, 
\be
||\wv||_Q^2=||\wv_d||_Q^2+||\wv_u||_Q^2\ ,
\ee
where the norms are with respect to the general metric, $g_{mn}=q_{mn}$.
Since the norm is a sum of two positive numbers, to find the smallest norm we should set the unconstrained coordinates to zero, so that $||\wv_u||_Q^2=0$. Since the constrained coordinates are fixed, we are thus left to scan the semi-constrained subspace spanned by $\env_{\muw}$, $\muw\in \cS$, or vectors of the form
\be
\wv=y^{\mud}\env_{\mud}+\eta^{\muw}\env_{\muw}=\yv+\sv\ ,
\label{sc-decomposition}
\ee
where $\sv$ now lives in the semi-constrained subspace.

Geometrically, the semi-constrained subspace is an $\cS$-dimensional hyperplane where all the constrained and unconstrained coordinates are fixed, $\eta^{\mud}=y^{\mud}$ and $\eta^{\mub}=0$, see Fig.~\ref{fig:nonlinear}A. On the other hand, a $Q$-surface is an $\cN$-dimensional (hyper)ellipse centered around the origin that grows as $Q$ increases. Our goal is to find the smallest ellipse that touches the solution space defined via $\eta^{\muw}\leq 0\ \forall\ \muw$, and thus the minimum $Q$ value . 

Before the ellipse can touch the solution space, it first has to become large enough to touch the semi-constrained hyperplane, see Fig.~\ref{fig:nonlinear}A. This is nothing but the point obtained according to Eq.(\ref{wmin-boundary}) where all the semi-constrained dimensions are essentially treated as unconstrained, or in other words,  $\cB=\emptyset$. Let us now characterize this point in the semi-constrained subspace in terms of induced basis vectors. Because of the index placement and transformation properties in Eqs.~\ref{complementary-transform} and ~\ref{transformation-basis}, we use tensor terminology and refer to $\{\envw^{\,\muw}\}$ as contravariant basis vectors and $\{\envw_{\,\muw}\}$ as covariant basis vectors (see Appendix ~\ref{sec:tensors}). Note that we can compute the contravariant basis, $\{\envw^{\,\muw}\}$, in the semi-constrained subspace using the induced metric from the covariant basis:
\be
\envw^{\,\muw}=\w{\ga}^{\muw\nuw}\env_{\nuw}\im \env_{\muw}=\ga_{\muw\nuw}\envw^{\,\nuw}\ ,
\ee
where $\w{\ga}^{\muw\nuw}$ is the inverse of the induced metric in the semi-constrained subspace:
\be
\w{\ga}^{\muw\nuw}\w{\ga}_{\nuw\w{\rho}}=\da^{\muw}{}_{\w{\rho}}\ ,\mand \w{\ga}_{\nuw\w{\rho}}=\ga_{\nuw\w{\rho}}\ .
\ee

We note that the norm of any vector in this subspace, Eq.~(\ref{sc-decomposition}), can be simplified as
\ba
||\wv||_Q^2&=&y^{\mud}y^{\nud}\ga_{\mud\nud}+\eta^{\muw}\eta^{\nuw}\ga_{\muw\nuw}+2y^{\mud}\eta^{\nuw}\ga_{\mud\nuw}\non
&=& ||\yv||_{\cC}^2+||\sv-\cv||_{\cS}^2-||\cv||_{\cS}^2\ ,
\label{w-gen-norm}
\ea
where we have now defined the center vector (Fig.~\ref{fig:nonlinear}A, B, D, purple dot) as, 
\be
\cv\equiv c^{\muw}\env_{\muw}= c_{\muw}\env^{\,\muw}\ ,
\ee
with
\be
c^{\muw}=\w{\ga}^{\muw\nuw}c_{\nuw}\ \mand  c_{\muw}\equiv -\ga_{\muw\nud}y^{\nud}\ ,
\ee
and the norms,  $||\dots||_{\cC}$ and $||\dots||_{\cS}$, are defined with respect to the induced metrics, $\ga_{\mud\nud}$ and $\ga_{\muw,\nuw}$, in the constrained and semi-constrained subspaces, respectively. 
By inspection of Eq. (\ref{w-gen-norm}), one finds that $||\wv||_Q$ is minimized at $\sv=\cv$, and so the growing $Q$-surface first touches the semi-constrained hyperplane at $\wv=\yv+\cv$. And, the problem of  minimizing $||\wv||_Q^2$ reduces to the problem of minimizing $||\sv-\cv||^2_{\cS}$, subject to the inequalities arising from semi-constrained patterns.
\sssn{Characterizing the minimizing boundary using covariant and contravariant basis vectors}
We are now going to characterize the boundary where the expanding $Q$-surface  first touches the solution space.
Let us consider a mutually exclusive partition of the semi-constrained dimensions, $\cS=\cB\cup\bar{\cB}$. Note that for each such partition we can construct a basis for the semi-constrained subspace by including natural (covariant) basis vectors $\env_{\muw}\ \forall\ \muw\in \bar{\cB}$ and dual (contravariant) basis vectors  $\envw^{\,\muw}\ \forall\ \muw\in \cB$.
Since by construction the basis vectors $\envw^{\,\muw}$'s are all orthogonal to the  $\env_{\muw}$'s, and they are linearly independent, they must together form a basis of the semi-constrained space. 

\begin{figure*}[!htbp]
	\centering
	\includegraphics[width=0.7
 \textwidth,angle=0]{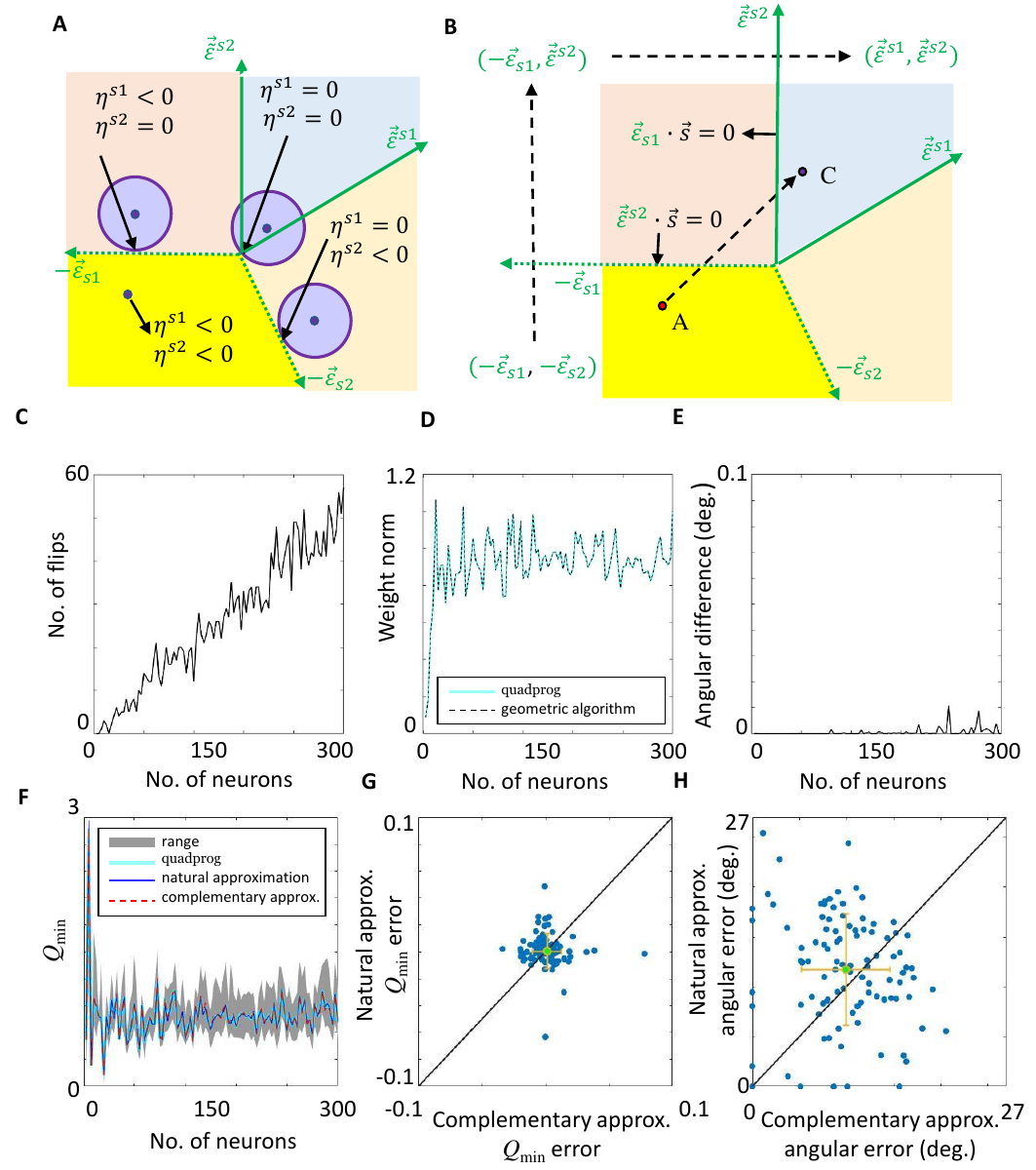}
\caption{{\bf Geometric algorithm to determine $Q$-minimum and $Q$-critical in threshold linear network and  related approximations.} {\small  ({\bf A}) We depict how the co- and contravariant basis vectors can be used to partition the semi-constrained subspace into the deep-yellow, orange, blue and yellow wedges spanned respectively by positive components of $(-\env_{s1},-\env_{s2})$, $(-\env_{s1},\envw^{s2})$, $(\envw^{s1},\envw^{s2})$, and $(\envw^{s1},-\env_{s2})$. ({\bf B}) We illustrate the geometric algorithm to identify the correct boundary. We start from a point in the solution space, A, and move towards the location of the center, C. In the process we cross two wedge-boundaries, the first one flips $-\env_{s2}\ra \envw^{s2}$, while the second  $-\env_{s1}\ra \envw^{s1}$, so that we know that C must be inside the blue wedge. ({\bf C-E}) We simulate input-output transformations in feedforward networks of increasing size. ({\bf C}) We plot the number of flips that the geometric algorithm required to reach the correct wedge. It seems to approximately grow linearly with the size of the network. ({\bf D}) We depict the norm of the optimized weight vectors obtained using the {\texttt{quadprog}} function in Matlab (light blue) and the geometric algorithm (dashed black), as the network size increases.  ({\bf E}) We plot the small angular difference between the two optimized vectors. ({\bf F}) We plot the $\Qm$ values corresponding to the optimized weight vectors obtained using {\texttt{quadprog}} function (light blue) and the approximate weight vectors (blue and dashed red) obtained using natural, Eq.~(\ref{conjecture}), and check complementary approximation  Eq.~(\ref{conjecture2}), schemes respectively, as the network size increases. The grey region depicts the range of values $\Qm$ must be restricted to according to the analytically obtained bound given by Eq.~(\ref{range}). ({\bf G}) The difference in the corresponding $Q$-minimum values between the optimized (obtained numerically using {\texttt{quadprog}} function) and approximate vectors are plotted. To facilitate comparison between the two different approximations we plot the  difference in $Q$ values coming from the two approximate vectors against one another. ({\bf H}) Same as in ({\bf G}), except that we compare the Euclidean angular error corresponding to the natural and complementary approximation.}}
\label{fig:simulations}
\end{figure*}

Suppose therefore that $\cv$ is expressed as 
\be
\cv=\cv_{\cB}+\cv_{\cB\perp}\ ,
\ee
where 
\be
\cv_{\cB}\equiv-\sum_{\muw\in\cBb}c_{\cB}^{\muw}\env_{\muw}\ \mand \cv_{\cB\perp}\equiv \sum_{\muw\in\cB}c_{\cB\muw}\envw^{\,\muw}\ 
\label{cB}
\ee
are by construction perpendicular to each other. 
We are  going to prove that if all the coordinates are positive ($c_{\cB\mu}, c_{\cB}^{\mu}>0$), then the boundary where $\Qm$ occurs is given by setting 
\be
\eta^{\muw}=0\ ,\forall\ \muw\in\cB\ ,
\label{boundary}
\ee
in Eq.~(\ref{w-decomposition}), see Fig.~\ref{fig:simulations}A for an illustrative example.
Further, the point that minimizes $\Qm$ is simply given by
\be
\sv_{\min}=\cv_{\cB}\ .
\ee
To prove this, consider an arbitrary point that resides in the solution space,
\be
\sv=\cv_{\cB}+\dav\mx{ with }\dav=\sum_{\mu\in\cS}\da^{\muw}\env_{\muw}\ .
\ee
The fact that $\sv$ belongs to the solution space means that the coordinates along all $\env_{\muw}$'s must be non-positive $\forall\ \muw$. Rewriting $\sv$ as 
$$
\sv=\sum_{\muw\in\cBb}(\da^{\muw}-c_{\cB}^{\muw})\env_{\muw}+\sum_{\muw\in \cB}\da^{\muw}\env_{\muw}\ ,
$$
we see that while $\da^{\muw}$ could be either positive or negative if $\muw\in \cBb$,  $\da^{\muw}\le 0$ for all $\muw\in \cB$. 
Now,
$$
\sv-\cv=\dav-\cv_{\cB\perp}\Ra ||\sv-\cv||^2=||\dav||^2+||\cv_{\cB\perp}||^2-2\dav\cdot\cv_{\cB\perp}\ .
$$
Further~\footnote{Note that $\envw^{\,\muw}$ is orthogonal to all $\env_{\nuw}$ $\forall\ \nu\neq \mu$.},
\be
\dav\cdot\cv_{\cB\perp}=\sum_{\muw\in \cBb}c_{\muw}\da^{\muw}\leq 0\ ,
\ee
since all $\da^{\mu}\le0$ and $c_{\cB\mu}>0$. Thus, the distance from the center is minimized when $\dav=0$. 

\sssn{Geometric prescription to address the combinatorial challenge in identifying the boundary}
In the previous section we have provided a geometric way to visualize the problem of identifying the correct boundary to solve the optimization problem for the threshold-linear network. In particular, we have seen that there is a way to divide the semi-constrained subspace into $2^{\cS}$ wedges\footnote{For every $\muw\in \cS$, we can either choose the natural basis vector, $-\env_{\muw}$, or the complementary basis vector, $\envw^{\,\muw}$, as part of the wedge basis leading to $2^{\cS}$ basis whose positive components span $2^{\cS}$ wedges.}, each of which is spanned by a combination of the covariant ($-\env_{\muw}$) and contravariant ($\envw^{\,\muw}$) basis vectors with positive components. We have also seen that if we know, in which of these wedges the center vector lies, then we can determine the boundary, \ie, where the expanding $Q$-surface touches the solution space. The problem is that finding in which wedge the center lies is a combinatorial challenge. This is because one has to find the components of $\cv$ associated with the basis vector for each of the wedges separately because the basis is different for each wedge.

We are now going to provide an intuitive and geometric prescription to identify this boundary that avoids the combinatorial search.  To understand how the geometric prescription works, let us consider a simple example where we have two semi-constrained dimensions so that the solution space within the semi-constrained subspace is spanned by
\be
\sv=\eta^{1}\env_1+\eta^{2}\env_2\ ,\with \eta^{1},\ 
 \eta^{2},\ \leq 0\ .
\ee
In Fig.~\ref{fig:simulations}A we have depicted  the different wedges that lead to different intersection boundaries for an $L^2$ optimization. If two of such wedges are separated by a line ($(\cS-1)$-dimensional hyperplane in general) then we will refer to them as contiguous. An example of such a pair are the wedges spanned by  $\{ -\env_{1},-\env_{2}\}$ and $\{ -\env_{1},\envw^{\,2}\}$. We now notice that these two wedges share all the basis vectors, except one. For the example pair, $-\env_{2}\ra \envw^{\,2}$. This basic observation suggests a novel geometric way to determine the wedge where the center vector lies. 

Let us start with an arbitrary point (red dot in Fig.~\ref{fig:simulations}B) inside the solution space,
\be
\av=a^{\muw}\env_{\muw}\ ,\with a^{\muw}<0\ .
\ee
Now consider vectors parameterized by $\la$ that interpolates between $\av$ and $\cv$:
\be
\sv(\la)=\la\cv+(1-\la)\av\ , \la\in [0, 1]\ .
\label{path}
\ee
Our goal will be to keep track of all the hyperplanes, separating two wedges, that the straight line path crosses on its way to its destination, $\cv$, starting from $\av$. If we can accomplish this, we will know all the ``flips'' in the basis vectors and therefore the final set of basis vectors and, in turn, the wedge where $\cv$ is located. For orthogonal wedges one can check that such  interpolations can at most cross $\cS$ times, for the general nonorthogonal case we expect the number of crossings to be $\cO(\cS)$. Our simulations, which we will describe shortly, indeed bear this out (Fig.~\ref{fig:simulations}C) and demonstrate the accuracy of the method (Fig.~\ref{fig:simulations}D,E). We will also soon see that once inside a given wedge, one has to perform $\cO(\cS)$ computations to determine which contiguous wedge the path next goes to. Thus approximately, we expect our geometric prescription to identify the boundary in $\cO(\cS^2)$ steps.  

To complete the story, let us describe how one can determine the contiguous wedge that the  path travels to from a given wedge. Suppose for a given $\la=\la_0$ we know which wedge it belongs to. This means we know the spanning basis, $\{\sav_{\muw}\}$, where  $\sav_{\muw}$ is either $-\env_{\muw}$ or $\envw^{\,\muw}$. In other words, if we express $\sv(\la)$ in terms of this basis,
\be
\sv(\la)=\xi^{\muw}(\la)\sav_{\muw}\ , 
\ee
then, we must have  $\xi^{\muw}(\la)>0\ \forall\ \muw$ in the vicinity of $\la=\la_0$.
Now, as $\la$ increases, there comes a point when one of the $\xi$ coordinates become zero (or it can reach $\cv$, \ie, this occurs at a value of $\la\geq 1$). This is when the path enters a new wedge. To find when this occurs one has to first solve 
\be
\xi^{\muw}(\la)=0\ ,\forall\ \muw\ .
\label{boundary-condition}
\ee
We can then compare all the values of $\la$'s that one obtains for the different ${\muw}$, to find the $\muw$ for which we get the smallest $\la>\la_0$. The basis vector corresponding to this index must be the one to flip, either  $-\env_{\muw}\ra \envw^{\,\muw}$ or $\envw^{\,\muw}\ra -\env_{\muw} $. We can repeat this process until we reach $\cv$. 

Our goal thus is to solve Eq.~(\ref{boundary-condition}).  Now, for any set of basis vectors, one can always find the dual basis which in turn can be used to find the components corresponding to the original basis. Explicitly, the dual basis to $\{\sav_{\muw}\}$ is obtained as follows:
\be
\sav^{\,\muw}\equiv \ta^{\muw\nuw}\sav_{\nuw}\ , 
\ee
where $\ta^{\muw\nuw}$ are the components of the inverse of the metric defined by 
\be
\ta_{\muw\nuw}\equiv \sav_{\muw}\cdot\sav_{\nuw}\ , \ \ta_{\muw\nuw}\ta^{\nuw\w{\rho}}=\da_{\muw}{}^{\w{\rho}}\ .
\ee
Accordingly, we have to solve the equations
\be
\sav^{\,\muw}\cdot \sv(\la)=0\ ,
\ee
to determine all the $\la$ crossings. 

The fact that the tensor formalism that was developed to describe curve space-time in physics could, in fact, be used to provide insights into a biological problem was somewhat surprising to us. We believe that the geometric tensor-based framework we have developed will enable us to go beyond quadratic programming and, for instance in a manner similar to what was discussed in the context of orthogonal patterns~\cite{Biswas22}, help us deal with noise and modifications of input-output transformation functions that are not threshold linear. Whether our geometric formalism based on distance minimization can be generalized to deal with optimizing functions more general than quadratic, remains an open question, but this was also an important motivation for proving a link between physics-based tensor formalism and biologically relevant optimization/ensemble-based predictions.

\sssn{An approximation to determine $\Qm$}
Our geometric algorithm provides us with a prescription to ascertain which semi-constrained dimensions behave as constrained for the purpose of solving  
a given quadratic optimization problem. This allowed us to calculate $\Qm$, the minimum value of $Q$ for which there exist a solution to the input-output transformation. The outcome of the algorithm cannot succinctly be characterized analytically. However, can one use geometric intuition to find an analytic approximation for $\Qm$ and $\wv_{\min}$? 

Consider first the case of minimizing the $L^2$ norm when the patterns are orthogonal. If there exists a semi-constrained direction, $\env^{\,\muw}$ which is positively aligned with the center vector, $\cv$, \ie,
\be
\cv\cdot \env^{\,\muw}>0 ,
\label{conjecture}
\ee
then $\cv$ doesn't belong to the solution space, as the  network specified by $\cv$ will then evoke a positive response, $\cv\cdot\env^{\,\muw}>0$, to the $\mu^{th}$ pattern.  One has to therefore consider weight vectors in the semi-constrained subspace that have norms larger than $||\cv||$. As argued previously, the most judicious choice is to keep increasing the norm until we find a vector, $\sv$, whose projection is zero along $\env^{\,\muw}$, effectively turning $\muw$ into a constrained dimension so the the network {\it just} starts to not respond to the pattern. We know that this argument breaks down when the patterns are non-orthogonal but we surmise that finding the $\Wm$ for a linear problem where all the semi-constrained dimensions with $\cv\cdot\env^{\,\muw}>0$ are treated as constrained along with the constrained patterns may provide a good approximation to the real $\Wm$ and $\wv_{\min}$. Further, we decided to extend our hypothesis to the general problem of finding $\Qm$ by covariantly generalizing the conjecture (\ref{conjecture}) to inner products given by $g_{mn}=q_{mn}$.

One could also argue that treating all the semi-constrained dimensions that are positively aligned with the natural basis vectors, $\env_{\muw}$, as constrained may also produce a good approximation. For instance, we know that the correct solution, $\wv_{\min}$, must be expressible as a non-positive linear combination of these semi-constrained basis vectors, suggesting that typically in order to find the lowest norm solution, one has to move away from $\cv$ such that one decreases the positive alignment between $\env_{\muw}$ and $\cv$. Accordingly we also decided to test a second approximation where we consider any  $\env_{\muw}$ that is positively aligned with the center vector, \ie,
\be
\cv\cdot \env_{\muw}>0 ,
\label{conjecture2}
\ee
to be a constrained pattern and then find the $\Qm$ corresponding to the linear problem.

To test these ideas we compute  two different approximate $Q$-minimum values that we term as natural and complementary approximations using Eq.~(\ref{wmin-gen}) by treating all the semi-constrained dimensions that respectively satisfy either Eq.~(\ref{conjecture}) or Eq.~(\ref{conjecture2}) as constrained, and the rest as unconstrained. Explicitly, 
\be
\wv_{\min}=\yv+\cv_{\cB_0}\ ,
\label{wvapp}
\ee
as given by Eqs.~(\ref{wmin}) and (\ref{cB}), where $\cB_0$ is now the set of all semi-constrained dimensions satisfying either Eq.~(\ref{conjecture}) or Eq.~(\ref{conjecture2}) corresponding to either the natural or complementary approximation respectively.

Let us point out that these approximations, especially the natural approximation, have a geometric interpretation. Suppose we are interested in finding whether the $\muw^{th}$ semi-constrained pattern behaves as a constrained or unconstrained pattern. If this input pattern is positively correlated with the constrained patterns, it makes it hard, \ie, require larger weights, for the target neuron to not respond the pattern. To keep the weight norm small, intuitively it then makes sense for the input drive corresponding to the $\muw^{th}$ pattern to be as large as possible, which is zero. In other words, one would expect the $\muw^{th}$ semi-constrained pattern to behave as a constrained pattern. If on the other hand, the $\muw^{th}$ semi-constrained pattern is anti-correlated with the constrained patterns, then one would naively expect its drive to be negative and the target neuron to not respond to the this pattern, thereby implying no further constraints. In other words, one then would naively expect the pattern to behave as unconstrained. Now, let us remind ourselves that the $\cv$ weight-vector is the vector with the minimum norm that satisfies all the constrained equations. Thus one way to assess whether the correlation of the $\muw^{th}$  pattern with the constraining patterns makes it more likely to behave as a constrained pattern is to compute the input drive, $(\env^{\,\muw}\cdot\cv)=(\xv^{\,\muw}\cdot\cv)$,  that the target neuron receives for this minimum norm weight-vector, $\cv$.
And, depending upon whether it is positive or negative the $\muw^{th}$ pattern is conjectured to behave as constrained or unconstrained.

Finally, let us discuss a lower and an upper bound for $\Qm$ that one can also readily obtain from two different choices of $\cB$. We remind the readers that the bounding ellipse has to become large enough to be able to touch the hyperplane defined via $\eta^{\mud}=y^{\mud}$. This is nothing but the point where all the semi-constrained dimensions are essentially treated as unconstrained. Or in other words,  $\cB=\emptyset$, so that
\be
\Qm^2\geq Q_{\mt{low}}^2=y^{\mud}y^{\nud}\ga_{\emptyset,\mud\nud}\ ,
\label{lower}
\ee
where we remind the readers that $\ga_{\emptyset,\mud\nud}$ corresponding to $\cB=\emptyset$, is the inverse of the sub metric-inverse in the constrained subspace,
\be
\ga_{\emptyset,\mud\nud}\ga^{\nud\rhod}=\da_{\mud}{}^{\rhod}\ ,
\ee
and $Q_{\mt{low}}^2$ can be interpreted as the average response squared to patterns in the constrained space.

Next, we note that since the origin of the semi-constrained subspace, $\eta^{\muw}=0$, is a solution, we can also obtain an upper bound on $\Qm$, denoted $Q_{\mt{up}}$. This is equivalent to treating all the semi-constrained dimension as constrained and set to zero, or $\cB= \cS$. The value of this radius can be calculated from the data:
\be
\Qm^2\leq Q_{\mt{up}}^2\equiv \ga_{\mud\nud}y^{\mud}y^{\nud}\ .
\ee
We emphasize that  $\ga_{\mud\nud}$ refers to the components of the original metric.

To summarize, we have arrived at the following bounds and approximation  for $\Qm$:
\be
y^{\mud}y^{\nud}\ga_{\emptyset,\mud\nud}\leq \Qm^2\approx \ga_{\cB_0,\mud\nud}y^{\mud}y^{\nud}\leq \ga_{\mud\nud}y^{\mud}y^{\nud}\ ,
\label{range}
\ee
where $\ga_{\mud\nud}$,  $\ga_{\cB_0,\mud\nud}$, and  $\ga_{\emptyset,\mud\nud}$  are the metric components corresponding to  the full weight-space, the induced subspace containing all the constrained dimensions and the semi-constrained dimensions used by the natural or complementary approximations (see text below Eq.~(\ref{wvapp}), and the constrained subspace, respectively.  
We'll see (Fig.~\ref{fig:simulations}F-H) that these approximations and bounds can be quite accurate.
\subsubsection{Numerical verification}
To verify the accuracy of our geometric algorithm, as well as track how the number of flips grows with the network size, we performed simulations involving $\cN=3,6,\dots,300$ input neurons in a feedforward network. For each simulation we selected, $\cP= 2\cN/3$, patterns where each input response was chosen from a uniform distribution in the interval $(-1,1)$. We assumed that for $\cS=\cN/3$ of these patterns the output neuron doesn't respond at all, while for the other  $\cC=\cN/3$ patterns the output responses were chosen from a uniform distribution in the interval $(0,1)$. In other words, for each of the input-output transformation, we had equal number of constrained, semi-constrained and unconstrained dimensions, each equalling $\cN/3$. 

Next, we generated a quadratic convex function as follows: we first constructed an antisymmetric matrix, $A$, from a matrix, $M$, with entries chosen from a uniform distribution between 0 and 1, via
\be
A\equiv M-M^T\ .
\ee
This yielded an orthogonal transformation matrix, $S$, via
\be
S\equiv e^A\ .
\ee
We also selected a diagonal matrix, $Q_d$, with entries chosen from a uniform distribution between 0 and 2, to finally obtain the matrix $Q$ whose elements define the loss function via,
\be
Q\equiv S^{-1}Q_dS\ .
\ee
We then solved for the weight vector that has the minimum $Q$ value, Eq.~(\ref{Qfunction}), using our analytic formula Eq.~(\ref{wmin-gen}) in conjunction with the geometric algorithm.

In  Fig.~\ref{fig:simulations}C we have plotted the number of flips that was required to arrive at the solution as the size of the network increased. As we suspected, the growth seems linear, strongly suggesting that our algorithm doesn't suffer from a combinatorial explosion. To verify the accuracy of our geometric approach, we decided to compare our results with solutions that can be obtained using standard convex optimization approaches to quadratic programming~\cite{boyd2004convex,wright1999continuous}. In Fig.~\ref{fig:simulations}D,E, we compare our results using the two different methods. In Fig.~\ref{fig:simulations}D we have plotted the norm of the optimum weight vectors, the norms from geometric and quadratic programming are plotted in light blue and dashed black respectively, and seem indistinguishable. Fig.~\ref{fig:simulations}E shows the angular difference between the two vectors, it is $\cO(10^{-2})$ degrees, further confirming the agreement between the two methods. Our geometric approach thus might provide a new and insightful way to solve quadratic programming problems, which have wide applicability~\cite{mccarl1977quadratic}. 

In the previous subsection we also obtained $\wv_{\min}$, Eq.~(\ref{wvapp}), and $\Qm$ Eq.~(\ref{range}) using natural and complementary approximation schemes. 
In Fig.~\ref{fig:simulations}F-H, we try to assess how well the approximation schemes work. In ~\ref{fig:simulations}F, we plot the $\Qm$ values obtained using quadratic programming and the two approximation schemes along with the $\Qm$ range given by the upper and lower bounds. In Fig.~\ref{fig:simulations}H we compute the error or difference in the $\Qm$ values obtained using quadratic programming and the two approximations.  The average discrepancy is about a few percent.  
Fig.~\ref{fig:simulations}H is similar, except that
we compute the Euclidean angles, $\ta_{\mt{diff}}$'s, between the optimized vector and the approximate vectors according to,
\be
\ta_{\mt{diff}}\equiv \cos^{-1}\LF{\sum_{m=1}^{\cN}w_{\mt{qp}}^{m}w_{\mt{app}}^{m}\over \sqrt{\sum_{m=1}^{\cN}(w_{\mt{qp}}^{m})^2\sum_{m=1}^{\cN}(w_{\mt{app}}^{m})^2}}\RF
\ee
 and plot these angular errors against one another. In the above expression, $w_{\mt{qp}}^{m}$ and $w_{\mt{app}}^{m}$ refers to the synapse values of the numerically optimized and approximate $\wv_{\min}$ respectively. The average angular difference is about ten degrees.  These comparisons show the value of the geometric intuition our formalism provides to the optimization problems considered here. For most configurations, the errors are small suggesting that both the natural and complementary schemes are providing a good approximation to the actual $\wv_{\min}$.

In Fig.~\ref{fig:runtimes} in Appendix \ref{app:runtimes}, we also compare approximate run times of a standard quadratic programming implementation (the \verb+quadprog+ function with default parameters in MATLAB) to our geometric algorithm and analytical approximations (also in MATLAB). We found that our analytical approximation was fastest for all tested problems. The geometric algorithm's run times were faster than \verb+quadprog+ up to a network size of $\cN\sim \cO(100)$, but \verb+quadprog+ ran faster than the geometric algorithm as the network size increased. 
\section{Generalizations: Multiple synapses, sign constraints, and Predicting neuronal responses}
\label{sec:activity}
While we have mostly focused on predicting connectivity at the level of a single synapse from the activity in the ensemble modeling context, we can apply our framework to more general situations. 

For instance, one can extend our framework to predict connectivity patterns involving more than one synapse. In the ensemble modeling approach one can ask whether at least one, out of a subset of synapses, must be present. The problem is equivalent to setting all the synapses in the subset to zero and then computing $\Qm$ and comparing it with the bound on $Q$. 

We can also impose known biological constraints on synapse signs. For instance, if the sign of a given synapse, $w^p$, is known to be negative, that information can be recast as a fictitious semi-constrained pattern,
\be
\eta^{\mt{fic}}\equiv x^{\mt{fic}}{}_m w^m\leq 0\ , \where\ x^{\mt{fic}}{}_m\equiv \da^p{}_m \ .
\ee
Positive synaptic weights can be imposed by simply flipping the sign of input pattern.

As a final application of our formalism, we are going to try and predict the activity state of a neuron to a new input pattern. As before, suppose we have access to $\cP$ activity patterns involving $\cN$ input neurons and one target neuron. We have seen so far that optimization of $L^2$ norm type quadratic functions yield unique weight vectors. Once the weight vector is known, we can of course predict the response of the target neuron to any new input pattern, $\xv^{\,\new}$, according to
\be
y^{\new}=\Phi(\xv^{\,\new}\cdot \wv_{\min})\ .
\ee
Response to such new patterns can be used to test predictions of the optimization principle. 

Can one also make predictions of the neuronal response in the ensemble modeling approach? To see how such predictions can indeed be made, let us revisit the problem of predicting synapse signs. We realized that the problem of finding $Q_{\mt{cr},p}$ can be converted to the problem of finding $\Qm$ with the $p^{th}$ synapse being absent. This is because $Q=Q_{\mt{cr},p}$, at the moment the expanding $Q$-surface just touches the $w^p=0$ hyperplane. This suggest an equivalent way of formulating the problem: $Q_{\mt{cr},p}$ is the minimum $Q$ needed when an additional constraint, 
\be
w^p=0\ ,
\ee
is satisfied. In other words, we do not reduce the number of synapse dimensions, as prescribed in Section~\ref{sec:ensemble-modeling}, but rather increase the number of patterns, with the new fictitious constrained pattern\footnote{Although according to our general formalism if the output to a pattern is zero, this should be considered a semi-constrained pattern, we will here impose the new pattern to be a constrained pattern.} being given by, 
\be
x^{\cP+1}{}_n=\da^p{}_n\ ,\mand y^{\cP+1}=0\ .
\ee

We now note that whether the target will respond to a new pattern or not depends on the sign of its input drive, $(\xv^{\,\new}\cdot \wv)$, and in fact, the hyperplane given by,
\be
\xv^{\,\new}\cdot \wv=0\ ,
\ee
demarcates regions in weight space where the neuron is turned on or turned  off. Thus one can determine a $Q_{\mt{cr},\xv^{\,\new}}$ such that if $Q<Q_{\mt{cr},\xv^{\,\new}}$ then one can predict whether the target neuron will respond to the pattern or not. This is the same as determining $\Qm$ for the network with one additional constrained pattern,
\be
\xv^{\,\cP+1}=\xv^{\,\new}\ ,\mand y^{\new}=0\ .
\ee 
In this language, the $Q_{\mt{cr},p}$ corresponding to the $p^{th}$ synapse is just given by $Q_{\mt{cr},\ev^{\,p}}$. Finally, to determine whether the neuron is turned on or off, one can simply compute the sign of its drive for the weight vector, $\wv=\wv_{\min}$, which is always present in the solution space. Depending on whether this sign is positive or negative, the neuron will respectively respond or not until $Q<Q_{\mt{cr},\xv^{\,\new}}$. 

\section{Applications to optokinetic and optomotor stimulus responses in larval zebrafish}
\label{sec:zebrafish}
It is useful to illustrate how our abstract theory applies to biological problems with an example. Consider binocular integration in the zebrafish pretectum~\cite{Naumann,kubo2014functional}, which originally motivated the development of our theory. The pretectum is a retinorecipient brain area~\cite{kramer2019neuronal,robles2014retinal} that integrates behaviorally relevant visual motion signals~\cite{yildizoglu2020neural,wang2020parallel} over space~\cite{zhang2022robust} and time~\cite{bahl2020neural,dragomir2020evidence} to stabilize the fish's eyes~\cite{kubo2014functional,portugues2014whole,wu2020optical} and its physical position in space~\cite{Naumann,yang2022brainstem}. Functional imaging experiments have identified neuronal populations that respond with non-orthogonal patterns to monocular and binocular motion cues~\cite{Naumann,kubo2014functional,wang2019selective}. Previous models have hypothesized inter-hemispheric patterns of synaptic connectivity that could account for these responses~\cite{Naumann,kubo2014functional}, and multiple groups are interested to test such models against connectomics data~\cite{Hildebrand,svara2022automated}. Our theory could be applied to go beyond the particularities of individual models to find the synaptic connections that are most indispensable for generating the functional responses. In this context, we note that although our previous work \cite{Biswas22} determined the space of networks consistent with observed neuronal patterns, the framework and results presented in this paper make it possible to obtain optimization/ensemble-based predictions with {\it correlated} biological neuronal patterns. 

To illustrate and test our formalism, here we focus on verifiable activity predictions that we can make by analyzing neuronal activity data in larval zebrafish in response to optokinetic ~\cite{kubo2014functional,svara2022automated} and optomotor ~\cite{Naumann} visual cues. Future work will make and test anatomy predictions using function-linked connectomics data ~\cite{svara2022automated}. 
We will first discuss how we plan to model the neuronal circuit, which involves the retina and the pretectum, in the context of our framework. Next we will focus on the circuit responsible for optokinetic gaze stabilization in larval zebrafish. Finally, we'll discuss how a very similar analysis can be applied to investigate the optomotor response, a behavior that enables a larval zebrafish to maintain its location by swimming against water currents. 
\ssn{Neural network model of binocular integration in the pretectum}\label{GeneralPrectumProcedure}
Both~\cite{kubo2014functional} and~\cite{Naumann}, studying optokinetic and optomotor responses respectively, envisioned a recurrently connected network of pretectal neurons receiving four different input patterns from direction-selective retinal ganglion cells ~\cite{robles2014retinal,kramer2019neuronal}. 
Both studies identified  pretectal neurons that tended to respond similarly to each other to all the different visual stimuli in their experiments and accordingly assigned them to belong to the same neuronal population. To apply our formalism, we therefore decided to model the pretectal circuit  as a recurrent network of $\cD$ \emph{driven} neuronal populations receiving feedforward connections from $\cI = 4$ \emph{input} direction-selective retinal ganglion cell populations\footnote{Each of these populations should be conceptualized as the distributed response of cells responding to one of the four patterns.}. The optokinetic study identified $\cD=11$ such functionally-defined neuronal populations, while the optomotor study found that the population response types were highly lateralized and identified $\cD=2\times 8=16$ neuronal populations, 8 in each hemisphere\footnote{It is possible that there is a different classification scheme of neuronal populations that is more appropriate for our formalism in linking structure and function, but we leave this investigation for future.}. 

To summarize the network's response properties, we constructed a $\cP\times \cN_{\tot}$ dimensional response matrix for each experiment, $z=\{z^{\mu}{}_m,\  \mu=1,\dots,\cP,\mand m=1,\dots,\cN_{\tot}\}$, where $\cP$ represents the number of stimulus conditions for the experiment, which was 8 in both cases, and $\cN_{\tot}$ represents the sum total number of inputs that the pretectal population receives (we will often refer to ``populations'' as ``neurons'' for brevity). Apart from the $\cD$ pretectal neurons and the $\cI$ retinal inputs, we allowed $\cN_{\tot}$ to also include a bias that can alter the threshold of the neurons to become active. Thus, $\cN_{\tot}=\cD+\cI+1$.  
Each pretectal population response was quantified as the average fluorescence measured during the stimulus presentation. Since fluorescence data were not available for the retinal inputs, we modeled each retinal population as an indicator function for whether the corresponding stimulus component was present. In other words, each population's activity took some nonzero value whenever a corresponding monocular stimulus component was present in the binocular stimulus and was zero otherwise. We chose the response value to equal the average of all the positive responses of the pretectal populations over stimulus conditions\footnote{Choosing other values for this constant is equivalent to altering the {\it cost} of connectivity between retina and pretectum relative to the cost of recurrent connections within the pretectum. Our formalism certainly allows us to consider such elliptic bounding/optimization functions.  
However, since our primary goal here is to present a proof-of-concept demonstration of the utility and applicability of our methods, we will stick to the above assignment of retinal input values as a way of approximately equalizing costs of different types of connections.}. Finally, we modeled the bias term as a constant input that was present for all stimulus conditions, and we chose this constant to be the same as the value we chose for the retinal inputs\footnote{It could also be interesting to explore the implications of changing this bias value in the future.}.

While comprehensive connectivity measurements that would allow us to test structural predictions based on ensemble modeling or optimization are not yet available, the general framework we have developed in this paper also allows us to predict the activity of neurons in response to a new input pattern. Therefore, we decided to test our formalism in the following way. We considered one pretectal neuron at a time  (referred to as the target neuron) and treated its $\cN$ input pathways from other pretectal populations, direct retinal inputs, and its bias as feedforward inputs to the target neuron. All the input-output response patterns for the target neuron were modeled as fixed points of neural network dynamics and can be read off from the $z$ matrix\footnote{See Methods and \cite{Biswas22} for more details on how predictions can be made in a recurrent network by decomposing the network's fixed point response patterns into $\cD$ feedforward network response patterns, one for each target neuron.}. We next used $\cP-1=7$ input-output patterns to predict the response of the target neuron (on or off) to the held-out eighth pattern. For a given target neuron we repeated this procedure $\cP=8$ times, where each time we predicted the response to a different held-out pattern. 

More explicitly, we computed $W_{\mt{cr},\xv^{\nu}}^p$, the weight norm below which all the incoming weight vectors onto the $p^{th}$ target neuron consistently predict an on or an off response to the $\nu^{th}$ input pattern when constrained to reproduce the $\mu\ne\nu$  response patterns. As prescribed in the previous section, it is the weight vector with the smallest norm that satisfies
\be
y^{\mu}=\Phi\LF\sum_{m = 1}^\cN x^{\mu}{}_m w_p{}^m\RF\ \forall\ \mu\neq \nu\ 
\label{p-equivalent}
\ee
and 
\be
\sum_{m=1}^\cN x^{\nu}{}_m w_p{}^m=0 \ ,
\label{nu-pattern}
\ee
where $y^\mu$ denotes the activity of the target neuron in stimulus condition $\mu$, $x^{\mu}{}_m$ denotes the activity of the input pathway $m$ in condition $\mu$, both $y^\mu$ and $x^{\mu}{}_m$ can be read off of $z$, and $\wv_p$ is the vector of weights onto the target neuron. We also remind the readers that the predicted activity state (whether the target neuron responds or not to the $\nu^{th}$ stimulus) is given by the sign of the input drive from the minimum norm weights that can reproduce the target neuron's responses to the other seven stimulus conditions, 
\be
d^{\nu}\equiv \sum_{m=1}^\cN x^{\nu}{}_m w_{\min}{}^m\ ,
\label{drive}
\ee
where we notationally neglect the dependence of $
\wv_{\min}$ on $p$ and $\nu$ for brevity. We would predict the target neuron to respond if $d^{\nu}>0$ and to not respond if $d^{\nu}\leq 0$. Having calculated the $W$-critical values for all the patterns, we ranked them in decreasing order for each target neuron. We hypothesized that patterns with higher ranks (higher $W$-criticals) will be consistent with observations. 

\begin{figure*}[!htbp]
	\centering
	\includegraphics[width=0.85
 \textwidth,angle=0]{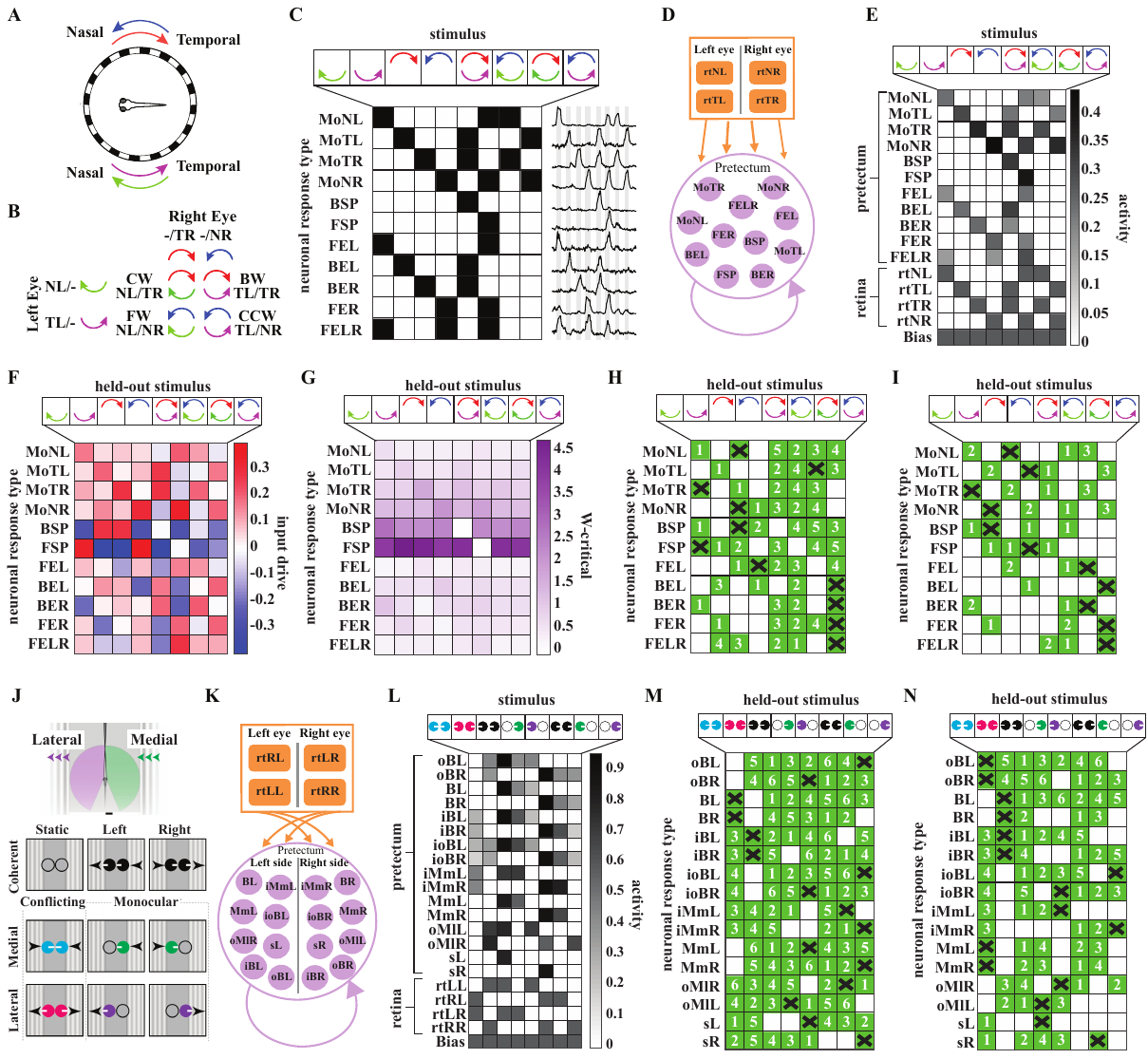}
	\caption{{\bf Applying the theoretical framework to  neuronal activity data.} 
{\small  ({\bf A-I}) Analysis of optokinetic response data from Svara et al.~\cite{svara2022automated}. ({\bf A}) Larval zebrafish were surrounded by LED monitors showing nasalward or temporalward motions to each eye. This experimentally elicited the optokinetic response. Adapted with permission from Kubo et al. ~\cite{kubo2014functional}. ({\bf B}) Eight monocular and binocular stimulus conditions were used in the experiment. Monocular: NL/- nasalward left eye, TL/- temporalward left eye, -/NR nasalward right eye, -/TR temporalward right eye; binocular: CW clockwise, CCW counterclockwise, FW forward, and BW backward. ({\bf C}) Neuronal responses in pretectum were measured from the eleven most common response types identified by Kubo et al.~\cite{kubo2014functional}. This included four monocular types (MoNL, MoTL, MoTR, MoNR) and seven binocular types (FEL, BEL, FER, BER, FSP, BSP, and FELR). Example calcium imaging traces shown at right. Each response type is identified with the barcode showing whether the neuron responded to each stimulus. For better visualization, here we have interchanged the pattern (here columns) and neuronal (here rows) indices as compared to the text. ({\bf D}) Model architecture. All response types could receive feedforward inputs from left and right retinal neurons that respond to temporalward or nasalward motion (rtTL, rtNL, rtTR, rtNR) and recurrent inputs from other pretectrum response types. ({\bf E}) We depict the $z$ matrix, which encodes the activity patterns of the pretectal response types, retinal inputs, and bias. ({\bf F}) For each response type, we made activity predictions for a held-out stimulus condition based on minimum norm model that explains the other seven. Red indicates a positive input drive and neural response. Blue indicates a negative input drive that leads the neuron to not respond. 
({\bf G}) We plot $W^p_{cr,\xv^\nu}$ values for each pairing of target neuron and held out stimulus. Larger values indicate more consistent responses across the solution space. 
({\bf H}) We ranked activity predictions based on the $W^p_{cr,\xv^\nu}$ magnitudes (highest to lowest). The green boxes with the values on top indicate correct predictions and their ranking. The green boxes with the crosses show the first incorrect predictions. No predictions were assessed after the first incorrect one, indicated by white boxes. 
({\bf I}) Similar to {\bf H}, we ranked activity predictions by the absolute values of the input drives until the first incorrect prediction. 
({\bf J-N}) Analysis of optomotor response data from Naumann et al.~\cite{Naumann}. ({\bf J})  Moving monocular gratings were presented medially (towards the midline) and/or laterally (away from the midline) below larval zebrafish in their left and right visual fields (top). The stimulus set consisted of all monocular and binocular combinations of these components (bottom). Circle icons represent the motion shown to each eye. Black arrows represent the motion direction of each grating. Adapted with permission from Naumann et al. \cite{Naumann}. ({\bf K}) Model architecture. Retinal cells send projections contralaterally (rtLL - left eye, leftward motion; rtLR - left eye, rightward motion; rtRL - right eye, leftward motion; rtRR - right eye, rightward motion).  In pretectum, neural response types were recurrently connected and lateralized to one side of the brain. Left-motion selective cells were usually localized to the left hemisphere, and right-motion selective cells were usually localized to the right hemisphere. oMIL and oMIR were lateralized oppositely, reflecting Naumann et al.'s hypothesis that these neurons relay lateral retinal signals from the contralateral eye across the pretectum. ({\bf L}) Optomotor variant of ({\bf E}). ({\bf M}) Optomotor variant of ({\bf H}). ({\bf N)} Optomotor variant of ({\bf I}).
		\label{fig:zebrafish}}	
}
\end{figure*}

We performed these calculations for every pretectal neuron and every (held-out) pattern. Our predictions for all the pretectal neurons, $m=1,\dots,\cD$, and for all the stimulus conditions, $\nu=1,\dots,8$ for both the optokinetic and optomotor experiments, are depicted in Fig.~\ref{fig:zebrafish}.  We also decided to compare these ensemble modeling based predictions with naive predictions that rank predictions based on the absolute values of the input drives without accounting for the full geometry of the solution space. These comparative predictions are also shown in Fig.~\ref{fig:zebrafish}. Note that the details of Fig.~\ref{fig:zebrafish} will be explained panel-by-panel in subsections below, and here we merely mean to convey the overall logic of the analyses.

\ssn{Including self coupling terms}
Before discussing our results for optokinetic and optomotor activity data, there is an important biological issue related to self couplings that we wish to address. Since we are dealing with neuronal populations and not individual neurons,  synapses between neurons within a given population would effectively manifest as self connections. However, we must ignore such self connections when computing held-out activity predictions, as including them would amount to using the response of the target neuron itself to ``predict'' its response to the held-out pattern. Nevertheless, we will now explain why these self-connections, even if they exist, do not affect the predictions ma when ignoring them.

Suppose $\wv'_{\min}$ is the weight-vector including self-connections that has the minimum norm among all weight vectors satisfying Eqns. \ref{p-equivalent} and \ref{nu-pattern}, and recall that the norm of $\wv'_{\min}$  corresponds to $W_{\mt{cr},\xv^{\nu}}^p$.
Now, we can always decompose $\wv'_{\min}$ into a synapse vector without self-couplings, $\uv$, and a self-synapse part, $u^\cN$,
\be
\wv'_{\min}=\uv+u^\cN\eh_\cN\Ra ||\wv'_{\min}||^2=||\uv||^2+(u^\cN)^2\ ,
\label{decomposition}
\ee
where we've assigned the self-synapse to index $\cN$ for notational convenience. Now for any given value of $u^\cN$, there may be many different $\uv$'s that give rise to weight vectors satisfying Eqns. \ref{p-equivalent} and \ref{nu-pattern}, but the $\uv$ defined by Eqn. \ref{decomposition} must be such that the resulting $\wv'$ has the smallest norm. 
Since the right hand side of Eqn.(\ref{decomposition}) is a sum of two positive terms, this means that, for the given value of $u^\cN$, $||\uv||$ must be minimized. We will therefore try to find this minimum norm $\uv$ for a given value of $u^\cN$.

For any pattern imposing an equality constraint\footnote{This includes $\nu$ because of Eqn. \ref{nu-pattern}.}, we note that $\uv$ must satisfy
\ba
y^{\mu}&=&\sum_{m = 1}^\cN x^{\mu}{}_m w_p{}^m = \sum_{m=1}^{\cN-1} x^{\mu}{}_m u^m+ y^{\mu} u^\cN\ , 
\ea
where we've noted that $x^{\mu}{}_\cN = y^\mu$. This implies
\ba
y^{\mu}&=&\sum_{m = 1}^{\cN-1} x^{\mu}{}_m \LF{u^m\over 1-u^\cN}\RF=\sum_{m =1}^{\cN-1} x^{\mu}{}_m u'{}^m\ ,\label{lowD-C}
\ea
where we have defined the rescaled input weights to the target neuron as
\be
u'{}^m={u^m\over 1-u^\cN}\ \label{w-prime}
\ee
and assumed $u^\cN<1$ to ensure that Eqn. \ref{w-prime} is well defined and avoid unstable amplification of neural activity\footnote{Recall that $u^\cN$ notates the amplitude of the target neuron's self-synapse within a recurrent neural network.}. 

For the semi-constrained patterns, we have $y^\mu=0$, which implies
\ba
\sum_{m=1}^{\cN} x^{\mu}{}_m w_p{}^m= \sum_{m=1}^{\cN-1} x^{\mu}{}_m u^m \le 0
\ea
and
\ba
\sum_{m = 1}^{\cN-1} x^{\mu}{}_m \LF{u^m\over 1-u^\cN}\RF=\sum_{m=1}^{\cN-1} x^{\mu}{}_m u'{}^m\ \le 0\ ,\label{lowD-S}
\ea
where we've used $u^\cN<1 \Rightarrow 1-u^\cN>0$ to ensure that the weight rescaling does not flip the inequality constraint. 

Eqns. \ref{lowD-C} and \ref{lowD-S} together imply that $\uv'$ obeys the same nonlinear equation as the network weights without self-couplings (Eqns. \ref{p-equivalent} and \ref{nu-pattern}). Moreover since, $||\uv||\propto ||\uv'||$ for given $u^\cN$, the former is minimized when the latter is minimized, so that
\be
||\uv||_{\min}= (1-u^\cN)||\uv'||_{\min}=(1-u^\cN)||\wv_{\min}||\ ,
\ee
where $\wv_{\min}$ now denotes the minimum norm weight vector without self-synapses that satisfies all of the constraints. Thus for a given value of $u^\cN$, the minimum norm with self-synapses is given by
\be
||\wv'_{\min}||^2=F(u^\cN)=(1-u^\cN)^2||\wv_{\min}||^2+(u^\cN)^2\ .
\ee
To find the value of $u^\cN$ that minimizes this function we can equate 
\be
{dF\over du^\cN}=0\ ,
\ee
which after straightforward algebra yields
\be
u^\cN={||\wv_{\min}||^2\over 1+||\wv_{\min}||^2}\ ,\mand ||\wv'_{\min}||^2={||\wv_{\min}||^2\over 1+||\wv_{\min}||^2}\ .
\ee
We notice that the above relationship between $||\wv'_{\min}||^2$ and $||\wv_{\min}||^2$ is monotonic, so the ranking of non-self-synapse $W$-critical values based on $||\wv_{\min}||$ is identical to that based on $||\wv'_{\min}||$.

\ssn{Optokinetic response predictions}
In this subsection, we will discuss our analysis of the optokinetic neuronal response data in larval zebrafish and explain the results from our predictions. We used the Svara et al. dataset~\cite{svara2022automated} because its functional calcium imaging data are aligned to electron-microscopy-based structural data that could facilitate future anatomical tests of the theory. Larval zebrafish were stimulated in their left and/or right visual fields by visual gratings that moved in the nasal (i.e., tail-to-nose) or temporal (i.e., nose-to-tail) directions (Fig.~\ref{fig:zebrafish}A). 
Each of these four stimulus components (i.e., nasal motion in left visual field, temporal motion in left visual field, nasal motion in right visual field, temporal motion in right visual field) was presented as a monocular stimulus, and these components were also binocularly combined to mimic clockwise, counterclockwise, forward, and backward motions (Fig.~\ref{fig:zebrafish}B). The total number of stimulus conditions was thus $\cP=8$. Earlier work from Kubo et al. ~\cite{kubo2014functional} identified $\cD=11$ over-represented functional neuronal response types  based on the responses of pretectal neurons to these stimuli (Fig.~\ref{fig:zebrafish}C), see Methods and ~\cite{kubo2014functional} for details. We therefore considered a retina-pretectum neural network model involving $\cD=11$ pretectal populations and $\cI=4$ monocular retinal input patterns (Fig.~\ref{fig:zebrafish}D). 

Our ensemble modeling framework allowed us to make accurate predictions for neuronal responses to optokinetic stimuli. 
As explained previously and detailed in the Methods, we used the Svara et al. dataset~\cite{svara2022automated} to construct a response matrix, $z$, that included the stimulus responses of each pretectum response type, the four retinal inputs, and a stimulus-independent bias (Fig.~\ref{fig:zebrafish}E). 
We then considered each pairing of target neuron and held-out stimulus condition to compute the minimum norm input weights needed to explain the other $\cP-1$ stimulus responses of the target neuron.
Fig.~\ref{fig:zebrafish}F plots the corresponding input drive (Eqn.~\ref{drive}), and the sign of each matrix element determines the target neuron's predicted activity state (active or inactive) in the held out stimulus condition based on either ensemble-modeling or weight-norm optimization. 
We then computed $W^p_{cr,\xv^\nu}$ for each pairing of target neuron and held-out stimulus (Fig.~\ref{fig:zebrafish}G), 
with higher $W^p_{cr,\xv^\nu}$ values indicating more consistency of the target neuron's activity state across the ensemble of network models. Thus, for a given target neuron we ranked the stimulus conditions according to their $W^p_{cr,\xv^\nu}$ values.  We found that the top three ranked predictions were all correct across neuronal response types (Fig.~\ref{fig:zebrafish}H). For some neurons (e.g., MoNL), all of the correct predictions were rank-ordered to the front of the list.

This success of ensemble modeling is not guaranteed from individual models. For instance, out of the $11\times 8 =88$ activity state predictions that could be made, 55 were accurately predicted by the minimum norm model. This is less than the chance level, as 63 correct predictions could be made by always predicting that the target neuron does not respond. The strongest magnitude predictions of the minimum norm model were also not reliable. When we ranked stimuli according to the absolute value of the input drive, we found that only the top-ranked predictions turned out to be consistently correct (Fig.~\ref{fig:zebrafish}I). 

Achieving as many correct predictions as ensemble modeling without error is rather unlikely by random chance. 
The total number of ways one can rank-order the stimuli is $\cP!$.
For a given target neuron, suppose that $L^2$ optimization correctly predicts the activity state for $\cT$ out of $\cP=8$ patterns. 
If the rank-ordering of ensemble modeling made $\cE$ correct predictions without making any mistakes, the number of ways this could happen is $(^{\cT}_{\cE})\cE!(\cP-\cE)!$, where the first term represents the number of ways to select $\cE$ correct predictions from the available set of $\cT$, the second term represents the number of ways to order these correct predictions, and the third term represents the total number of ways to order the remaining $\cP-\cE$ correct and incorrect predictions.
Thus, the probability that ensemble modeling would get at least $\cE$ correct predictions by chance is given by
\be
P_{\mt{ensemble},i}={(^{\cT}_{\cE})\cE!(\cP-\cE)!\over \cP!}={\cT!(\cP-\cE)!\over \cP!(\cT-\cE)!}.
\ee
The probability is tabulated as:
\begin{center}
\begin{tabular}{ |c|c|c|c|c|c|}
\hline
MoNL&MoTL&MoTR&MoNR&BSP&FSP\\
\hline
0.018&0.071&0.071&0.071&0.018&0.018
\\ \hline\hline
FEL&BEL&BER&FER&FELR& -
\\ \hline
0.071&0.071&0.071& 0.21& 0.21& -\\
\hline
\end{tabular}
\end{center}
The median chance probability is thus 7.1\%. 

\ssn{Optomotor response predictions}
We similarly analyzed prectectal responses during the optomotor response, a visual behavior wherein larval zebrafish maintain a stable position by orienting to the optic flow and swimming against it \cite{randlett2015whole,Naumann,yang2022brainstem}. We used an experimental dataset from Naumann et al. \cite{Naumann}, which drove optomotor behavior by presenting $\cP = 8$ monocular and binocular combinations of leftward and rightward motion in the fish's lower visual field (Fig.~\ref{fig:zebrafish}J)\footnote{The static stimulus did not activate pretectal neurons and was shown between the other stimuli.}. Naumann et al. used these stimuli to identify $\cD = 16$ over-represented functional response types that were grouped into 8 mirror-symmetric left-right pairs. Since pretectal responses to optomotor stimuli are highly lateralized with leftward (rightward) motion activating the left (right) pretectum ~\cite{Naumann,yildizoglu2020neural}, we assumed that each pretectal response type appears only on one side of the brain. 
Moreover, we imposed the biological constraint that the retina projects contralaterally in zebrafish~\cite{robles2014retinal}. So in our model neural network (Fig.~\ref{fig:zebrafish}K), each pretectal population received inputs from $\cN = 18$ circuit elements, with the first 15 corresponding to the other pretectal populations, two corresponding to direction-selective retinal ganglion cells from the contralateral eye, and one bias term. The $z$ matrix collecting the responses of all $\cN_\tot=21$ circuit elements is shown in Fig.~\ref{fig:zebrafish}L. 

We again found that ensemble modeling made many accurate predictions. As with the optokinetic model, we rank-ordered our activity predictions either based on the magnitudes of $W$-critical or the absolute value of the input drive. The ensemble modeling predictions using $W$-critical perform remarkably well (Fig.~\ref{fig:zebrafish}M), predicting at least four activity states correctly before making an error. These predictions were better than the input-drive based predictions of the minimum norm model (Fig.~\ref{fig:zebrafish}N), which could only consistently make three correct predictions. The minimum norm model correctly predicting 94 out of 128 responses. Unlike the optokinetic model, this is better than chance levels, as only 78 correct predictions could be made by assuming that the neurons never respond. The chance probabilities of achieving the accuracy of ensemble modeling's rank ordering was,
\begin{center}
\begin{tabular}{ |c|c|c|c|c|c|c|c|}
\hline
oBL&oBR& BL&BR&iBL&iBR&ioBL&ioBR
\\ \hline
0.036&0.036&0.036&0.107&0.036&0.036&0.25&0.25
\\ \hline\hline
iMmL&iMmR&MmL&MmR&oMlR&oMlL&sL&sR
\\ \hline
0.018&0.018&0.036&0.036&0.036&0.036&0.018&0.018
\\
\hline
\end{tabular}
\end{center}
yielding a median chance percentage of only 3.6\%.
\section{Discussion}
\label{sec:discussions}
Here we developed a tensor-based geometric formalism to infer structure-function relationships in  threshold-linear neural network models. Our primary goal was to provide a framework for combining neuronal activity measurements with hypothesized principles and constraints from biology to make predictions about neuronal connectivity. Using an ensemble modeling approach, we illustrated how to identify synapses that must be present to generate the observed activity patterns, as well as their signs, assuming that synaptic weight vectors are constrained within an elliptic region. Taking an optimization approach, we could further find the specific synapse vectors that optimize various quadratic functions of  weights. From a technical point of view, we solved both problems by recasting them as distance minimization problems with an objective-dependent and data-dependent metric. We then used tensor machinery to solve the general problem from known solutions to simpler problems via coordinate transformations. A crucial part of the derivation was converting the nonlinear problem to an equivalent linear one, which we interestingly found using a geometric algorithm that constructed mixed bases of contravariant and covariant vectors. This distinction between contravariant and covariant vectors is central to tensor calculus but rarely encountered in neuroscience. 


Ensemble modeling and optimization have complementary strengths for predicting experimental data. Our ensemble modeling approach is conservative, only making predictions for the subset of synapses that were found to be critical for function. We expect ensemble-modeling-based predictions to be particularly useful for testing fundamental model assumptions with limited data~\cite{transtrum2015perspective,Biswas20,Biswas22}. In particular, because these predictions are theoretically key, any disagreement with data points to a core problem in the model's formulation and assumptions. In contrast, the optimization approach makes predictions for each and every synapse in the system. These predictions will be highly biased by the optimization objective in the low data regime. However, once there is sufficient data to sample the full functionality of the system, optimization-based predictions can be used to probe candidate design principles. Such principles could be fundamental to many neural systems~\cite{sterling2015principles,bialek2012biophysics}.

We illustrated and tested our formalism by applying it to neuronal activity in the circuit responsible for binocular integration in the pretectum of larval zebrafish~\cite{kubo2014functional, Naumann, svara2022automated}. 
Our successful predictions demonstrate that our formalism is applicable to biological data and can provide useful information about how structure and function are linked in biological networks. Future work should explore whether even more accurate activity predictions can be made by considering alternate norms and neuron-type classifications. Future work should also apply our methods to make synapse-resolution anatomical predictions for pretectal connectivity that can be tested with emerging connectomics datasets~\cite{Hildebrand,svara2022automated}. 
For instance, we can now use our formalism to compute $W$-critical values for every synapse, thereby identifying the synapses that are predicted with certainty for any maximally allowed weight norm. Since this bound is \emph{a priori} unknown, we can rank how indispensable the synapses are according to their $W$-critical values~\cite{Biswas22}, just as we did while predicting activity-states. 
If conservative predictions from ensemble modeling are successful, then it would be interesting to consider whether hypothetical optimization criteria can account for more of the experimentally observed structure. For instance, we can assess how close the pretectal implementation is to the one that minimizes the Euclidean weight norm.

It is important to extend our analyses to account for noise. 
Neuronal activity is stochastic~\cite{shadlen1994noise, van1996chaos}, and experimental measurements of it are noisy~\cite{wilt2013photon,theis2016benchmarking}. It is thus unrealistic to demand an exact fit between a model and the data. Our previous work examined how the solution space changes when it is expanded to consider models that imperfectly match the data~\cite{Biswas22}.
We found that noise has the largest effect on the solution space when some output responses were lower than the noise level. Such responses could be zero within noise tolerance, thereby enabling constrained dimensions to become semi-constrained. These \emph{topological transitions} open flexible directions in the solution space, and we found that they could significantly change the predictions derived by ensemble modeling ~\cite{Biswas22}. Some of these analyses apply directly to the more general ensemble modeling and optimization problems that we consider here. For instance, a simple way to estimate these effects is to compare the results when all transition-prone patterns are treated as semi-constrained versus constrained. 
More work is necessary to generalize the precise formulas we derived for orthogonal patterns to non-orthogonal patterns, and treating arbitrary quadratic norms could become very complicated. 
For instance, it is possible that allowing for noise changes the relevant linear problem that must be solved. We leave the investigation of such effects for the future.

Our tensor-based analyses and resulting geometric algorithm relate interestingly to core concepts and algorithms from optimization theory. Most directly, the optimization problems we studied correspond to quadratic programming problems~\cite{boyd2004convex,wright1999continuous}, and we've shown how various predictions from ensemble modeling can be cast as quadratic programming problems as well. Here we were able to provide an insightful geometric algorithm via our comprehensive understanding of the solution space geometry~\cite{Biswas22}, as well as obtain useful analytical approximations. These  were possible because we considered the $\cP\leq \cN$ case, a regime  of biological interest, and it would be interesting to investigate how much of our framework generalizes to $\cP>\cN$. Standard quadratic programming algorithms~\cite{boyd2004convex,wright1999continuous} operate on the dual formulation of the optimization problem, taking advantage of the Karush-Kuhn-Tucker (KKT) conditions~\cite{kuhn1951proceedings,karush1939minima} to introduce multiple auxiliary variables. Our algorithm is able to bypass such constructions and obtain results that are more directly interpretable in terms of the correlations of pre- and post-synaptic neuronal activities. Our formalism thus provides a new approach to solving a rather popular class of optimization  problems that have found wide applicability in science and technology~\cite{mccarl1977quadratic}. A comparison of the run times of the different algorithms and approximation methods revealed that our analytical solutions (e.g., the “complementary approximation”) were always fastest. Our geometric algorithm was faster than the standard MATLAB quadratic programming solver for small network sizes, but \verb+quadprog+ eventually overtook it. Our geometric algorithm remained a practical option for all network sizes considered. We emphasize that we value the geometric algorithm and analytical approximations more for the insights they provide than for their speeds.

Generalizing our analysis from quadratic functions to other convex functions could be another fruitful direction, as convex optimization objectives appear in many scientific problems~\cite{boyd2004convex}. Firstly, our correspondence between  $\Qc$ in an ensemble modeling approach with $\Qm$ 
obtained from an appropriate optimization generally holds true.
Second, the insight that semi-constrained dimensions behave either as  constrained or unconstrained pattern when minimizing an objective function carries over from quadratic functions to more general functions, as can be seen from the KKT conditions~\cite{kuhn1951proceedings,karush1939minima}. Thus, we can reduce the nonlinear optimization problem to an appropriate linear optimization problem. Since we have a complete parameterization of the solution space in terms of the $\eta$ coordinates~\cite{Biswas22}, it should be possible to use standard techniques from calculus to find minima of the convex function for the different possible equivalent linear problems. Note that a local minimum of a convex function is its unique global minimum. One could then in principle find the correct minimum using the brute force combinatorial approach described in Section~\ref{sec:nonlinear}. However, it would  be better to leverage our geometric understanding of the solution space to generalize our geometric algorithm. Whether this is possible for arbitrary convex functions remains an open question.

An important future direction is to assess whether our framework could allow us to directly minimize the $L^0$ loss, which in the current context simply counts the number of non-zero synaptic weights. 
Here we have argued that $\Wc$ provides a measure of how indispensable a synapse is for the observed input-output function. The larger its value, the more likely the synapse is to be necessarily non-zero. Could it be that synapses with low values of $\Wc$ are in turn the most dispensable, such that if one is interested in constructing the most sparsely connected network one could simply eliminate the synapses with the lowest $\Wc$ values? From the structure of the solution space~\cite{Biswas22}, we already know that we have $\cU+\cS$ continuous flexible dimensions, and we may be able to use this flexibility to set some synaptic weights to zero. Preliminary explorations of random input-output patterns suggest that it is indeed often possible to reach a minimal $L^0$ norm network by eliminating the $\cU+\cS$ synapses with the lowest $\Wc$ values, sometimes all at once but usually sequentially upon removing synapses and recomputing  $\Wc$ values. Further work is needed to discern how reliably the solution space geometry allows us to build sparse models.

Minimizing the $L^0$ loss is of wide interest in statistics~\cite{hastie2009elements,tibshirani1996regression}, because it produces a model that explains the data with as few parameters as possible. This can powerfully regularize the problem and significantly aid interpretabilty. 
The current standard approach to minimizing the $L^0$ loss is to do so indirectly by minimizing the $L^1$ loss~\cite{hastie2009elements, tibshirani1996regression}, and the theory of compressed sensing explains why this is often a good approach~\cite{donoho2006compressed, ganguli2012compressed, candes2006robust, ganguli2010statistical}. For instance, compressed sensing theory shows that a sparse set of parameters can be exactly reconstructed from the minimization of the $L^1$ cost with sufficiently many measurements. 
Both compressed sensing and our geometric solution space approach apply to the high-dimensional case where $\cP < \cN$~\cite{advani2013statistical, buhlmann2011statistics}, and they may have overlapping domains of application. It would be particularly exciting if our geometric approach can mimimize the $L^0$ loss with less data than needed by $L^1$ minimization.\vs

\section{Methods}\label{Methods}
\ssn{Feedforward decomposition of recurrent neural network responses}
Although in this paper we have exclusively focused on feedforward networks, Biswas et. al. (2022) described how one can apply the formalism to recurrent networks by decomposing fixed point responses into equivalent feedforward networks~\cite{Biswas22}. For the sake of completeness, here we briefly discuss the procedure in the context of the retina-pretectal network.  
Let $\cN_{\tot}=\cD + \cI + 1$ be total number of elements in the neural circuit, including $\cD$ pretectal neuronal response types, $\cI$ retinal inputs,  and a bias. 
We assume that the activity of the pretectal populations change over time according to the firing rate equations of a threshold-linear recurrent neural network 
\be
\tau_p \frac{dz_p}{dt}=-z_p+\Phi\LF\sum_{m=1}^{\cN_{\tot}}z_m w_p{}^m\RF\\,
\label{threshold-linear-dynamics}
\ee
where $p=1,\dots,\cD$ indexes pretectal neurons, $\tau_p$ denotes the single neuron time constants, $\zv$ is a $\cN_{\tot}$-dimensional vector of activity levels (e.g., firing rates) of the circuit elements, and the weights $w_p{}^m$ connect the $m^{th}$ pre-synaptic element to the $p^{th}$ pretectal neuron\footnote{Recall that the circuit elements include a constant drive, so one element of $w_p{}^m$ corresponds to the bias of $p^{th}$ pretectal neuron.}. To make ensemble-modeling-based predictions, we follow our previous work~\cite{Biswas22} and focus on the network's steady-states. As described in the main text, we define a matrix of stimulus-response activity levels, $z^{\mu}{}_m$, with $m=1,2,\dots,\cN_{\tot}$ indexing the number of elements in the circuit and  $\mu = 1,2\dots,\cP$ indexing the stimulus patterns. Assuming that each $\zv^\mu$ is a fixed point of the network dynamics, this yields the following $\cP\times \cD$ non-linear equations
\be
z^{\mu}{}_p=\Phi\LF\sum_{m=1}^{\cN_{\tot}}z^{\mu}{}_m w_p{}^m\RF\\.
\label{p-patterns}
\ee
We now notice that the steady-state activity of the $p^{th}$  neuronal population depends only on the recurrent network weights through its incoming synapses, $w_p{}^m$, $m=1,\dots,\cN_{\tot}$. 

Hence, the recurrent network gives rise to steady state equations matching those of a feedforward network (Eqn.~\ref{SteadyStateEqn}), with the $p^{th}$ pretectal neuron acting as the target neuron, $y^{\mu}= z^{\mu}{}_p$, and $z^{\mu}{}_m$ playing the role of the input patterns. 
Thus, the recurrent network can be broken down into $\cD$ feedforward networks, one for each pretectal neuron, and the solution space of the feedforward networks can be translated back into the space of recurrent networks~\cite{Biswas22}. \emph{A priori}, all the circuit elements, including the $p^{th}$ neuron, should be considered as providing inputs to the target neuron. However, as discussed in Section~\ref{sec:zebrafish}, we can often ignore the self-coupling and only consider the other $\cN_{\tot}-1$ inputs. Furthermore, it's sometimes appropriate to exclude other synapses on biological grounds (e.g., ipsilateral connections from retina to pretectum). It's therefore convenient to let $x^{\mu}{}_m$ denote the input response matrix, with $m=1,\cdots,\cN\le\cN_\tot$ indexing only those circuit elements allowed to provide input to the target neuron. As such, the input matrix implicitly depends on $p$. Each column of $x$ can be read off as a columns of $z$, with $m=1,\cdots,\cN$ indexing columns of $x$, $m=1,\cdots,\cN_\tot$ indexing columns of $z$, and no general correspondence assumed between these two distinct set of indices.

\ssn{Data analysis, optokinetic response}
Our first goal was to obtain the average population level activities from the single neuron activities that were provided in Svara et al.~\cite{svara2022automated}. In these datasets, experimenters recorded dynamics using genetically encoded fluorescent calcium indicators as a proxy for neural activity. The time series of calcium imaging reflects neural responses to a set of 8 different stimuli, each presented three times with 10-second pauses between sets. To account for background fluorescence, 
for each cell we calculated the baseline activity by averaging the neural response over 8-second time windows prior to the start of each stimulation set. We then computed the average fluorescence activity over all sets and subtracted the baseline. This gave us mean-subtracted fluorescence traces for each cell. 

We characterized each cell’s responsiveness to 8 stimuli as a binary ``barcode”, where 1 indicates that cell responds to a given stimulus and 0 otherwise. Two schemes were used to identify the barcode. In the original scheme used by Kubo et al.~\cite{kubo2014functional} and Svara et al.~\cite{svara2022automated}, 8 binary steps were convolved with kernels that simulate fluorescence dynamics. This process generated 256 regressors (since there are $2^8$ possible combinations of binary states). Each of the regressors was then correlated with the fluorescence trace of each cell, and the regressor that exhibited the highest correlation with the cell's fluorescence trace was selected as the barcode. 
In another scheme adapted from Naumann et al.~\cite{Naumann}, if the average activity of the cell during stimulus presentation for one of the 8 stimuli exceeded 1.8 standard deviations from its baseline activity, we assigned a `1' to the corresponding component of the barcode and a `0' otherwise. We focused our analysis on the subset of 97 cells for which both schemes assigned the same barcode. All of these cells belonged to one of the most common 11 response types identified by Kubo et al.~\cite{kubo2014functional}, as Svara et al. targetted their analyses to these neurons~\cite{svara2022automated}. For all stimulus conditions for which a given neuronal type had a bar code of 1, we averaged over the activities of all the neurons belonging to the response type as a proxy for the population-level response to the corresponding stimulus. When the barcode was 0, we assumed the population activity to be zero as well. 
Finally, we constructed a $\cP\times \cN_{\tot}$ matrix $z$ for the optokinetic response dataset as described in the main text.
\ssn{Data analysis, optomotor response}
The Naumann et al. data were already background subtracted and classified with the Naumann et al. convention~\cite{Naumann}, so we skipped the pre-processing steps used to analyze the optokinetic response data. The other procedures matched those described in the main text and previous Methods section.
\vs
{\bf Acknowledgments:}  The authors would like to thank Herwig Baier, Eva Naumann, and Florian Engert for sharing data from the optokinetic and optomotor experiments and relevant expertise needed to analyze the data. The authors also thank Larry Abbott, Bill Bialek, members of the Fitzgerald Lab, and members of Janelia's Computation and Theory Research Area for input and feedback over the course of this project. This research was funded by the Howard Hughes Medical Institute and the Janelia Undergraduate Scholars Program. JEF acknowledges support from the
National Institute for Theory and Mathematics in Biology through the National Science Foundation (grant number DMS-2235451) and the Simons Foundation (grant number MPTMPS-00005320).
\appendix
\renewcommand{\arraystretch}{1.5}
\onecolumngrid
\newpage
\section{Tables}\label{sec:Table}
{\bf Table 1. Coordinate systems in metric spaces and transformations between them\vspace{1cm}\\}
\begin{tabular}{ |c| p{12.7cm}|}
\hline
\multicolumn{2}{| c |}{Coordinates and basis}\\ \hline
\hline
 Coordinate transformation & $z^{'i'}=\La^{i'}{}_iz^i\im z^{i}=(\La^{-1})^i{}_{i'}z^{'i'}$ \\ \hline
 Basis transformations & $
 \ev^{\,'i'}=\La^{i'}{}_i \ev^{\,i}\im \ev^{\,i}=(\La^{-1})^{i}{}_{i'} \ev^{\,'i'}\,\mand\ 
 \ev^{\,'}_{i'}=\ev_i(\La^{-1})^i{}_{i'}\im \ev_{i}=\ev^{\,'}_{i'}\La^{i'}{}_{i} $\\  \hline
 Vector representation & $\av=a^i\ev_i=a_i\ev^{\,i}= a^{'i'}\ev^{\,'}_{i'}=a_{i'}\ev^{\,'i'}$ \\  \hline
 Vector transformations & $a^{'i'}=\La^{i'}{}_ia^i\im a^{i}=(\La^{-1})^i{}_{i'}a^{'i'}, \mand  a'_{i'}=a_i(\La^{-1})^i{}_{i'}\im a_{i}=a_{i'}\La^{i'}{}_{i}=a^{'i'}\ev^{\,'}_{i'}$\\ 
 \hline
 Metric and its inverse & $g^{ij}g_{jk}=\da^i{}_j\ ,\mand g_{ij}g^{jk}=\da_i{}^k$\ , $g_{ij}$ elements of a positive-definite symmetric matrix. \\ 
 \hline
 Inner product definition & $\ev_i\cdot \ev_j\equiv g_{ij} \Ra \ev^{\,i}\cdot \ev^{\,j}\equiv g^{ij}\,\mand\ \ev_i\cdot \ev^{\,j}\equiv \da_i^j$\\ \hline
 Inner product & $\av\cdot \bv=a_ib^i=a^ib_i=g_{ij}a^ib^j=g^{ij}a_ib_j$\\
 \hline
 Component relations & $a_i=g_{ij}a^j=\ev_i\cdot \av\,\mand  a^i=g^{ij} a_j=\ev^{\,i}\cdot \av$\\
 \hline
 Metric transformations & $g'_{i'j'}=g_{ij}(\La^{-1})^i{}_{i'}(\La^{-1})^j{}_{j'}$, and $g^{'i'j'}=\La^{i'}{}_{i}\La^{j'}{}_{j}g^{ij}$
 \\ \hline
 \hline
 \multicolumn{2}{| c |}{Submanifold and induced metric}
 \\ \hline
\hline
Submanifold & Spanned by linearly independent set of  vectors $\{\env_I=\ev_i(\La^{-1})^i{}_{I}\}$.\\ \hline
Induced metric & $g_{\mt{in},IJ}=g_{ij}(\La^{-1})^i{}_{I}(\La^{-1})^j{}_{J}$ and $g_{\mt{in }}^{IJ}g_{\mt{in }JK}=\da^{I}{}_K$ defines $g_{\mt{in }}^{IJ}$\\ \hline
Dual basis vectors & $\env^{\,I}\equiv g_{\mt{in }}^{IJ}\env_{\mt{in},J}= \ev_i(\La^{-1})^i{}_{J}g_{\mt{in }}^{JI}\neq \La^{I}{}_j\ev^{\,j}$
\\ \hline \hline
 \multicolumn{2}{| c |}{Data-constrained subspace}
\\ \hline\hline
 Natural coordinates & $\eta^{\mu}=x^{\mu}{}_{m}w^m$, where $\mu=\mud,$ or $\muw$\\ \hline
 Natural subspace bases & $\env^{\,\mu}=x^{\mu}{}_{m}\ev^{\,m}$, and $\env_{\mu}=\ga_{\mu\nu}\env^{\,\nu}$ where $\mu=\mud,\muw$\\ \hline
 Induced metric & $\ga_{\mu\nu}\ga^{\nu\rho}=\da_{\mu}{}^{\rho}$ defines $\ga_{\mu\nu}$, where $\ga^{\mu\nu}\equiv \env^{\,\mu}\cdot\env^{\,\nu}=x^{\mu}{}_mx^{\nu}{}_ng^{mn} $, $\mu=\mud,\muw$  \\ \hline
Computing $g^{mn}$ &$g^{mn}g_{np}=\da^m{}_p$ defines $g^{mn}$,  where $g_{mn}\equiv q_{mn}$
\\ \hline
Decomposition & Data-constrained subspace = Constrained $\oplus$ Semi-constrained subspaces\\ \hline
Constrained subspace & Spanned by $\{\env^{\,\mud}\}$\\ \hline
Induced metric  & $\dot{\ga}^{\mud\nud}=\ga^{\mud\nud}$, and $\dot{\ga}^{\mud\nud}\dot{\ga}_{\nud\rhod}=\da^{\mud}{}_{\rhod}$, defines $\dot{\ga}_{\nud\rhod}$\\ \hline 
Basis vectors & $\vec{\dot{\en}}^{\,\mud}=\env^{\,\mud}$, and $\vec{\dot{\en}}_{\mud}=\dot{\ga}_{\mud\nud}\vec{\dot{\en}}^{\,\nud}=\dot{\ga}_{\mud\nud}\vec{\en}^{\,\nud}$\\ \hline
Semi-constrained subspace & Spanned by $\{\env_{\muw}=\ga_{\muw\nu}\env^{\,\nu} \}$\\ \hline
Induced metric  & $\w{\ga}_{\nuw\w{\rho}}=\ga_{\nuw\w{\rho}}$, and $\w{\ga}^{\muw\nuw}\w{\ga}_{\nuw\w{\rho}}=\da^{\muw}{}_{\w{\rho}}$, defines $\w{\ga}^{\muw\nuw}$\\ \hline
Basis vectors & $\vec{\w{\en}}_{\muw}=\env_{\muw}$, and $\vec{\w{\en}}^{\,\muw}=\w{\ga}^{\muw\nuw}\vec{\w{\en}}_{\nuw}=\w{\ga}^{\muw\nuw}\env_{\nuw}$\\ \hline
Orthogonality & Since $\env_{\muw}\cdot \env^{\,\mud}=0$, the constrained and semi-constrained subspaces are orthogonal to each other.
\\ \hline
\end{tabular}
\newpage
{\bf Table 2. Key results toward computing $\Qm$\vspace{5mm}\\}
\begin{tabular}{ |c| p{12.7cm}|} \hline
 \multicolumn{2}{| c |}{Minimizing a quadratic function in a linear theory}\\ \hline
\hline
Minimum of $Q^2=q_{mn}w^mw^n$ & $\Qm^2=||\yv||^2\equiv\yv\cdot \yv=y^{\mud}y^{\nud}\dot{\ga}_{\mud\nud}$, where $\yv\equiv y^{\mud}\vec{\dot{\en}}_{\mud}$\\ \hline 
Corresponding $\wv_{\min}$ & $\wv_{\min}=\yv=w^m\ev_m$, where $w^m=y^{\mud}x^{\nud}{}_{n}\ga_{\mud\nud}g^{mn}$ \\ \hline
\hline
\multicolumn{2}{| c |}{Minimizing a quadratic function in a threshold-linear theory}\\ \hline \hline
Step 1 & Working in the semi-constrained subspace, find the set of effectively constrained dimensions using geometric algorithm (see below).\\ \hline
Step 2 & Working in the {\it effectively} constrained subspace, compute $\Qm$ and $\wv_{\min}$ as in linear theory, where $\mud,\nud$ runs over all effectively constrained dimensions (see above).\\ \hline \hline
 \multicolumn{2}{| c |}{Geometric algorithm}\\ \hline
\hline
Center vector & $\cv\equiv c^{\muw}\env_{\muw}= c_{\muw}\env^{\,\muw}$, 
where $ c^{\muw}=\w{\ga}^{\muw\nuw}c_{\nuw}\ ,\mand  c_{\muw}\equiv -\ga_{\muw\nud}y^{\nud}$\\ \hline
Wedge & For every partition of semi-constrained dimensions, $\cS=\cB\cup\cB_{\perp}$, we consider all points spanned by positive components of basis vectors $\{\sav_{\muw}\}$, where
$\sav_{\muw}=-\env_{\muw}\ \forall\ \ \muw\in \cB_{\perp}$, and $\sav_{\muw}=\env^{\,\muw}\ \forall \ \muw\in \cB$\\ \hline
Algorithm goal & Identify the wedge where $\cv$ is located. All $\muw\in \cB$ are {\it effectively constrained} with $y^{\muw}=0$\\ \hline
Algorithm & Starting from $\sv_{\mt{ini}}\equiv -\sum_{\muw} \env_{\muw}$ in wedge $\cB=\emptyset$, consider the path $\sv(\la) = (1-\la)\sv_{\mt{ini}}+\la \cv$, moving towards $\cv$ ($\la=0$ to $\la =1$). We will keep track of the wedge identity as it crosses from one wedge to the next.\\ \hline
Potential border crossings& From any wedge along the path,  calculate $\cS$ potential wedge crossings that can occur next:
$ \la_{\muw}={\sv_{\mt{ini}}\cdot \sav^{\,\muw}\over (\sv_{\mt{ini}}-\cv)\cdot \sav^{\,\muw}}$\\ \hline
Calculating inner products&$\sav^{\,\muw}, \sv_{\mt{ini}}$ and $ \cv$ can be expressed in terms of $\env_{\muw}$ and $\env^{\,\muw}$, which in turn, can be expressed in terms of the synapses basis vectors: $ \env_{\muw}=\ga_{\muw\nu} x^{\nu}{}_m\ev^{\,m}$, and $\vec{\w{\en}}^{\,\muw}=\w{\ga}^{\muw\nuw}\ga_{\nuw\rho} x^{\rho}{}_n\ev^{\,n}$. In terms of the synapse basis, inner products can be calculated as  $\av\cdot\bv=a_mb_mg^{mn}$. \\ \hline
The border crossing & Choose the smallest $\la_{\muw}$ that is larger than previous crossing value. If this is greater than one, the current wedge is the correct wedge.\\ \hline
Identity of the next wedge & Once $\la_{\muw}<1$ has been identified, only one basis vector changes in the next wedge, either $(-\env_{\muw})\ra \vec{\w{\en}}^{\,\muw}$ or $ \vec{\w{\en}}^{\,\muw}\ra (-\env_{\muw})$, as appropriate. \\ \hline \hline
 \multicolumn{2}{| c |}{Approximate $\Qm$}\\ \hline
\hline
Effective constrained patterns & Includes all $\mud\in\cC$ and $\muw\in \cB$ satisfying, $\cv\cdot\env_{\muw}=-\ga_{\muw\nud}y^{\nud}>0$.\\ \hline
Approximate $\Qm$ & $\Qm^2=y^{\mud}y^{\nud}\ga_{\appr,\mud\nud}$\ , where $\ga_{\appr,\mu\nu}$ are the elements of the inverse of the submatrix $\ga^{\mu\nu}$, $\mu,\nu\in \cC\cup\cB$.\\ \hline
\end{tabular}

\section{Coordinate transformations and the metric, a brief review}
\label{sec:tensors}
\ssn{Dot products and norms}
Solving the general optimization or ensemble modeling problems that we are interested in becomes particularly easy if one introduces an appropriate metric in the synapse space and introduces the notion of covariant and contravariant basis vectors that transform differently under coordinate transformations. The main reason for this simplicity is that the problem of finding the solution space (as well as various quantities that we will optimize) can be stated in a coordinate invariant language and therefore one is free to choose the coordinate system where the problem is easiest to solve and then transform the results in any desired coordinate system. Here we provide a very brief overview of coordinate transformations of vectors and how they can be used to compute invariant quantities that we encounter in the paper, for a more detailed treatment see for e.g. \cite{wald2010general,synge1978tensor,misner1973gravitation}.

Let us consider an $N$-dimensional space labeled by coordinates $\{z^i\}$, $i=1,\dots,N$. Let us also assume that the space is endowed with a constant symmetric  metric tensor, $g_{ij}$, that can be used to compute the distance between any two points~\footnote{While in Einstein's special theory of relativity the metric is assumed to be a constant over space-time, similar to our case, in general theory of relativity (GR) the metric depends on the location and (\ref{distance}) is interpreted as the distance between {\it infinitesimally} separated points. The formalism we will use is actually more related to the GR framework as it deals with nonorthogonal basis vectors, but because our metric is a constant, we can use (\ref{distance}) to compute finite distances between points.}:
\be
d^2=g_{ij}\Da z^i\Da z^j\ ,\where \Da z^i\equiv  z_A^i- z_B^i\ ,
\label{distance}
\ee  
is the difference between the coordinates of the two points, $A$ and $B$. We have also assumed the Einstein summation convention according to which repeated indices, one lower and one upper, are assumed to be summed over their entire range unless otherwise stated. The metric can also be used to define and compute dot products between two vectors: Let us introduce a set of covariant basis vectors $\{\ev_i\}$ such that 
\be
\ev_i\cdot \ev_j=g_{ij}\ .
\ee
Accordingly, if we have two arbitrary vectors,
\be
\av=a^i\ev_i\ ,\mand \bv=b^j\ev_j\ ,
\ee
then
\be
\av\cdot \bv=g_{ij}a^ib^j\ ,||\av||_g^2\equiv \av\cdot \av=g_{ij}a^ia^j\ ,\mand ||\bv||_g^2=g_{ij}b^ib^j\ .
\ee
In particular, the distance of a point from the origin is given by
\be
||\zv||_g^2=g_{ij}z^iz^j\ .
\ee 
We note that if we have a Euclidean metric, $g_{ij}=\da_{ij}$, then $||\zv||_g$ is nothing by the conventional $L^2$ norm,
\be
||\zv||_g^2=\da_{ij}z^iz^j=\sum_{i=1}^N (z^{i})^2\ .
\ee

To contextualize the formalism, let us look at the synapse space where $z^i\ra w^m$, and $N\ra \cN$. Any synapse vector can now be represented as 
\be
\wv=w^m \ev_m\ ,
\ee
where $\ev_m$ can be thought of as a direction along the $m^{th}$ synapse. We will refer to these basis vectors as the physical basis as it is most intimately connected with biology of the brain. One of the problems we will solve in this paper is to find the synapse vector with the lowest $L^2$ norm that satisfies  the constraints (\ref{SteadyStateEqn}). Thus, if we choose an Euclidean metric on the synapse space this problem reduces to minimizing the norm, $ ||\wv||$, of the synapse vectors that belong to the solution space. In fact, this correspondence works for optimizing any quadratic function, 
\be
Q^2=q_{mn}w^mw^n\ ,
\ee
as long as  $q_{mn}$ defines a positive definite matrix.  If we identify $q_{mn}$ as the metric on the weight space, $g_{ij}\ra q_{mn}$, and $z^i\ra w^m$, then we can rewrite $Q^2$ as the square of the norm
\be
Q^2= ||\wv||_g^2\ .
\ee
\ssn{Coordinate transformations}
We are now going to look at how the components of a vector transforms under a general linear coordinate transformation:
\be
z^{'i'}=\La^{i'}{}_iz^i\ ,
\label{transformation}
\ee
where $\La$ is assumed to be a nonsingular matrix. In physics all physical quantities can be classified based on their transformation properties, collectively known as tensors. The way the components of a tensor transform is conveniently indicated by the position of their indices, upper or lower. So, if a tensor $T$ has an index structure, $T^{i_1,i_2,\dots}_{j_1,j_2,\dots}$, then under the  coordinate transformation (\ref{transformation}), the tensor components transform as
\be
T^{'i_i'i_2'\dots}_{j_i'j_2'\dots}=\La^{i_1'}{}_{i_1}\La^{i_2'}{}_{i_2}\dots(\La^{-1})^{j_1}{}_{j'_1}(\La^{-1})^{j_2}{}_{j_2'}\dots T^{i_ii_2\dots}_{j_ij_2\dots}\ .
\label{Ttransformation}
\ee
The upper and the lower indices are said to vary contravariantly and covariantly, respectively. Thus according to this convention the coordinates, $z^i$, and vector components $a^i$, are referred to as contravariant components~\footnote{Sometimes, they are also referred to as contravariant vectors, however to avoid confusion we will not use this terminology. A vector for us is an abstract quantity that doesn't depend on what basis one chooses to describe it.}, the latter transforming similar to (\ref{transformation}):
\be
a^{'i'}=\La^{i'}{}_ia^i\ .
\ee
Since coordinate transformations do not change the vector itself, the basis vectors must transform inversely:
\be
\ev^{\,'}_{i'}=\ev_i(\La^{-1})^i{}_{i'}\Ra \av=a^i\ev_i=a^{'i'}\ev^{\,'}_{i'}\ .
\ee
Thus, while the basis vectors themselves are said to vary covariantly, the components, $a^i$'s transform contravariantly (opposite to the basis vectors). 

One can check that according to the above definition, the components of the metric transform covariantly:
\be
g'_{i'j'}=\ev^{\,'}_{i'}\cdot \ev^{\,'}_{j'}=(\La^{-1})^i{}_{i'}(\La^{-1})^j{}_{j'}g_{ij}\ .
\ee 
And, indeed the dot product between two vectors remain invariant under coordinate transformations,
\be
\av\cdot \bv=g_{ij}a^ib^j=g'_{i'j'}a^{'i'}b^{'j'}\ ,
\ee
and thus can be computed in any coordinate system. In fact, since the framework of tensor calculus is designed such that components with lower and upper indices transform covarintly or contravariantly respectively,  any quantity that has no ``free'' indices, \ie, all its upper indices are summed with a lower index, is a {\it scalar} invariant and can be computed in any coordinate system. This property will prove crucial in some of our derivations. 
\ssn{Inverse metric and contravariant basis vectors} 
With the help of the inverse of the metric, $g^{ij}$, defined via
\be
g^{ij}g_{jk}=\da^{i}{}_k\ ,
\ee
it will prove convenient to also define a set of contravariant basis vectors by raising the indices of the covariant basis vectors:
\be
\ev^{\,i}\equiv g^{ij} \ev_j\ .
\ee
Indeed one can check that the inverse metric components transform contravariantly, consistent with its index structure. Further, one can lower or raise indices, thereby changing the transformation properties of vectors and tensors, with the metric and its inverse respectively. For instance, we can now write any given vector either in terms of covariant or contravariant basis vectors:
\be
\av=a^i\ev_i=a_i\ev^{\,i}\ ,\where a_i=g_{ij}a^j\Ra a^i=g^{ij} a_j\ .
\ee
The contravariant basis vectors and the corresponding covariant components transform as
\be
\ev^{\,'i'}=\La^{i'}{}_i \ev^{\,i}\ ,\mand a'_{i'}=a_i(\La^{-1})^i{}_{i'}\ ,
\ee
consistent with their index structure. The covariant and the contravariant basis vectors are also called dual to each other and obey the convenient orthogonality relation,
\be
\ev^{\,i}\cdot \ev_j=\da^{i}{}_j\ ,
\ee
because of which it is easy to extract the components of the vectors:
\be
a^i=\ev^{\,i}\cdot \av\ ,\mand  a_i=\ev_i\cdot \av\ .
\ee
Finally, we can also express dot products between two vectors succinctly as
\be
\av\cdot \bv=a_ib^i=a^ib_i=g_{ij}a^ib^j=g^{ij}a_ib_j\ .
\ee
\ssn{Induced metric in a subspace}
In our derivations, we often need to define metrics in weight subspaces as induced by the metric in the original space, \ie, define distances between points in the subspace by equating it to the distance between them according to the original metric. Let us say that we want to define an induced metric on subspace spanned by a subset of original basis vectors, $\{\ev_{I}\ ,I\in \cK\}$. If the remaining basis vectors corresponding to the complement of $\cK$, lets notate the complement as $\bar{\cK}$, spans an orthogonal subspace, then the subspace spanned by the covariant basis vectors, $\{\ev_{I}\ ,I\in \cK\}$, is the same as that spanned by the contravariant basis vectors, $\{\ev^{\,I}\ ,I\in \cK\}$. In this case, the induced metric and its inverse are the same as the original metric and its inverse:
\be
g_{\mt{in},IJ}=g_{IJ}\ ,\mand g^{IJ}_{\mt{in}}=g^{IJ}\ ,\forall\ I,J\in \cK\ .
\ee
The dot product of vectors in the subspace,
\be
\av=\sum_{I\in\cK}a^I\ev_I\ ,\mand \bv=\sum_{J\in\cK}b^J\ev_J\ ,
\ee
can be computed, as if, they belong to the original weight space:
\be
(\av\cdot\bv)_{\mt{in}}=\sum_{I,J\in\cK}a_Ib_Jg^{IJ}=\sum_{I,J\in\cK}a^Ib^Jg_{IJ}\ .
\ee
This is essentially because the original metric is block diagonal with respect to the $\cK$ and $\bar{\cK}$ indices. An example of this situation arises when one decomposes the natural basis into data-constrained indices (constrained and semi-constrained) and unconstrained indices. Since these basis vectors spans orthogonal spaces, the induced metric on the data-constrained subspace is the same as the original metric.

When the basis corresponding to indices belonging to $\cK$ and $\bar{\cK}$ are not orthogonal, then defining the induced metric becomes a bit more subtle as, $\{\ev_{I}\ ,I\in \cK\}$ and $\{\ev^{I}\ ,I\in \cK\}$ do not span the same subspace. In this case the induced metric in the subspace spanned by $\cK$ is still given by the original metric:
\be
g_{\mt{in},IJ}=g_{IJ}\ ,
\ee
so that
\be
(\av\cdot\bv)_{\mt{in}}=a^Ib^Jg_{\mt{in }IJ}=a^Ib^Jg_{IJ}=\av\cdot\bv\ .
\ee
However now the dual contravariant basis vectors within the subspace will not agree with the contravariant vectors in the original space as the induced inverse metric is different from the submatrix of the original inverse metric 
\be
g_{\mt{in }}^{IJ}\neq g^{IJ}\ ,
\ee
where $g_{\mt{in }}^{IJ}$ is, as usual, defined via
\be
g_{\mt{in }}^{IJ}g_{\mt{in }JK}=\da^{I}{}_K\ .
\ee
Accordingly, 
\be
\ev_{\mt{in}}^{\,I}\equiv g_{\mt{in }}^{IJ}\ev_{J}\neq \ev^{\,I}\ .
\ee
\section{From sphere to ellipse}\label{sec:spheretoellipse}
Here we are going to show how the $L^2$ optimization problem can be generalized to minimization of an arbitrary homogeneous quadratic function of the form 
\be
Q^2=q_{mn}w^mw^n\ ,
\label{Qquad}
\ee
subject to constraints,
\be
x^{\mud}{}_mw^m=y^{\mud}\ ,
\label{constraints-appnd}
\ee
in a linear setting. In other words we will derive Eqs.~(\ref{w-linear},\ref{Qlinear}). 

Let us identify $q_{mn}$ with the metric components in the synaptic coordinate system,
\be
g_{mn}=q_{mn}\ ,
\ee 
so that the optimization problem can be recast as optimizing the norm
\be
Q^2=||\wv||_Q^2\ ,\mx{ given }\ \env^{\,\mud}\cdot\wv=y^{\mud}\ ,
\ee
where, as usual, we have defined
\be
\env^{\,\mud}=x^{\mud}{}_m\ev^{\,m}\ .
\ee
We note that in this geometry, the synaptic basis vectors are no longer orthonormal, but rather satisfy,
\be
\ev_m\cdot\ev_n=q_{mn}=g_{mn}\ ,\mand \ev^{\,m}\cdot\ev^{\,n}=g^{mn}\ ,
\ee
where $g_{mn}$ and $g^{mn}$ are inverses of each other~\footnote{While the solution space was derived by constructing an $X$ matrix from the data matrix, $x$, that implicitly assumed an Euclidean metric (see discussion around (\ref{env-up}), $g_{mn}=\da_{mn}$, this is not necessary. For any constant metric, one can first define the $\env_{\mub}$ basis vectors as spanning the null space of $x$: $\env^{\,\nu}\cdot \env_{\mub}=0\ \forall\ \nu\in \cC\cup\cS$, and $X^{-1 m}{}_{\mub}=\en_{\mub}^m$. Note that irrespective of the metric the null vectors, by definition, always satisfy $x^{\nu}{}_m\en_{\mub}^m=0$. To complete the covariant basis, one can find dual to the contravariant vectors, $\env^{\,\mu}=x^{\mu}{}_m\ev^{\, m}$ within the data-constrained subspace spanned by $\{\env^{\,\mu}\ , \mu\in \cC\cup\cS\}$. These dual covariant basis vectors, $\{\env_{\mu}\ , \mu\in \cC\cup\cS\}$, satisfy the relation, $\env^{\,\nu}\cdot \env_{\mu}=\da^{\mu}{}_{\nu}$, and can then be used to specify the complete inverse transformation matrix, $X^{-1 m}{}_{\mu}=\en_{\mu}^m,\ \mu\in \cC\cup\cS$. On the other hand, the dual of the covariant basis vectors spanning the unconstrained subspace, $\env^{\,\mub}$, satisfying $\env^{\,\nub}\cdot \env_{\mub}=\da^{\mub}{}_{\nub}$, and forming another basis for the unconstrained subspace, can be use to define the extension of the $x$-matrix: $X^{\mub}{}_{m}=\en^{\mub}{}_m$.}. 

We are going to solve this problem by using coordinate transformations that reduce (\ref{Qquad},\ref{constraints-appnd}) into the problem of optimizing the $L^2$ norm for which we already derived the answer (\ref{Ww-quad}).
It is well known that there exists coordinate transformations that can take ellipses into spheres. Specifically, it is a combination of rotation and scaling, 
\be
w^{'m}=\La^m{}_nw^n \ , \mx{ such that }g _{mn}w^mw^n=\da_{mn}w^{'m}w^{'n}\ .
\ee
The constraint equations (\ref{constraints-appnd}) can now be rewritten as:
\be
x^{\mud}{}_mw^m=x^{\mud}{}_m(\La^{-1})^m{}_{n}  w^{'n}= x^{'\mud}{}_nw^{'n}=y^{\mud}\ ,
\ee 
We note that we have now written the constraint equations in terms of the $w^{'m}$ coordinates in terms of which $Q^2$ reduces to the Euclidean $L^2$ norm, and can therefore write down $w^{'m}$ values where this $L^2$ norm is minimized. According to (\ref{Ww-quad}),
\ba
w^{'m'}=y^{\mud}x^{'\nud}{}_{n'}\ga_{\mud\nud}\da^{m'n'}
\ea
where $\ga_{\mud\nud}$ is the inverse of  $\ga^{\mud\nud}$ that are defined via 
\be
\ga^{\mud\nud}=x^{'\mud}{}_{m'}x^{'\nud}{}_{n'}\da^{m'n'}\ .
\ee
We remind the readers that in terms of the primed coordinates we have an Euclidean metric. Further, since the $m', n'$ indices are summed over, the expression is invariant under coordinate transformation~\footnote{Note that $\mud,\nud$ indices don't transform, \ie, the $\env^{\mud}$ vectors, and therefore the dot products between them, $\ga^{\mud\nud}$, also don't change. More explicitly, the $x$'s and the metric transform inversely of each other to keep $\ga^{\mud\nud}$  invariant.}, so that
\be
\ga^{\mud\nud}=x^{\mud}{}_mx^{\nud}{}_ng^{mn}\ .
\ee
We can now obtain the original synapse coordinates applying inverse transformations:
\ba
w^m&=&w^{'m'}(\La^{-1})^m{}_{m'} =(\La^{-1})^m{}_{m'}y^{\mud}x^{'\nud}{}_{n'}\ga_{\mud\nud}\da^{m'n'}\non
&=&(\La^{-1})^m{}_{m'}y^{\mud}x^{\nud}{}_{n}(\La^{-1})^n{}_{n'}\ga_{\mud\nud}\da^{m'n'}\non
&=&y^{\mud}x^{\nud}{}_{n}\ga_{\mud\nud}g^{mn}\ ,
\ea
where the last equality again follows from the transformation properties of the metric. The expression for $\Qm$ can be obtained even more readily as
\be
\Qm^2=y^{\mud}y^{\nud}\ga_{\mud\nud}\ ,
\ee
where we have already explained how to obtain $\ga_{\mud\nud}$ from $g^{mn}$. Note that this expression doesn't change because $y^{\mud}$ and $\ga_{\mud\nud}$ remain invariant under coordinate transformations. So, does the $\wv_{\min}$ vector,
\be
\wv_{\min}=y^{\mud}\env_{\mud}\ .
\ee
\section{Average activity}\label{sec:average}
Consider a linear network with $\cN$ neurons feeding to a target neuron. Suppose the response of the target neuron to $\cP$ orthonormal patterns, $x^{\mu}{}_m$, are known and given by $y^{\mu}$, where $\mu=1,\dots,\cP$, as usual. Any weight vector that produces this specified input-output transformation can then be written as
\be
\wv=y^{\mud}\env_{\mud}+\eta^{\mub}\env_{\mub}\ .
\ee

Now consider an arbitrary input pattern, $z_m$, of unit norm,  
that can be expressed as a linear combination of the above input patterns:
\be
z_m=\za_{\mud}x^{\mud}{}_m\Ra \zv=\za_{\mud}\env^{\,\mud}\ ,\with 
\sum_{\mud=1}^{\cP} (\za_{\mud})^2=1\ .
\ee
The last equality follows from requiring $\zv$ be unit norm. Then the response to this pattern is given by
\be
r=y^{\mud}\za_{\mud}\ .
\ee
The average squared response to an arbitrary pattern with unit norm is then given by
\be
<r^2>={\int d^{\cP-1}\za\  (y^{\mud}\za_{\mud})^2\over \int d^{\cP-1}\za} \ .
\ee
The above integrals can be reduced to
\be
<r^2>={|\yv|^2\int_0^{\pi}d\ta\ \cos^2\ta \ S_{\cP-2}(\sin\ta) \over \int_0^{\pi}d\ta S_{\cP-2}(\sin\ta)}
\ee
where $\ta$ is the angle between $\yv$ and $\vec{\za}$, and $S_{n}(r)$ is the surface area of a $(n-1)$-sphere with radius, $r$,
\be
S_{n-1}(r)={2\pi^{n\over 2}r^{n-1}\over \Ga(n/2)}
\ee
After some simplifications, the above integrals yield
\be
<r^2>={||\yv||^2\over \cP+2}\propto\sqrt{\sum_{\mud}(y^{\mud})^2}\ .
\ee
In other words, the average  of squared responses in a network is proportional to the sum of the squared responses to an orthonormal basis set of input patterns. 
\section{$Q$-critical}
\label{sec:Qcr}
One way to obtain $Q_{\mt{cr},p}$ for the $p^{th}$ synapse from $\Qm$ is to consider an additional constrained~\footnote{Although the response for this pattern is assumed to be zero, we are going to treat this as a constrained pattern.} pattern, 
\be
x^{\new}{}_n=\da^{p}{}_n\ ,
\ee
for which the response is assumed to be zero:
\be
y^{\new}=0\ .
\ee
The new input-output transformation simply implies $w^p=0$. As argued in the main text finding the $\Qm$ that satisfies $w^p=0$ yields $Q_{\mt{cr},p}$ for the $p^{th}$ synapse. Suppose that we have identified the correct boundary according to our geometric algorithm so that we know all the semi-constrained dimensions that are effectively constrained. As before let $\mud$ denote all these dimensions along with the constrained ones. This means that $\wv_{cr}$ resides in the subspace spanned by the basis vectors, $\{\env^{\,\mud},\env^{\new}\equiv \ev^{\,p}\}$. As in our discussion in the main text, we will denote the dual to the basis $\{\env^{\,\mud}\}$ in the subspace spanned by it by $\{\env_{\cB,\mud}\}$. Let us introduce a new unit vector, $\env_{\perp}$ that is orthogonal to the subspace spanned by $\{\env^{\,\mud}\}$s, but that lies in the subspace spanned by the extended pattern vectors,  $\{\env^{\,\mud} s,\env^{\new}\equiv \ev^{\,p}\}$. In particular this means we can express $\wv_{cr}$ as
\be
\wv_{cr}=y^{\mud}\env_{\cB,\mud}+\eta^{\perp}\env_{\perp}\ .
\ee
We remind the readers that both the contravariant and covariant basis $\{\env^{\,\mud}\}$ and $\{\env_{\cB,\mud}\}$ span the same subspace, so $\wv_{cr}$ must be expressible as a linear combination of $\{\env_{\cB,\mud},\env_{\perp}\}$. Moreover, since we must satisfy $\env^{\,\mud}\cdot\wv_{cr}=y^{\mud}$ and $\env^{\,\mud}$'s are all orthogonal to $\env_{\perp}$, the coefficients of $\env_{\cB,\mud}$'s must be given by $y^{\mud}$.


Now, the new pattern vector, $\ev^{\,p}$, can also be decomposed as a projection onto the constrained subspace and a vector orthogonal to it:
\be
\ev^{\,p}=(\ev^{\,p}\cdot \env^{\,\mud})\env_{\cB,\mud}+a\env_{\perp}\ ,
\ee
Since $w^p=0$, for $\wv_{cr}$, we have
\ba
\ev^{\,p}\cdot\wv_{cr} &=& (\ev^{\,p}\cdot \env^{\,\mud})y^{\,\nud}(\env_{\cB,\nud}\cdot \env_{\cB,\mud})+a\eta_{\perp}\nonumber\\
&=& y^{\,\mud}(\ev^{\,p}\cdot \env_{\cB,\mud})+a\eta_{\perp}=0
\ea
where we have used the orthogonality relation,
\be
\env_{\cB,\mud}\cdot \env^{\,\nud}=\da_{\mud}{}^{\nud}\ .
\ee
Thus we can compute $\eta_{\perp}$ as
\be
\eta_{\perp}=-{ (\ev^{\,p}\cdot \env_{\cB,\mud})y^{\,\mud}\over a}=-{ (\ev^{\,p}\cdot \yv)\over \sqrt{g^{pp}-(\ev^{\,p}\cdot \env^{\,\mud})(\ev^{\,p}\cdot \env_{\cB,\mud})}}\ ,
\ee
where the index $p$ is not summed over and just refers to the synapse whose $Q$-critical we are trying to find.
Thus we have
\be
||\wv_{cr}||_Q^2=||\yv||_Q^2+{ (\ev^{\,p}\cdot \yv)^2\over Q^{pp}-(\ev^{\,p}\cdot \env^{\,\mud})(\ev^{\,p}\cdot \env_{\mud})}
\ee
The above can be re-expressed as
\ba
Q_{\mt{cr},p}^2&=&||\wv_{cr}||_Q^2\non
&=&||\yv||_Q^2\LT{ Q^{pp}-(\ev^{\,p}\cdot \env^{\,\mud})(\ev^{\,p}\cdot \env_{\mud})+ (\ev^{\,p}\cdot \yh)^2\over Q^{pp}-(\ev^{\,p}\cdot \env^{\,\mud})(\ev^{\,p}\cdot \env_{\mud})}\RT\ ,\non
\ea
which provides several insightful interpretations. For instance, consider first the case of $L^2$ norm, so that $q_{mn}=\da_{mn}$, and accordingly $W_{\mt{cr},p}$ can be expressed as, 
\be
W_{\mt{cr},p}^2={\ga_{\mud\nud}y^{\mud}y^{\nud}(1-\ga_{\mud\nud}x^{\mud}{}_{p}x^{\nud}{}_{p})+ \ga_{\mud\nud}x^{\mud}{}_{p}y^{\nud}\over 1-\ga_{\mud\nud}x^{\mud}{}_{p}x^{\nud}{}_{p}}\ .
\ee
We see that $W_{\mt{cr},p}$ depends on various weighted correlations. For instance, $\ga_{\mud\nud}x^{\mud}{}_{p}x^{\nud}{}_{p}$ and $\ga_{\mud\nud}y^{\mud}y^{\nud}$ are proportional to the autocorrelations of the pre- and post-synaptic neurons respectively, their responses given by $ x^{\mud}{}_{p}$ and $y^{\mud}$. In fact, as elaborated in Appendix \ref{sec:average}, the metric components in the weighted correlation ensures that the expressions represent the average activity of the neurons over all possible linear combinations of the observed response patterns. The last quantity that $W_{\mt{cr},p}$ depends on is the weighted correlation between the pre and post-synaptic neuron ($\propto  \ga_{\mud\nud}x^{\mud}{}_{p}y^{\nud}$). As the correlations increase, so does $W_{\mt{cr},p}$, as one would intuitively expect; the stronger the responses of the input and the target neurons, the more information one can hope to gain about the connectivity, and the more these two neurons are (anti-)aligned, the greater is the likelihood of being able to predict the sign of the synapse between them. This is exactly what we had observed for the special case of orthogonal patterns and indeed our expression trivially reduces to the results obtained in~\cite{Biswas22} for this special case. 
\section{Run time comparisons of different methods to compute $\Qm$}
\label{app:runtimes}
\begin{figure*}[!htbp]
	\centering
	\includegraphics[width=0.7
 \textwidth,angle=0]{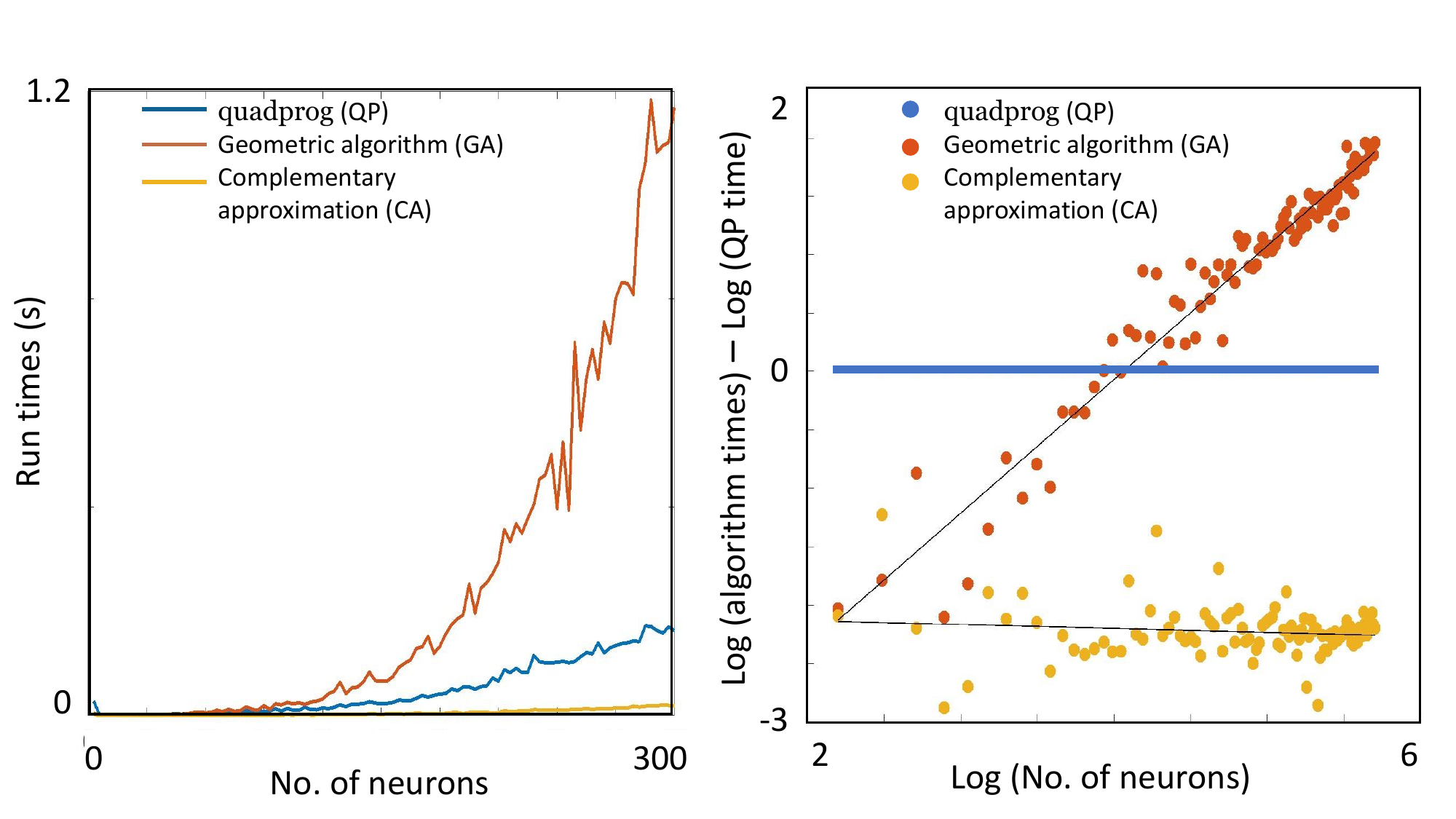}
	\caption{{\bf Comparing run times for different algorithms to calculate $\Qm$.} 
{\small  As in Fig.~\ref{fig:simulations}, we simulate $\cP=2\cN/3$ activity patterns involving $\cN$ input neurons and enforce that the target neuron responds positively to $\cC=\cN/3$ patterns and does not to the other $\cS=\cN/3$ patterns. The neuronal activity patterns and the coefficients for the quadratic function, $q_{mn}$, are chosen the same way as in the simulations in Fig.~\ref{fig:simulations}. ({\bf A}) We depict the run times required to compute $\Qm$ by three different algorithms: the default algorithm that the \texttt{quadprog}  function in Matlab uses, our geometric algorithm, and the analytic formula for the complementary approximation. \texttt{quadprog} starts to do better than the geometric algorithm once the network size reaches $\sim 100$, while the complementary approximation is faster than both of them. ({\bf B}) To compare how the run times scale with the number of neurons, we decided to plot the difference of the (natural) logarithms of the run times between the different algorithms and the \texttt{quadprog} implementation as we scale the network size. The slope of the difference between the geometric algorithm and  \texttt{quadprog} lies in the range  $1.14^{+0.06}_{-0.06}$ at 95\% confidence level, while the slope of the difference between the complementary approximation and \texttt{quadprog} lies within $-0.03^{+0.05}_{-0.01}$. We only considered  $\cN>10$ for this plot. The plot illustrates that geometric algorithm's run times scale less well than \texttt{quadprog} as the network size increases, and \texttt{quadprog} scales approximately the same way as the complementary approximation. 
\label{fig:runtimes}}	
}
\end{figure*}
\twocolumngrid
\bibliography{test}



\end{document}